\documentclass[reprint,amsmath,amssymb,aps]{revtex4-2}
\usepackage[colorlinks,bookmarks=true,citecolor=blue,linkcolor=black,urlcolor=black]{hyperref}% for figs, eqns, hrefs to have blue links
\usepackage{graphicx}% Include figure files
\usepackage{float} % Required for [H] option to force placement
\usepackage{dcolumn}% Align table columns on decimal point
\usepackage{bm}% bold math
\usepackage{diagbox}
\usepackage{ulem}
\usepackage{siunitx}

\begin{document}

\preprint{APS/123-QED}

\title{Controlled localization of anyons in a graphene quantum Hall interferometer}% Force line breaks with \\

\author{Christina E. Henzinger$^1$} 
\author{James R. Ehrets$^2$} 
\author{Rikuto Fushio$^2$}
\author{Junkai Dong$^2$}
\author{Thomas Werkmeister$^{1,3}$}
\author{Marie E. Wesson$^1$}
\author{Kenji Watanabe$^4$}
\author{Takashi Taniguchi$^5$}
\author{Ashvin Vishwanath$^2$}
\author{Bertrand I. Halperin$^2$}
\author{Amir Yacoby$^{1,2}$}
\author{Philip Kim$^{1,2}$}
\email{pkim@physics.harvard.edu}

\affiliation{$^1$John A. Paulson School of Engineering and Applied Science, Harvard University, Cambridge, MA, USA}
\affiliation{$^2$Department of Physics, Harvard University, Cambridge, MA 02138, USA}
\affiliation{$^3$Department of Applied Physics and Applied Mathematics, Columbia University, New York, NY 10027, USA}
\affiliation{$^4$Research Center for Electronic and Optical Materials, National Institute for Materials Science, 1-1 Namiki, Tsukuba 305-0044, Japan}
\affiliation{$^5$Research Center for Materials Nanoarchitectonics, National Institute for Materials Science, 1-1 Namiki, Tsukuba 305-0044, Japan}

\date{\today}% It is always \today, today,
             %  but any date may be explicitly specified

\begin{abstract}
Exchange statistics are a fundamental principle of quantum mechanics, dictating the symmetry of identical particle wavefunctions and thereby enabling emergent phenomena of many-body quantum states. The exchange-induced unitary transformation of both abelian and non-abelian anyonic wavefunctions can be probed using electronic fractional quantum Hall (FQH) interferometers, where quasiparticles propagating along the interfering FQH edge braid with those localized within the interferometer. Here, we add a gate-controlled dot/anti-dot in the center of a bilayer graphene FQH interferometer cavity to tune the number of enclosed anyons. We observe hundreds of controlled phase slips in the diagonal conductance across the interferometer for both abelian and non-abelian states, consistent with discrete changes in the localized quasiparticle population. For abelian anyons, the observed phase slips agree with the theoretically expected value. At half filling, our results suggest the interfering edge carries charge $|e^*/e| = 1/2$ abelian excitations, whereas charge $|e^*/e| = 1/4$ putative non-abelian anyons remain localized in the interferometer cavity. Controlling the population of localized $e/4$ anyons in an interferometer marks a significant milestone towards observing their non-local exchange statistics and building a fault tolerant topological qubit based on non-abelian anyon manipulation.
\end{abstract}

\maketitle

\section{Introduction}\label{sec1}
In two spatial dimensions, quasiparticles with new quantum exchange statistics emerge, beyond fermions and bosons \cite{leinaas1977theory, laughlin1, halperin1, wilczek1, jain1, stern1, halperin2}. These ``anyons" fall into two categories: the abelian anyon, where a quasiparticle exchange induces an additional nontrivial phase factor in the many-body wavefunction, and the more exotic non-abelian anyon, where exchanges of different pairs of quasiparticles lead to different ground states depending on the order of operations. Interferometry and collider experiments have both been proven fruitful methods of probing anyons in GaAs \cite{manfra1, feve1} and monolayer \cite{tom_james_braiding,andrea_third} and bilayer \cite{yuval_const_fill,yuval_even} graphene. Yet, despite experimental efforts \cite{willett, yuval_even}, observing unitary evolution from braiding non-abelian anyons remains elusive. Demonstrating non-abelian interference would be a significant step in the advancement of fault-tolerant topological quantum computing \cite{kitaev1, nayak1, nayak2, stern2, yazdani1}. 

The fractional quantum Hall (FQH) effect provides a robust experimental platform for the emergence of anyons. While most of the odd denominator FQH states host abelian anyons~\cite{jain_CF, wen2}, certain even denominator FQH states are predicted to host Ising-type non-abelian anyons. In particular, the orbital degeneracy of bilayer graphene (BLG) has lead to the observation of multiple even denominator FQH states for Landau level (LL) fillings $\nu$ satisfying $|\nu|<4$~\cite{zibrov, dean3, zhu2, assouline, yazdani2, ronen_quarter}. Graphene-based interferometers are particularly advantageous for measuring FQH states as adjacently spaced graphite gates allow tuning between FQH states at fixed magnetic fields while simultaneously acting as an efficient means to screen electrostatic coupling between bulk and edge quasiparticles~\cite{sacepe1, tom_yuval, yuval_const_fill, yuval_even, andrea_third, tom_james_braiding, finkelstein, zhu, tom_james_coupling, sacepe2, andrea_hard_soft, dean1, dean2}.

In experiments based on the Fabry-P\'erot interferometer (FPI) geometry, anyons are exchanged by performing braiding operations, whereby edge quasiparticles encircle the $N_{qp}$ localized quasiparticles in the the central cavity \cite{wen1, halperin_theory_of_fp, halperin_robustness, sorba1, eunah_kim}. An abelian quasiparticle on the interfering edge acquires two contributions to the phase as it winds around the cavity: an Aharonov-Bohm phase due to the magnetic field $B$ piercing through the cavity area $A$, as well as a contribution due to quasiparticle exchanges \cite{dassarma1, manfra1, kivelson1}, given by
\begin{align}\label{eq:1}
    \theta_{tot} = 2\pi\frac{e^*_{ie}}{e}\frac{AB}{\phi_0} - N_{qp}{\theta_{a}}
\end{align}
\noindent
where $e^*_{ie}$ is the charge of the interfering  quasiparticle that circulates  around the interferometer edge, $\phi_0=h/e$ is the flux quantum, $N_{qp}$ is the number of cavity bulk quasiparticles, and $\theta_{a}$ is the mutual braiding phase for winding the edge quasiparticle around a bulk quasiparticle. (See Supplementary Information (SI) for sign conventions employed.) For Ising-type non-abelian quasiparticles, the parity of $N_{qp}$ dictates whether the interference will be visible due to a possible rotation within the topologically degenerate ground state manifold, resulting in an even--odd effect~\cite{halperin_evenodd, shtengel, slingerland, nayak1, willett, nayak3}. Importantly, the charge of the interfering quasiparticle $e^*_{ie}$ can differ from those localized in the cavity $e^*_{loc}$.  In general, an edge  mode can carry  more than one type of  quasiparticle, but one type will produce the dominant signal in an interference experiment. This  will depend on which type tunnels most readily across the quantum point contacts (QPCs) at the ends of the interferometer region, but it may also be affected by decoherence effects along  the edges. Recent BLG experiments~\cite{yuval_even} indeed find that in even-denominator states the interfering charge can be $|e^*_{ie}/e|= 1/2$, twice the fundamental quasiparticle charge $|e^*/e|=1/4$. 
 
Properties of localized anyons can only be revealed based on Eq.~(\ref{eq:1}) if their number is changed in discrete steps while preserving coherent interference of the FQH edge states that are coupled through the pair of QPCs. To date, most anyon-interferometer measurements have relied on localized anyons trapped by disorder in unintended, random potential extrema within the cavity~\cite{manfra1, manfra2,tom_james_braiding, andrea_third}. Anyon trap occupancy is tuned by sweeping the magnetic field and gate voltages and may switch stochastically on experimental timescales~\cite{tom_james_braiding}. To control the FQH potential landscape, anti-dots (ADs) have been used in resonant tunneling experiments~\cite{goldman1, goldman2, west1, du1, diluca}. Recently, a larger AD was placed in an optical Mach-Zehnder interferometer in GaAs, leading to the observation of phase slips only in a subset of the measured abelian FQH states~\cite{heiblum1, heiblum2}.

In this work, we demonstrate highly controlled tuning of localized anyon populations using a gate-defined dot/AD structure (referred to inclusively as AD below) integrated into a BLG FPI. By adjusting the local electrostatic potential in the AD region, we demonstrate hundreds of reproducible phase slips when the interferometer cavity has fractional fillings $\nu=-1/2,-2/5,-1/3,1+1/3,1+1/2$. The slips are induced as an AD in the center of the cavity is filled one-by-one with fractional quasiparticles. In all cases, the charge of the quasiparticles extracted through the phase slip events is close to the fundamental anyon quasiparticle charge, culminating in the direct measurement of charge $|e^*_{loc}/e|=1/4$ for the presumed non-abelian anyon quasiparticles in even-denominator states. We further find that tuning the AD only generates observable phase slips at certain AD density profiles, hinting at scenarios where edges between density plateaus in the cavity are strongly renormalized by interactions.

\begin{figure*}
\centering
 \makebox[\textwidth][c]{\includegraphics[width=1\textwidth]{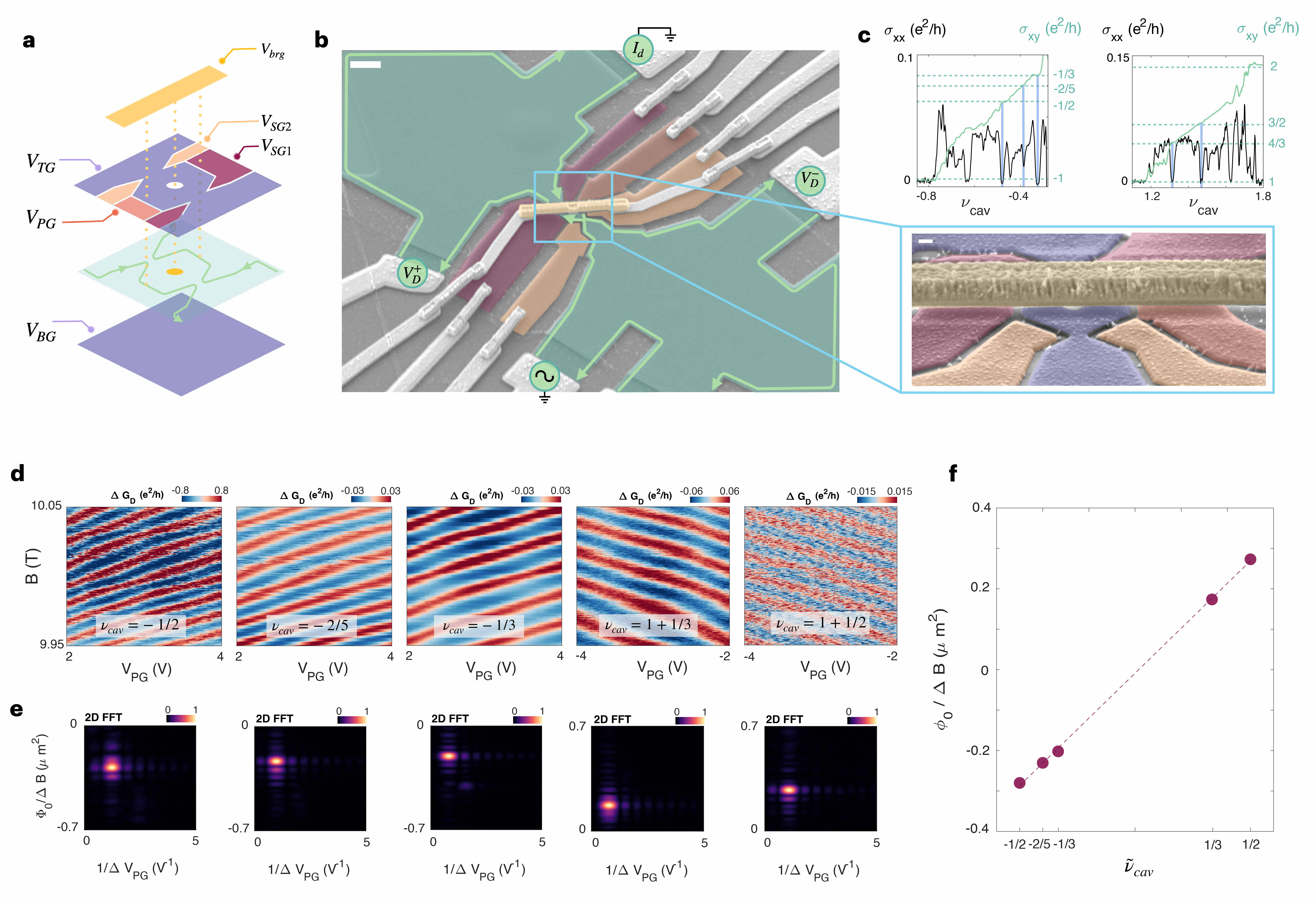}}
\caption{\textbf{Fabry-P\'erot fractional quantum Hall interferometry in bilayer graphene.} \textbf{a,} Schematic of device structure showing overhanging bridge gate, lithographically patterned top gates, graphene, and bottom gate. \textbf{b,} False-color scanning electron microscopy image of a similar device to the one measured. The 4-probe measurement setup across the interferometry cavity is illustrated in green to measure diagonal voltage $V_D$. Scale bar is $1\, \si{\mu m}$ on large scan, and $100\, \si{nm}$ on zoom-in. \textbf{c,} Hall conductivity $\sigma_{xy}$ and longitudinal conductivity $\sigma_{xx}$ measured with the contacts on the left reservoir of the device, showing states at $\nu = -1/2, -2/5, -1/3, 1+1/3, 1+1/2$. \textbf{d,} Two-dimensional map of the conductance across the interferometer $\Delta G$ as a function of magnetic field $B$ and plunger gate $V_{PG}$ showing clear Aharonov-Bohm oscillations in the five fractional states observed in (\textbf{c}). \textbf{e,} 2D FFT analysis of each panel in (\textbf{d}) indicating the magnetic field periodicity $\Phi_0/\Delta B$ and the plunger gate periodicity $1/\Delta V_{PG}$. \textbf{f,} Magnetic field periodicity $\Phi_0/\Delta B$ in each of the five fractional states used to extract the charge carried by quasiparticles on the interfering edge. The line fit yields a cavity area $A=0.56\, \si{\mu m^2}$, close to the lithographic area of $0.7\, \si{\mu m^2}$. All measurements were performed at $T=20\, \si{mK}$ and $B=9.95\, \si{T}$ unless otherwise specified.}
\label{FIG:1}
\end{figure*}

\section{\label{sec:level1} Fractional Aharonov-Bohm interference}

The BLG FPI was fabricated following the device fabrication process described previously (see Methods for detail) \cite{tom_yuval, tom_james_coupling}. As shown in Fig.~\ref{FIG:1}a, The device is operated by biasing several top graphite gates: a global top gate ($V_{TG}$), a plunger gate ($V_{PG}$), and split gates ($V_{SG}$) to define the interferometer cavity. Most importantly, the separate bridge gate ($V_{brg}$) can tune an isolated central region of the FPI cavity through a hole on the top gate region (Fig.~\ref{FIG:1}b, Methods).

\begin{figure*}
\centering
 \makebox[\textwidth][c]{\includegraphics[width=1\textwidth]{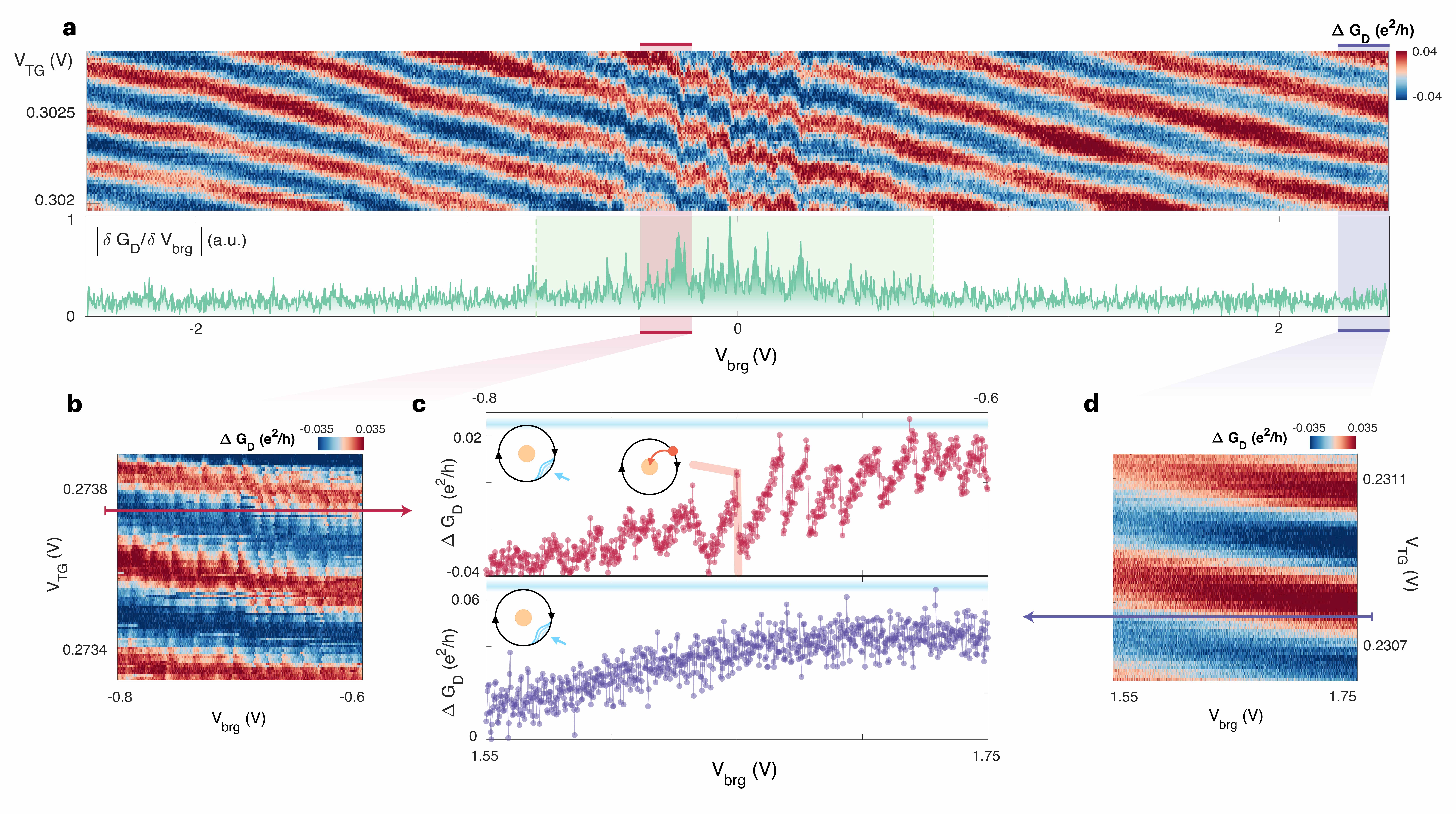}}
\caption{\textbf{Observation of discrete phase slips with the bridge gate tuning. }\textbf{a,} Conductance $\Delta G_D$ across the interferometer as a function of $V_{TG}$ and $V_{brg}$ showing Aharonov-Bohm (AB) oscillations at a fixed $V_{BG} = 0.2425\, \si{V}$. In the center of the scan, the AB oscillations turn into rapidly changing wiggles. Absolute value of the change in conductance with respect to $V_{brg}$ shows low magnitude when the oscillations change slowly, and larger magnitude when the oscillations become non-uniform. We identify two regions: one with high $|\delta G_D/\delta V_{brg}|$ in a green colored background, and the other with low $|\delta G_D/\delta V_{brg}|$ in a white background. \textbf{b,} High resolution scan of a section of the green region in (\textbf{a}), showing discrete phase slips. This scan is taken around the same anti-dot (AD) density as in (\textbf{a}) indicated in pink at fixed $V_{BG} = 0.2706\, \si{V}$. \textbf{c,} Top: linecut from (\textbf{b}) showing phase slips are induced as the bridge gate is swept. Bottom: linecut from (\textbf{d}) showing smooth AB oscillation, which is the zoomed in region marked by purple window in (\textbf{a}). This smooth background oscillation indicates that that it arises from a change in area, where the bridge gate overhangs the trenches defining the interferometry cavity. \textbf{d,} High resolution scan of a section of the white region in (\textbf{a}), showing no phase slips. This scan is taken around the same AD density as in (\textbf{a}) indicated in purple at fixed $V_{BG} = 0.3133\, \si{V}$. All measurements were performed at $T=20\, \si{mK}$ and $B=9.95\, \si{T}$.}
\label{FIG:2}
\end{figure*}

In this study, we focus on well developed BLG FQH states, in the hole-like states $\nu=-1/2, -2/5, -1/3$, and electron-like states $1+1/3, 1+1/2$ (Fig.~\ref{FIG:1}c). We first characterized the FPI with the bridge-gate voltage $V_{brg}$ held fixed. Figure~\ref{FIG:1}d shows the diagonal conductance variation $\Delta G_D$ as a function of $B$ and $V_{PG}$, while we populate the FPI cavity with FQH filling fraction $\nu_{cav}$. For each fixed $\nu_{cav}$, we observe periodic oscillations, corresponding to constructive and destructive interference, respectively. The oscillation slope reverses from positive, for $\nu_{cav}<0$, to negative, for $\nu_{cav}>0$, reflecting that fact that increasing $V_{PG}$ will decrease the cavity area for $\nu_{cav}<0$ but increase it for $\nu_{cav}>0$. These interference patterns measure the charge of each interfering edge $e^*_{ie}$, as seen in the first term of Eq.~\ref{eq:1}. 

To determine the AB period, we compute a two-dimensional fast Fourier transform (FFT) of the $B$--$V_{PG}$ maps for each $\nu_{cav}$ (Fig.~\ref{FIG:1}e), from which we extract the magnetic-field periodicity $\Delta B$ (expressed in units of $\phi_0$). As summarized in Fig.~\ref{FIG:1}f, the resulting $\phi_0/\Delta B$ values across different $\nu_{cav}$ follow a linear relationship when plotted versus the fractional edge's contribution to the filling factor $\tilde\nu_{cav}=\nu_{cav}-\text{sign}(\nu_{cav})\lfloor|\nu_{cav}|\rfloor$, agreeing with recent work~\cite{yuval_even}. Combined with the AB term in Eq.~\ref{eq:1}, this linear scaling implies $e^*_{ie}/e=\tilde \nu_{cav}$ with the estimated area $A=0.56\, \si{\mu m^2}$, comparable to the lithographically defined cavity area $0.7\, \si{\mu m^2}$ as finite edge widths and confining potential electrostatics tend to slightly decrease the area. (The sign of $e^*_{ie}$ is arbitrary, and we choose it to have the same sign as $\nu_{cav}$.) Charge $e^*_{ie}/e=-2/5$ for the hierarchical FQH state $\nu_{cav}=-2/5$ in BLG is in contrast to GaAs interferometers, where interference is consistent with the fundamental quasiparticle charge $|e^*/e|=1/5$ \cite{manfra2, heiblum1}. Discrepancies in the nature of quasiparticle tunneling at the QPCs, for example through the transmission of multi-anyon (clustered) excitations rather than single fundamental quasiparticles~\cite{jain_molec}, may account for the differences in observed charge between the two materials.

\section{\label{sec:level1} Controlled anti-dot phase slips}

\begin{figure*}
\centering
\makebox[\textwidth][c]{\includegraphics[width=1\textwidth]{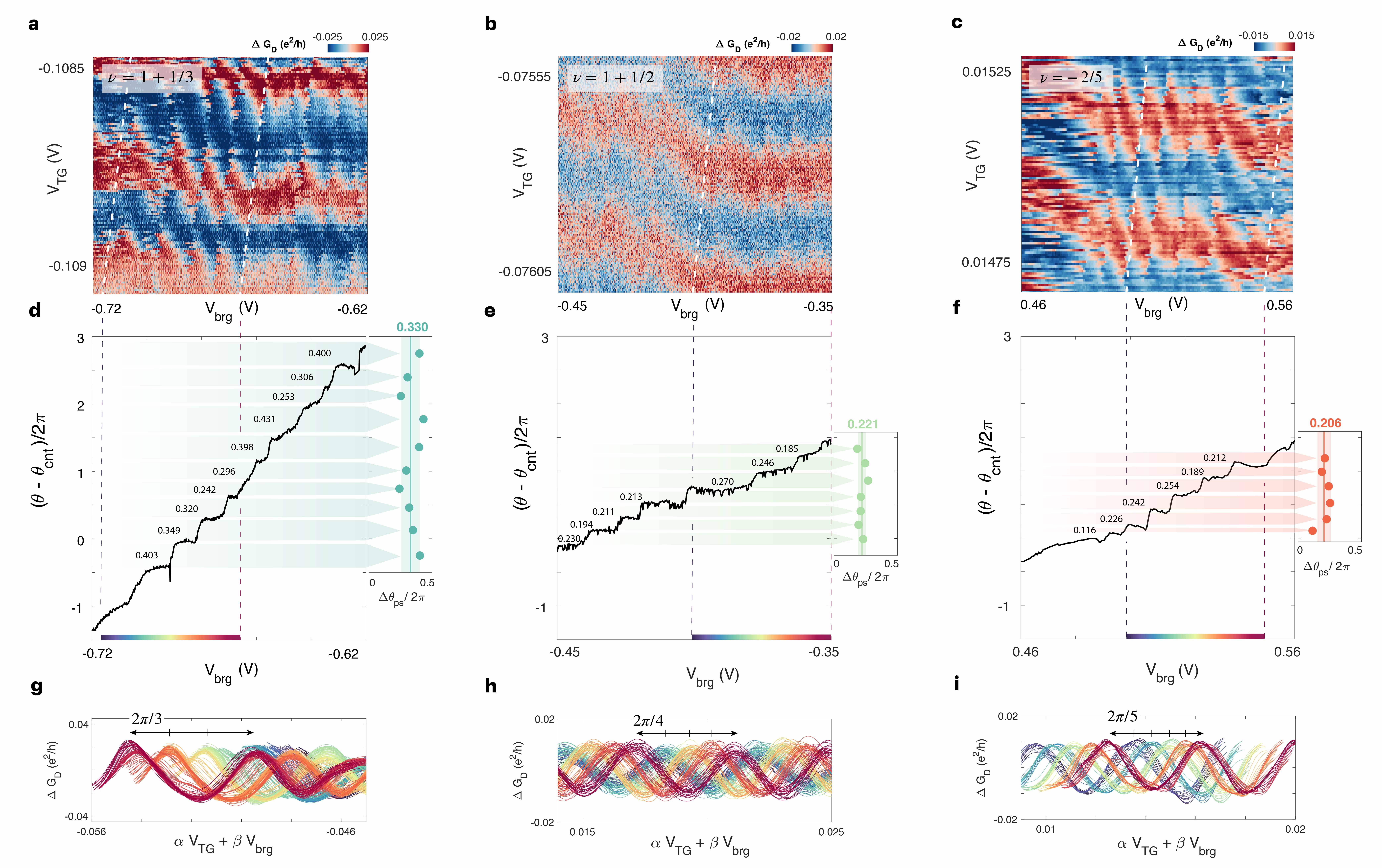}}
\caption{\textbf{Statistical phase slips in three different fractional states.} \textbf{a--c,} Two-dimensional map of the conductance oscillations $\Delta G$ of the fractional edge in (\textbf{a}) $\nu_{cav} \approx \nu_{AD} = 1+1/3$, (\textbf{b}) $\nu_{cav} \approx \nu_{AD} = 1+1/2$, and (\textbf{c}) $\nu_{cav} \approx \nu_{AD} =-2/5$ at $T =20\, \si{mK}$ and $B = 9.95\, \si{T}$ showing multiple phase slips as the bridge gate voltage is swept. \textbf{d--f,} Phase $\theta-\theta_{cnt}$ of the the 1D FFT extracted along linecuts parallel to the phase jumps in (\textbf{a--c}), respectively. A linear change in the phase $\theta_{cnt}$ corresponding to the continuous phase evolution between the phase slips is subtracted to highlight the magnitude of the discrete phase slips. \textbf{g--i,} Conductance oscillations parallel to the phase slips along $\Delta V_{brg} = 0.05\, \si{V}$. Each trace is plotted versus the combined gate coordinate $V=\alpha V_{TG}+\beta V_{brg}$, where $\alpha=\sin\varphi_s$ and $\beta=\cos\varphi_s$, with $\varphi_s=\tan^{-1}(dV_{TG}/dV_{brg})$ chosen to align the cuts with the slip trajectories. The color coding corresponds to different values of $V_{brg}$, selected to include a sufficient number of phase slips. Again, the linear phase $\theta_{cnt}$ was subtracted. We observe in (\textbf{g}) three evenly spaced sinusoids corresponding to the three phase slips in (\textbf{d}), showing phase jumps of size $\Delta \theta_{ps} = 2\pi/3$. In (\textbf{h}), four sinusoids are observed in, showing phase slips of size $\Delta \theta_{ps} = 2\pi/4$. In (\textbf{i}), five sinusoids are observed, showing phase slips of size $\Delta \theta_{ps} = 2\pi/5$.} 
\label{FIG:3}
\end{figure*}

After extracting the edge charge $e^*_{ie}/e=\tilde\nu_{cav}$ for fixed $\nu_{cav}$, we now sweep the bridge gate to tune the charge density in the AD region. Geometrically, the bridge gate overhangs the hole in the graphite top gate, thereby creating an AD of different filling $\nu_{AD}$ in the center of the cavity. In addition, the bridge gate covers part of the trenches of the patterned graphite top gates defining the FPI cavity (see Fig.~\ref{FIG:1}b). Since the edge channels are located close to these trenches, tuning the bridge gate also tunes the cavity area. These two effects can be easily disentangled as they have two different experimental signatures: (i) the modulation of the cavity area should generate continuous oscillations in the smooth AB phase similar to the $V_{PG}$ sweep we discussed in Fig.~\ref{FIG:1}d, and (ii) increasing the density of the AD should create discrete phase slips $\Delta \theta_{ps}$ as the quasiparticles populate the AD region. 

Fig.~\ref{FIG:2} shows a representative case for  $\nu_{cav} =1+1/3$ ($\tilde\nu_{cav} =1/3$). Here, we keep both $B$ and $V_{BG}$ fixed and vary $V_{TG}$ over a narrow range within $\nu_{cav} =1+1/3$ while sweeping $V_{brg}$ to tune a wide range of $\nu_{AD}$. The upper panel of Fig.~\ref{FIG:2}a shows the interference signal $\Delta G_D$ as a function of $V_{brg}$ and $V_{TG}$. Two distinct regimes appear in relation to (i) and (ii) above. In the first, $\Delta G_D$ exhibits smooth, broad oscillations as $V_{brg}$ is swept. In the second, additional fast wavy features appear on top of the same broad oscillatory background, most prominently near $V_{brg}\approx 0$. 
The contrast between the two regimes is more clearly seen in the lower panel of Fig.~\ref{FIG:2}a, which plots $|d G_D/dV_{brg}|$, where the region of abrupt fringe evolution is highlighted by the green shading. 

To resolve the underlying behavior, we perform higher-resolution scans in the rapid-change regime (ii) (Fig.~\ref{FIG:2}b) to compare with the slow-change regime (i) (Fig.~\ref{FIG:2}d). Strikingly, the apparent “rapid” variations in Fig.~\ref{FIG:2}b decompose into a sequence of discrete phase jumps separated by intervals of rapid linear variation, such that the overall  negative slope of the colored stripes appears  to be the same as the small  slope  observed in the slow-change regime. Representative line cuts (Fig.~\ref{FIG:2}c) indeed show that the sharp change of the interference phase in regime (ii) contains two components (upper panel): discrete phase slips and a continuous phase winding between slips. Their sum reconstructs (on average) the smooth AB phase evolution observed in regime (i) (lower panel). Since a phase slip region (ii) coincides with $\nu_{AD}\approx \nu_{cav}$ (see Extended Fig.~\ref{fig:many_jumps}), this strongly suggests that the discrete phase slips arise from quasiparticles entering (or exiting) the AD by tuning $V_{brg}$.

The phase slips are reproducible (Extended Fig.~\ref{fig:repeated_sweep}) and evenly spaced, with an average spacing of $|\Delta V_{brg}| = |\Delta s| =  8.7 \pm 5.5\, \si{mV}$. We find that the capacitance extracted from this spacing is consistent with individual fractional charges $|e^*_{loc}/e| = 0.2-0.4$ entering an AD of estimated radius $r_{AD} = 150 - 200\, \si{nm}$, close to the expected fractional $|e^*_{loc}/e|=1/3$ (Methods, Extended Fig.~\ref{fig:many_jumps}). This spacing is also in reasonable agreement with an estimated capacitance from electrostatic calculations between a bridge gate overhanging an etched graphite hole assuming $|e^*_{loc}/e|=1/3$ (Extended Fig.~\ref{fig:rik1}g). Furthermore, by counting the number of $|d G_D/dV_{brg}|$ peaks in regime (ii), we obtain the number of phase slips $N_{ps}$. For $\nu_{AD} \approx \nu_{cav} = 1+1/3$, we find $N_{ps}\sim 300$ consecutive phase slips (Methods, Extended Fig.~\ref{fig:many_jumps}). We therefore estimate the net number  of quasiparticles injected over the  range of $\nu_{AD}$ associated with the $\nu = 4/3$ plateau as $\Delta N_{qp} \sim 300$. On the other hand, the number of electrons forming the FQH state is $\tilde{\nu}_{AD} \Phi_{AD}$, in which $\Phi_{AD}=A_{AD}B/\phi_0$ is the total magnetic flux through the AD in units of $\phi_0$. Interestingly, we find that $|e^{*}_{loc}N_{qp}|\approx e\tilde{\nu}_{AD}\Phi_{AD}$ where we assume that $|e^{*}_{loc}/e|=1/3$ and $r_{AD} = 150 - 200\, \si{nm}$: the amount of charges it takes to destabilize the topological order in the AD is roughly equal to the amount forming the topological order itself.

Similar experiments were performed for $\nu_{AD}\approx\nu_{cav} =$ $-2/5$ and $1+1/2$, where we find series of multiple phase slips. The best estimate of the corresponding quasiparticle charge $e^*_{loc}$ jumping into the AD region are consistent with the expected elementary quasiparticle charges, where $|e^*_{loc}/e|=1/5$ and $1/4$ respectively, and inconsistent with larger charges such as $|e^*_{loc}/e|=1$ (Extended Fig.~\ref{fig:many_jumps_25}, ~\ref{fig:many_jumps_32}). Thus, we conclude that the observed phase slips originate from the AD filling with quasiparticles one-by-one, with each new quasiparticle causing a phase slip. The size of these phase slips, $\Delta \theta_{ps}$, should, therefore, allow us to measure exchange statistics of anyons in various filling.

\section{\label{sec:level1} Statistical angle measurements}

We now turn to a quantitative extraction of the phase-slip magnitude $\Delta\theta_{ps}$ for three representative FQH states, $\nu=\nu_{cav}\approx \nu_{AD}=1+1/3$, $1+1/2$, and $-2/5$, for which the interference visibility is high to allow for a robust analysis. Fig.~\ref{FIG:3}a--c show zoomed-in regions of the $\Delta G_D(V_{brg},V_{TG})$ data obtained for $\nu=1+1/3$, $1+1/2$, and $-2/5$, respectively. Specifically, we concentrate on regions that contain the most prominent $\Delta\theta_{ps}$ to analyze the jumps closest to the statistical value. Further statistical analysis of a wide range of $V_{brg}$ shows less prominent phase slips with smaller $\Delta\theta_{ps}$ consistent with effects of bulk-edge coupling (Extended Fig~\ref{fig:many_jumps}). We first define a continuous phase evolution $\theta_{cnt}$ as the average change  in phase over the linearly varying interval between successive jumps. Then, after subtracting the continuous phase evolution $\theta_{cnt}$ from the phase $\theta$ obtained via FFT of the interference signal, we extract $\Delta\theta_{ps}$ associated with each phase slip (see Methods). Fig.~\ref{FIG:3}d--f plot the resulting $\theta-\theta_{cnt}$ versus $V_{brg}$ for $\nu=1+1/3$, $1+1/2$, and $-2/5$, respectively. In each case we observe a step-like increase in the phase; $\Delta\theta_{ps}$ is extracted directly from the step height (right panels of Fig.~\ref{FIG:3}d--f).

The quantitative analysis of the measured phase slips $\Delta\theta_{ps}$ enables us to estimate the anyon braiding statistical angle $\theta_a$ using Eq.~(1). Because we have established that sweeping $V_{brg}$ changes the AD occupancy by single quasiparticles of the fundamental charge $e^*$ for each FQH state (i.e., $N_{qp}\rightarrow N_{qp}\pm 1$), the extracted $\Delta\theta_{ps}$ values can be directly mapped onto $\theta_a$.

In our FPI with an AD in the cavity, $\theta_a$ probes the mutual braiding  statistics
$\theta_a(e^*_{ie}, e^*_{loc})$, mod $2\pi$, between a localized quasiparticle of charge $e^*_{loc}$ in the AD region and an interfering edge quasiparticle of charge $e^*_{ie}$. In the case  where the two quasiparticles are identical, this determines their exchange phase, $\theta_{ex}= \theta_a/2$, mod $\pi$.    
Intuitively, this statistical angle can be viewed as an additional AB phase contributed by the localized anyons when they are treated as charge-flux composites \cite{jain_AB}. Since the flux attachment differs among Laughlin states, hierarchy states, and Moore-Read states, we analyze these cases separately below. 

For the Laughlin state at $\nu=1+1/3$, our measurement indicates that both the interfering edge and localized quasiparticles carry the fundamental charge $|e^*_{ie}/e|=|e^*_{loc}/e|=|e^*/e|=1/3$. In this case, theory predicts a braiding  phase of $|\theta^{th}_a(e/3, e/3)|/2\pi = 1/3$, in good agreement with the experimentally extracted mean phase-slip magnitudes $|\Delta\theta_{ps}|/2\pi \approx 0.330(72)$ (Fig.~\ref{FIG:3}d, left panel). This value is also consistent with prior interferometer and collider experiments in both GaAs and graphene for $|e^*/e|=1/3$ quasiparticles~\cite{manfra1, tom_james_braiding, andrea_third, heiblum1, feve1}. 
However, there are questions about the sign of $\Delta \theta_{ps}$, as will be discussed below. 

For the hierarchy FQH state at $\nu=-2/5$, the fundamental quasiparticle charge is expected to be $|e^*/e|=1/5$~\cite{jain_fqh}. However, the AB oscillations in our interferometer (Fig.~\ref{FIG:1}d) indicate that the interfering-edge excitation carries $|e^*_{ie}/e|=2/5$. Within the K-matrix framework (Methods), we therefore evaluate two possibilities for the localized quasiparticles entering the AD region: $|e^*_{loc}/e|=1/5$ or $|e^*_{loc}/e|=2/5$. These yield the theoretical exchange phases $|\theta^{th}_a(-2e/5, -e/5)|/2\pi = 1/5$ and $|\theta^{th}_a(-2e/5, -2e/5)|/2\pi = 2/5$, respectively. From Fig.~\ref{FIG:3}e, we experimentally extract $|\Delta \theta_{ps}|/2\pi = 0.206(50)$, which is consistent with the $-1/5$ expectation and therefore suggests that quasiparticles with charge $|e^*_{loc}/e|= 1/5$ anyons are being added to (or removed from) the AD region. This conclusion is also in line with recent reports of  $|e^*/e|=1/5$ populating random traps in both GaAs FPI\cite{manfra2} and Mach-Zehnder interferometers \cite{heiblum1}. 

For the non-abelian state at $\nu=1+1/2$, as we discussed in Fig.~\ref{FIG:1}e, the AB oscillations indicate an interfering-edge charge $|e^*_{ie}/e|=1/2$, even though the fundamental non-abelian quasiparticle is expected to carry $|e^*/e|=1/4$. Importantly, even when the interfering excitation is the abelian $|e^*_{ie}/e|=1/2$ mode, adding either a non-abelian quasiparticle of charge $\pm e^*$ or an abelian charge $\pm 2e^*$ excitation to the AD can still generate a phase slip, with theoretical exchange phases $|\theta^{th}_a(2e/4, e/4)|/2\pi = 1/4$ and $|\theta^{th}_a(2e/4, 2e/4)|/2\pi = 1/2$, respectively~\cite{slingerland}. Our measurement value, $|\Delta \theta_{ps}|/2\pi =0.221(30)$, is indeed consistent with the 1/4 expectation, leading us to conclude that the AD charge changes discretely by $|e^*_{loc}/e|=1/4$ in this state. Evidence for the fundamental $e/4$ charge in even-denominator FQH states has previously been obtained from single-electron transistor~\cite{yacoby_e4} and shot-noise measurements in GaAs~\cite{heiblum3} and has also been inferred from statistical analyses of interferometric phase fluctuations in BLG~\cite{yuval_even}. Achieving gate-controlled population of such non-abelian anyons in our AD-embedded interferometer is therefore an important step toward directly accessing their non-local exchange statistics.

We further illustrate the phase slips directly by plotting interference traces that span multiple slip events. Figures~\ref{FIG:3}g--h show a set of color-coded line cuts of $\Delta G_D$ taken parallel to the phase-slip direction (dashed lines) with the continuous phase subtracted. The resulting groups of color coded traces provide a clear visual demonstration of discrete phase jumps in our interferometer. For the abelian states $\nu=1+1/3$ and $-2/5$, the phase evolution can be read off as groups of shifted sinusoids: within one $2\pi$ period, the traces separate into three and five interleaved families, respectively, corresponding to phase steps of $2\pi/3$ and $2\pi/5$, respectively (Figs.~\ref{FIG:3}g and Fig.~\ref{FIG:3}i). For the putative non-abelian state at $\nu=1+1/2$, the same analysis yields four interleaved families within one period (Fig.~\ref{FIG:3}h), indicating an anomalous $\pi/2$ phase step, consistent with the discussion above.

We find that, in our experiments, the observed phase-slip sign consistently corresponds to removing quasiparticles from the interferometer cavity (equivalently, adding quasiholes), even though sweeping $V_{brg}$ tunes the local electrostatic potential in a way that loads additional quasiparticles into the AD region (Extended Fig.~\ref{fig:rik1}). A similar sign inversion has been reported in GaAs interferometer experiments in certain cases. \cite{manfra2, heiblum1}. By contrast, earlier monolayer-graphene studies saw disorder-induced phase slips in both directions \cite{tom_james_braiding, andrea_third}. In our device, however, the phase slips are consistently in the same direction, supporting the interpretation that each slip reflects the same underlying, repeatable charging event.

\begin{figure*}
\centering
\makebox[\textwidth][c]{\includegraphics[width=1\textwidth]{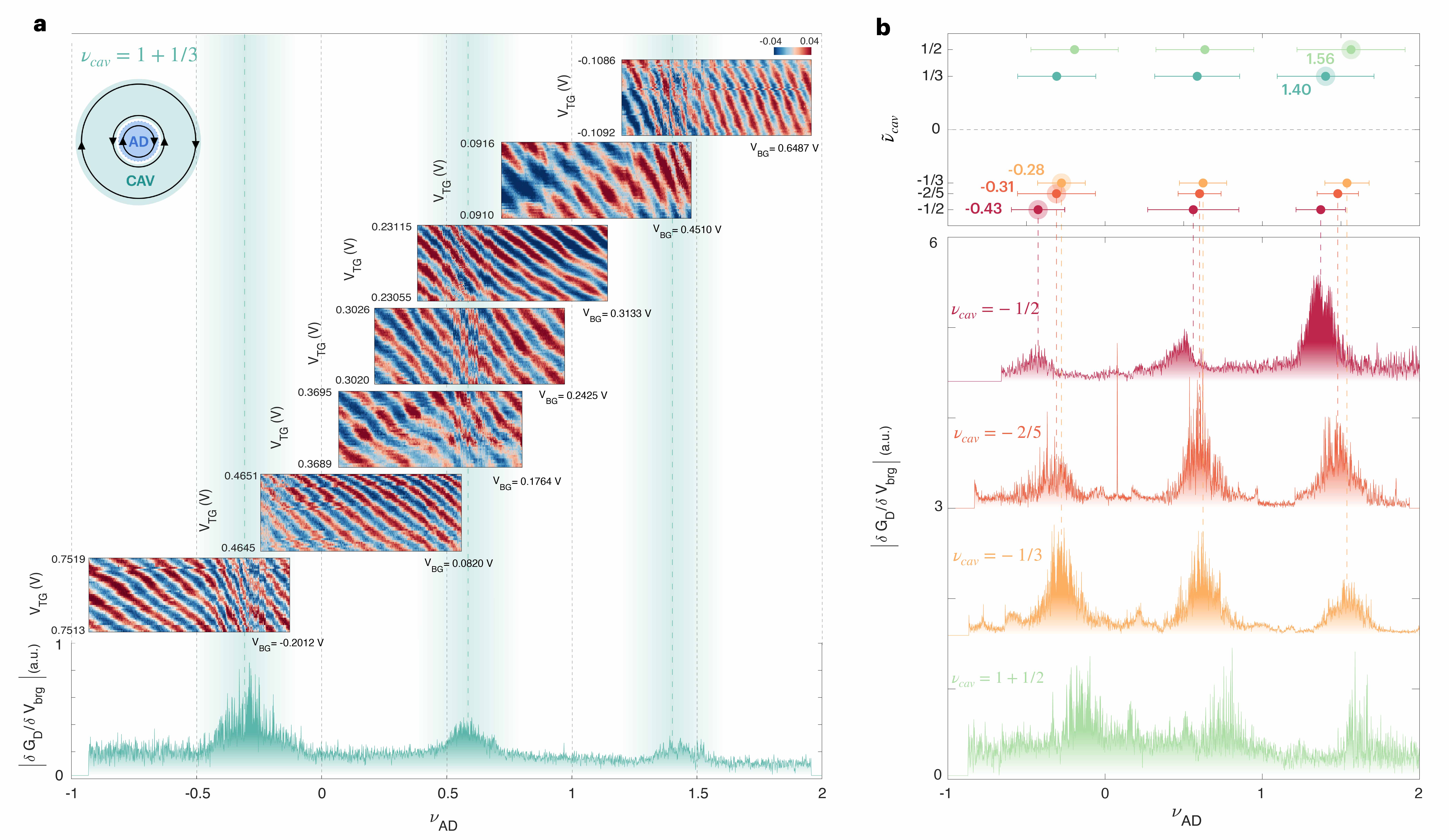}}
\caption{\textbf{Repeated filling of the anti-dot in each Landau level.} \textbf{a,} Each panel presents conductance oscillations $\Delta G_D$ of the fractional edge in $\nu_{cav} = 1+1/3$ while sweeping $V_{brg}$ and $V_{TG}$ at fixed $V_{BG}$. Multiple scans are performed at different $V_{BG}$ values, which enables additional tuning of the AD filling $\nu_{AD}$ over a wide range. We horizontally align these panels to express the abscissa in terms of the corresponding $\nu_{AD}$. Over $-1<\nu_{AD}<2$, we observe three distinctly grouped phase slip regions, one associated with each LL crossed by the AD. To highlight where the phase slip occurs, we plot $|\delta G_D/\delta V_{brg}|$ at the bottom of the graph. The inset cartoon illustrates a possible edge-reconstruction structure around the AD when $\nu_{AD}\neq\nu_{cav}$. \textbf{b,} (lower panel) $|\delta G_D/\delta V_{brg}|$ measured for four additional fractional states $\nu_{cav} =$ 1+1/2, -1/3, -2/5, and -1/2 over the same range $-1<\nu_{AD}<2$. (Upper panel) The mean (solid symbols) and standard deviation (horizontal bars) of the $\nu_{AD}$ values at which $|\delta G_D/\delta V_{brg}|$ peaks within each grouped region. These data delineate the $\nu_{AD}$ ranges where phase slips occur for each $\tilde\nu_{cav}$. Larger symbols highlight the regions near $\nu_{AD}\approx\nu_{cav}$, which are the focus of Fig.~\ref{FIG:3}. The ordering of the phase-slip regions across a given AD LL tracks the ordering of $\nu_{cav}$. All measurements were performed at $T=20\, \si{mK}$ and $B=9.95\, \si{T}$.}
\label{FIG:4}
\end{figure*}

\section{\label{sec:level1} Filling of anti-dot per Landau level}

So far, we kept $\nu_{AD}\approx \nu_{cav}$. By tuning $V_{brg}$ to a large range, we can access a wider range of $\nu_{AD}$ and eventually populate different LLs in the AD region. The filling at the center of the AD $\nu_{AD}$ can be computed from applied $V_{BG}$, $V_{TG}$, and $V_{brg}$ considering their capacitance coupling (Methods). Fig.~\ref{FIG:4}a shows $\Delta G_D$ obtained by sweeping $V_{brg}$ versus $V_{TG}$ at fixed $V_{BG}$ for each panel, maintaining $\nu_{cav}=1+1/3$. The horizontal axis ($V_{brg}$) is shifted to align the corresponding $\nu_{AD}$ axes for different panels. Within $-1<\nu_{AD}<2$, we observe two additional grouped phase-slip regions---identified as peaks in $|\delta G_D/\delta V_{brg}|$---centered near $\nu_{AD}\approx -0.3$ and $\nu_{AD}\approx 0.6$, in addition to $\nu_{AD}\approx\nu_{cav}=1+1/3$. The appearance of three distinct grouped phase-slip regions strongly suggests that they are associated with successive integer LL fillings of the AD across $-1<\nu_{AD}<2$.

Similarly, the same wide range $V_{brg}$ sweep protocol reveal three grouped phase-slip regions for other FQH states. The lower panel of Fig.~\ref{FIG:4}b summarizes the resulting phase-slip regions for $\nu_{cav}=-1/2$, $-2/5$, $-1/3$, and $1+1/2$, shown as the distribution of peaks in $|\delta G_D/\delta V_{brg}|$ versus $\nu_{AD}$ over the range $-1<\nu_{AD}<2$. As in the $\nu_{cav}=1+1/3$ case, each state exhibits three well-separated groups of peaks, indicating that the AD undergoes three successive ``fillings'' as its chemical potential traverses the three LLs in this interval. The upper panel of Fig.~\ref{FIG:4}b quantifies these locations by plotting the mean (solid symbols) and standard deviation (horizontal bars) of the $\nu_{AD}$ values at which $|\delta G_D/\delta V_{brg}|$ peaks within each grouped region. Together, these curves delineate the $\nu_{AD}$ windows where phase slips occur for each $\nu_{cav}$. Notably, the ordering of the phase-slip regions across a given LL tracks the ordering of $\nu_{cav}$: for example, the phase slips for $\nu_{cav}=-1/2$ consistently occur at lower AD fillings than those for $\nu_{cav}=-2/5$, which in turn are lower than those for $\nu_{cav}=-1/3$.

We also extract $\Delta\theta_{ps}$ at each $\nu_{AD} \neq \nu_{cav}$ and find that they are comparable to, or smaller than, the statistical phase obtained near $\nu_{AD} \approx \nu_{cav}$ (Extended Fig.~\ref{fig:all_jumps}). A natural interpretation is that these phase slips still originate from exchange statistics of the same anyon species. Here the interferometer cavity is held fixed, while tuning the AD through different LL fillings reshapes the local electrostatic potential, which can modify the edge reconstruction at both the inner and outer AD boundaries. In this picture, phase slips occur only when charging events involve the outer AD edge, which can share the same filling (and thus the same anyon type) as the surrounding interferometer cavity (see the inset in Fig.~\ref{FIG:4}a). Consequently, only within certain $\nu_{AD}$ intervals does the outer-edge structure support the appropriate quasiparticles and yield measurable exchange statistics induced phase slips.
Our observations  are  consistent with  the expected shapes of the density profiles for different combinations of $\nu_{AD}$ and $\nu_{cav}$ discussed in the  SI.

\section{\label{sec:level1} Conclusion}
In this work, we demonstrate controllable  loading of localized anyons in a bilayer-graphene Fabry–P\'erot interferometer by using a gate-defined AD embedded within the interferometer cavity. We find that the charge of the interfering quasiparticles in BLG agrees with the filling factor instead of the fundamental charge through measurements in Laughlin, hierarchy, and even-denominator states. Discrete changes in the localized anyon number produce reproducible phase slips in the FQH interference signal across five fractions, including two even-denominator fractions. The phase slips indicate that the charge of the quasiparticles is close to the fundamental anyon quasiparticle charge in all measured states: we find $|e^*/e|$ is consistent with $1/3, 1/4, 1/5$ jumping into the AD for $\nu=1+1/3, 1+1/2, -2/5$, respectively. These findings are especially encouraging for experiments targeting non-abelian braiding, since $|e^*/e|=1/4$ quasiparticles are predicted to realize non-abelian Ising anyons. Furthermore, by sweeping the AD chemical potential through successive LLs, we observe repeated anyon ``filling" behavior, pointing to a more intricate AD edge structure.

While our experiments provide insight into previously unexplored aspects of localized quasiparticles in the FQH liquid, they also raise several important open questions, showing that the underlying physics of localized anyons in the AD region is not yet fully understood. In particular, further work is needed to better understand the microscopic structure of the AD, which would help illuminate why phase slips occur at various AD fillings, as well as the sign of the phase slips. Also lacking is a proper understanding of the large negative slopes of the phase evolution observed between phase slips in the regions where the  jumps are seen.

Looking ahead, our results provide a promising route toward topological-qubit architectures. Once single-quasiparticle control of a non-abelian anyon is achieved at the edge-partitioning QPC, then even-denominator interference demonstrated here could be used as a sensitive readout probe of the quantum state of the anyonic topological-qubit.

\newpage
\section*{Methods}

\subsection{Device fabrication}

The BLG heterostructure is constructed with a standard polycarbonate dry transfer method. The final stack consists of bilayer graphene encapsulated by two hBN flakes (of 25, 30 nm), then two graphite top and bottom gates, and a top-covering hBN layer used to assist stacking. After picking up all six layers, the stack is deposited onto an SiO\textsubscript{2}/Si substrate patterned with alignment marks. The stack is subsequently annealed in vacuum for 3 hours at $300\si{\celsius}$ to remove polymer residue. The device is then etched using a $\text{CHF}_3/\text{O}_2$ reactive ion etch, and we deposit metallic contacts of Cr/Pd/Au to the graphene and graphite gates in separate steps in a thermal evaporator. The trenches defining the interferometry cavity as well as the AD hole are then etched through the cover BN to make contact with the top graphite gate with a $\text{SF}_6/\text{O}_2$ etch, and a metallic bridge gate is thermally evaporated over the hole.

\subsection{Measurement setup}
The device is measured in an Oxford MX400 dilution refrigerator with base temperature $T = 20\, \si{mK}$, achieved via custom electronic filtering as described in a previous work~\cite{tom_james_braiding}. The plunger gate, split gates, and bridge gate are all controlled via a custom-built 16-bit DAC, while the silicon backgate is set to $\pm 35\, \si{V}$ via a Keithley 2400 for measurements performed at positive and negative filling factors respectively. The top and bottom gates are controlled with two custom built 20-bit DACs for increased tuning precision and reduced noise. The graphene source contact is voltage biased with $20 \, \si{\mu V}$ via a voltage divider with a 10-20~Hz excitation generated by a SRS SR830 lock-in amplifier. The device is grounded through an Ithaco 1211 current preamplifier which measures the drain current, and voltage measurement contacts are connected to SRS SR560 voltage preamplifiers. All channels are simultaneously recorded with SR830 lock-in amplifiers synced to the excitation lock-in. See Extended Data Fig.~\ref{fig:measurement_setup} for detailed setup for various measurement schemes.

\subsection{Quantum Hall device characterization}
Bilayer graphene has an eight-fold degenerate zeroth Landau level (ZLL) due to the degeneracies of spin ($\sigma = \pm$), valley ($K,K'$) and orbital ($N=0,1$). These eight states can be tuned between by modifying the density $n_{cav} \propto \nu_{cav}$ and the displacement field $D$. Both density and displacement field can be controlled via the TG and BG following:
\begin{align*}
    n_{cav} &= V_{TG}\left[\frac{C_{TG}}{eA} \right] + V_{BG}\left[\frac{C_{BG}}{eA} \right]\\
    D &= \frac{-1}{2\epsilon_0/e}\left(V_{TG}\left[\frac{C_{TG}}{eA} \right]-V_{BG}\left[\frac{C_{BG}}{eA} \right] \right)
\end{align*}
The eight states are measured for the parameters of this experiment of $-1<\nu<2$ (Extended Fig.~\ref{fig:BG_states}). The transitions between each state are induced by the displacement field. Note that the even-denominator states $\nu=-1/2, 1+1/2$ are only present when $N=1$, where the wavefunction is distributed across both layers. We therefore choose to focus our interferometry measurements within the $N=1$ orbital index regions.

\subsection{Aharonov-Bohm interferometry measurements}
The top graphite was etched into multiple regions: the adjoining Hall bars are directly connected to the interferometry cavity (lithographically defined area $\sim0.7\, \si{\mu m^2}$) gated by the voltages applied to top gate ($V_{TG}$), the adjacent plunger gate ($V_{PG}$), and two sets of QPC-defining split gates ($V_{SG1,2}$) on either side of the cavity (see Fig.~\ref{FIG:1}a,~\ref{FIG:1}b). This allows for separate gating of the BLG layer in each region. The top gate was made continuous across each QPC in order to motivate screening of local disorder in the critical transmission regions, promoting smoother electrostatic saddle point potentials. The interferometer was initially measured at cryogenic temperatures without the lithographically defined AD to ensure its functionality. Then, a hole ($r_{AD} \sim 100\, \si{nm}$) was etched into the top gate in the center of the interferometer cavity, and a metal bridge gate was added for electrostatic tunability by the applied voltage $V_{brg}$ (Fig.~\ref{FIG:1}a, ~\ref{FIG:1}b). The density of the BLG under the hole can therefore be changed independently of the density in the rest of the interferometry cavity, creating a highly tunable AD structure.\\
\\
A change in the magnetic field populates the interferometry cavity with quasiholes as the degeneracy of the LL increases. In monolayer graphene, this manifests itself as individual phase slips puncturing a smooth AB phase \cite{tom_james_braiding, andrea_third}, similar to phase slips observed in GaAs FP interferometer~\cite{manfra3, manfra1, manfra2}. In contrast, magnetic field-induced quasiholes are added more continuously in BLG, modifying the global phase smoothly without explicit phase slips \cite{yuval_const_fill, yuval_even}. This continuous increase in quasiholes in the BLG cavity with increasing magnetic field must be compensated by an equal increase in quasiparticles by tuning the density of the cavity $n_{cav}$ with the TG to keep the total number of quasiparticles $N_{qp}$---and thus the cavity filling $\nu_{cav}$---constant according to $\nu_{cav} = n_{cav}\,\phi_0/B$ \cite{wen1, rosenow1}. The AB interferometer measurements in this work are therefore performed at constant filling $\nu_{cav}$.\\
\\
To extract the charge of the interfering edge, we first determine the periodicity  $\phi_0/\Delta B = Ae^*_{ie}/e$ by performing a 2D FFT in each filling $\nu_{cav}$ (Fig.~\ref{FIG:1}e). The linear scaling between the five states in Fig.~\ref{FIG:1}f gives us a measured area of $0.56\, \si{\mu m^2}$, which is slightly smaller than the lithographic area of $0.7\, \si{\mu m^2}$ due to the width of edge states. By performing 2D FFTs (Fig.~\ref{FIG:1}e), we observe the periodicities of $(\phi_0/\Delta B)^{\nu_{cav}}= 0.174 \, \si{\mu m^2} , 0.273\, \si{\mu m^2},  -0.202 \, \si{\mu m^2}, -0.230 \, \si{\mu m^2}, -0.280 \, \si{\mu m^2}$ for $\nu_{cav} = 1+1/3, 1+1/2, -1/3, -2/5, -1/2$, respectively.
By using  $A=0.56\, \si{\mu m^2}$ we obtain charges that follow $e^*_{ie}=\tilde{\nu}_{cav}e$, namely $e^*_{ie}/e =  0.31, 0.49, -0.36, -0.41, -0.50$ for $\nu_{cav} = 1+1/3, 1+1/2, -1/3, -2/5, -1/2$, respectively. We note that for fractional fillings $\nu$ where the fractional component is $\tilde{\nu}$ as defined in the main text, the Aharonov-Bohm phase can be modified by bulk-edge coupling  according to~\cite{manfra4, halperin3}:
\begin{align*}
    \frac{\phi_0}{\Delta B} = A\frac{e^*_{ie}}{e}\left(1-\zeta \frac{\nu}{\tilde{\nu}} \right)
\end{align*}
where the bulk-edge coupling parameter $\zeta = K_{IL}/K_I$ defines the ratio of the energy cost required to vary the area of the interfering edge $K_{IL}$ and the stiffness of this interfering edge $K_{I}$.
Since the charges we extract are very close to the theoretical charges, we conclude that there is minimal bulk-edge coupling ($\zeta$ small) when the filling is close to constant.

\subsection{Anti-dot radius and populating charge $e^*_{loc}$ relations}
 In $\nu_{cav}\approx\nu_{AD}=1+1/3$, we measure the peak-to-peak spacing in the derivative of the FFT in Extended Fig.~\ref{fig:many_jumps} to obtain $|\Delta s| = 0.0087 \pm 0.0055 \, \si{V}$. We know that the lithographically defined AD radius in the top gate is $100\, \si{nm}$, but it is expected to be larger in the BLG due to spreading of the fringe electric field lines through the $25\, \si{nm}$ thick top BN \cite{diluca}. The extracted spacing is in good agreement with our simulations of the device if charge $|e^*_{loc}/e|=1/3$ is jumping into the AD of size $r_{AD} = 150-200\, \si{nm}$ (Extended Fig.~\ref{fig:rik1}). Moreover, the slope along the phase slip line $m_{TG} = dV_{TG}/dV_{brg}$ allows us to find the capacitance per area with respect to the known capacitance per area of the TG. From bulk quantum Hall measurements, we know that $\frac{C_{TG}}{eA}$. Therefore, $\left(\frac{C}{eA}\right)_{brg} = m_{TG}\left(\frac{C}{eA}\right)_{TG} = 3.80\times 10^{14} \, \si{Fm^{-2}C^{-1}}$. Using $C_{brg} = |e^*_{loc}/\Delta s|$ from the spacing of the phase slips, we can write: $|e^*_{loc}/e |=A\cdot|\Delta s|\cdot m_{TG}\left(\frac{C}{eA}\right)_{TG}$. Again, we plug in $r_{AD} = 150-200\, \si{nm}$ to find $|e^*_{loc}/e| = 0.2-0.4$, in good agreement with quasiparticles of fractional charge $|e^*_{loc}/e| = 1/3$ populating the AD.\\
\\
 As we discussed in the text, we found that $e^*_{loc}$ has an interesting relation to the number of phase slips $N_{ps}$ corresponding to quasiparticles jumping into the AD region. To demonstrate this relation, we set the cavity to $\nu_{cav}=1+1/3$ and sweep the AD across $1.445 < \nu_{AD}<1.52$ (Extended Fig.~\ref{fig:many_jumps}c). To count the number of phase slips in this regime, we perform a 1D FFT parallel to the phase slip slope ($m_{TG}$) and take its derivative (second panel in Extended Fig.~\ref{fig:many_jumps}c). Peaks in the derivative indicate sharp changes in the phase, \textit{i.e.} a phase slip. A moving trimmed mean over a smoothing window significantly smaller than the spacing of the peaks of $0.001\, \si{V}$ is applied before the derivative is taken to remove noise between the peaks. We count $N_{ps} \geq 70$ consecutive peaks (phase slips) in this range. We see in Fig.~\ref{fig:many_jumps}a and Fig.~\ref{FIG:4} that phase slips take place over the range of fillings $1.20 < \nu_{AD}<1.64$. Assuming that the phase slips are equally spaced in this range with the spacing estimated from $1.445 < \nu_{AD}<1.52$, we thus estimate a total of $N_{ps} \sim 300$ jumps across the entire range.

 As explained in the main text, the number of electrons forming the FQH state is $\tilde{\nu}_{cav} \Phi_{AD}$, in which $\Phi_{AD}=A_{AD}B/\phi_0$ is the total magnetic flux through AD. If we assume $|e^{*}_{loc}/e|=1/3$ and $r_{AD} = 150 - 200\, \si{nm}$, we find that $|e^{*}_{loc}N_{ps}|\approx e\tilde{\nu}_{cav}\Phi_{AD}$.\\
\\
We repeat both of these analyses for $\nu_{cav} \approx \nu_{AD} = -2/5\text{ and } 1+1/2$ and find populating charges consistent with $|e^*_{loc}/e| = 1/5, 1/4$, respectively, and inconsistent with larger charges such as $|e^*_{loc}/e|=1$. (Extended Fig.~\ref{fig:many_jumps_25}, ~\ref{fig:many_jumps_32}).

\subsection{Phase Slip size Fourier analysis}
The interference phase $\Delta \theta$ may be separated into the continuous $\Delta\theta_{cnt}$ and discrete $\Delta \theta_{ps}$ changes in phase as $\Delta \theta = \Delta \theta_{cnt} + \Delta \theta_{ps}$. As discussed in the main text, the discrete phase slips are expected to arise from the statistical contributions of single quasiparticles jumping into the AD according to Eq.~(\ref{eq:1}), meaning that $\Delta\theta_{ps} =\theta_{a}$. To extract the statistical contrbution $\theta_{a}$, we must therefore follow three steps: (1) determine the total change in phase $\Delta \theta$ across a scan, (2) subtract the continuous phase to reveal the discrete steps as $\Delta \theta_{ps} = \Delta \theta - \Delta \theta_{cnt}$, and (3) measure the size of the jumps $\Delta \theta_{ps} = \theta_a$. In each combination of $\nu_{cav}$ and $\nu_{AD}$, we follow the same procedure.

\begin{enumerate}
    \item For each dataset, we first determine the slope of the phase-slip trajectories, $m_{TG}=dV_{TG}/dV_{brg}$ (white dashed lines in Fig.~\ref{FIG:3}a--c). We then perform a one-dimensional Fourier transform (FT) of $\Delta G_D$ along line cuts taken parallel to these trajectories, for each value of $V_{brg}$, following the methods in \cite{manfra1, manfra2, andrea_third, tom_james_coupling}. This ensures that each phase step is as sharp as possible. 
    \item Next, we find the average continuous phase $\langle\theta_{cnt}\rangle$ over each scan and subtract this value from the scan. To calculate $\langle\theta_{cnt}\rangle$, note that the change in the continuous phase in-between the slips is always negative while the change in discrete phase at a slip is positive. Therefore, we average over the negative values of $\Delta \theta$, after applying a moving trimmed mean (window size of $0.0005-0.002$~V, much smaller than $|\Delta s| \approx 0.01$~V, and trim percentage of 10) to smooth out noise.
    \item The phase steps obtained are not perfectly flat due to noise and slight variations in the local $\theta_{cnt}$. To extract a step height, we fit each plateau to the average phase within their respective regions. The plateau fit is performed by calculating the derivative of  $(\theta-\langle\theta_{cnt}\rangle)/2\pi$ and labeling the regions with the smallest derivatives (less than $0.5-0.25$ of the maximum derivative value) as the plateaus. The differences between these plateaus are the phase slip sizes we report in Fig.~\ref{FIG:3}d--f.
\end{enumerate}

\subsection{K-Matrix calculation of braiding phase}

In $\nu = -2/5$, quasiparticles are labeled by a vector of integers $\ell = \begin{bmatrix}
    \ell^A\\\ell^B
\end{bmatrix}$, such that their charge is given by
    $e^*/e = t^TK^{-1}\ell = \begin{bmatrix}
    1&1
\end{bmatrix} \begin{bmatrix}
    -3/5&2/5\\
    2/5&-3/5
\end{bmatrix}\begin{bmatrix}
\ell^A\\\ell^B\end{bmatrix}= -\frac{1}{5}(\ell^A+\ell^B)$. Assuming that our interfering charge is $e^*_{ie}/e = -2/5$, we consider two cases for the charge jumping into the AD: $e^*_{loc}/e = -2/5$ and $e^*_{loc}/e = -1/5$ (their antiparticles will generate the opposite phase). The braiding phase is given by:
\begin{align*}
    \theta_a(e^*_{ie}, e^*_{loc}) &= 2\pi\ell_{ie}^TK^{-1}\ell_{loc}\\ 
    &= 2\pi/5 (-3 (\ell^A_{ie}\ell^A_{loc} + \ell^B_{ie} \ell^B_{loc}) \\&+ 2 (\ell^A_{ie} \ell^B_{loc} + \ell^B_{ie} \ell^A_{loc})) \text{ mod } 2\pi\\
    &= 2\pi/5 (2 (\ell^A_{ie}\ell^A_{loc} + \ell^B_{ie} \ell^B_{loc}) \\&+ 2 (\ell^A_{ie} \ell^B_{loc} + \ell^B_{ie} \ell^A_{loc})) \text{ mod } 2\pi\\
    &= 4\pi/5 (\ell^A_{ie}+\ell^B_{ie}) (\ell^A_{loc}+\ell^B_{loc}) \text{ mod } 2\pi
\end{align*}
\textbf{Case 1: $e^*_{loc}/e = -2/5$}\\
We have: $\ell^A_{ie}+\ell^B_{ie} = 2$ and $\ell^A_{loc}+\ell^B_{loc} = 2$. Then, $\theta_a(e^*_{ie}, e^*_{loc}) = 4\pi/5\cdot 4 \text{ mod } 2\pi= -2\pi(2/5) \text{ mod } 2\pi$.\\
\textbf{Case 2: $e^*_{loc}/e = -1/5$}\\
We have: $\ell^A_{ie}+\ell^B_{ie} = 2$ and $\ell^A_{loc}+\ell^B_{loc} = 1$. Then, $\theta_a(e^*_{ie}, e^*_{loc}) = 4\pi/5\cdot 2 \text{ mod } 2\pi= -2\pi(1/5) \text{ mod } 2\pi$.\\
\\
Our experimental results agree with $|\theta_a| = 2\pi(1/5) \text{ mod } 2\pi$, leading us to conclude that $|e^*_{loc}/e| = 1/5$.

\subsection{Anti-dot filling estimation}

The filling of the anti-dot is given by:
\begin{align}\label{eq:nuAD}
    \nu_{AD} = \left(\left[\frac{C}{Ae}\right]_{brg} V_{brg} + \left[\frac{C}{Ae}\right]_{BG} V_{BG} + f(V_{TG})\right)\frac{\phi_0}{B}
\end{align}
\noindent
where $(C/eA)_{brg}$ is the capacitance per area and charge of the bridge gate to the BLG AD, $(C/eA)_{BG}$ is the capacitance per area and charge of the BG to the BLG AD, and $f(V_{TG})$ is a function of $V_{TG}$, which contains the fringe field effect of the TG to the AD region.  We know $(C/eA)_{BG} = 5.854 \times 10^{15} \, \si{Fm^{-2} C^{-1}}$ by fitting the QH plateaus in DC transport measurements. We find $(C/eA)_{brg}\sim 3.8\times 10^{14} \, \si{Fm^{-2} C^{-1}}$ from both the phase slip spacing and phase slip slope $m_{TG}$ (Methods). We know that the effect of $f(V_{TG})$ should be minimal, since geometrically the TG is not parallel to the AD and should therefore not impact the filling of the center of the AD significantly. To verify if $f(V_{TG})$ can be neglected such that the regular parallel-plate capacitor model (\ref{eq:nuAD}) holds in the AD region between the graphene, the BG, and the bridge gate, we employ three separate checks: (1) we experimentally measure by how much we need to move both the bridge and back gates $\Delta V_{brg}$ and $\Delta V_{BG}$ to keep the filling $\nu_{AD}$ constant, (2) we we ramp to a different magnetic field and repeat our experiments (Extended Fig.~\ref{fig:mult_field}), and (3) we numerically simulate the device geometry and measure the simulated AD filling $\nu_{AD}$ via a Thomas-Fermi calculation combined with a finite element method (see SI).\\
\\
\textbf{Check 1.} We want to keep $\nu_{AD}$ constant by changing $\Delta V_{brg}, \Delta V_{BG}$. We therefore have the following equation:
\begin{align}
\label{fvtg} 
    \left[\frac{C}{Ae}\right]_{brg} \Delta V_{brg} + \left[\frac{C}{Ae}\right]_{BG} \Delta V_{BG} = -f(\Delta V_{TG})
\end{align}
As the geometry of the device prevents directly measuring electrical transport through the AD, we utilize the phase slips present in the data to infer the filling factor. We begin by performing a scan along $V_{brg}$ at a constant $V_{BG}$ (Extended Fig.~\ref{fig:AD_filling_det}a). The BG is stepped by $\Delta V_{BG}$ and the scan is repeated, from which we observe that the location of the phase slips have shifted by $\Delta V_{brg}$. The location of the slips is extracted by averaging over the absolute value of the derivative of each row in each scan, and fitting this signal to a Gaussian (Extended Fig.~\ref{fig:AD_filling_det}b). The mean of the Gaussian is interpreted as the center of the phase slips. We repeat this one more time. We plot the effects of the TG on the density as $f(V_{TG}) - n_{AD} = -\left\{\frac{C}{Ae}\right\}_{brg} V_{brg}-\left\{\frac{C}{Ae}\right\}_{BG} V_{BG}$, of the BG on the density as $\left\{\frac{C}{Ae}\right\}_{BG} V_{BG}$, and of the bridge gate on the density as $\left\{\frac{C}{Ae}\right\}_{brg} V_{brg}$ (Extended Fig.~\ref{fig:AD_filling_det}c). Since our observation indicates that $n_{AD}$ remains constant across the scans and also $f(V_{TG})-n_{AD}$  remains flat, the effects of the TG on the density of the AD are significantly less than that of the BG and of the bridge. Therefore we can neglect $f(V_{TG})$ in Eqn.~\ref{eq:nuAD}.\\
\\
\textbf{Check 2.} To verify that the filling we extract is correct, we ramp to a magnetic field $B=7.95 \, \si{ T}$ that is significantly different from that of the experiments in the main text, $B=9.95 \, \si{ T}$, and repeat our experiments (Extended Fig.~\ref{fig:mult_field}). The filling $\nu_{AD}$ should depend on field $B$ according to $\nu_{AD} = n_{AD}\phi_0/B$. We tune the device to $\nu_{cav}=-1/3, -2/5, -1/2$ and sweep the density in the AD by changing the bridge gate in each scan, and the BG between scans, similarly to the method used to obtain Fig. 4. We observe the phase slips do occur at similar ranges of $\nu_{AD}$ at $B = 7.95 \, \si{ T}$ (computed by the above method) as at $B = 9.95 \, \si{ T}$. Since the data taken at $B = 7.95$ are in different combination of $V_{brg}$, $V_{TG}$ and $V_{BG}$ scan ranges compared to those of $B = 9.95 \, \si{ T}$, similar scaling behaviors observed in $\nu_{AD}$ confirms our AD filling assignments discussed above.\\
\\
\textbf{Check 3.} Our numerical Thomas-Fermi simulations of the device combined with a finite element method further confirm our filling assignment, as detailed in SI.

\section*{End notes}
\textbf{Acknowledgments.} We thank Alexandre Assouline and Andrea Young for providing data from their thermodynamic measurements for our simulations. \\

\textbf{Funding.} The major part of the experimental work for measurement, characterization and analysis is supported by the US Department of Energy (DOE) (DE-SC0012260). C.E.H and J.R.E. acknowledge support from ARO MURI (N00014-21-1-2537) for sample preparation and device fabrications. A.V. and J.D. acknowledge funding from NSF DMR-2220703 and the Simons Collaboration on Ultra-Quantum Matter,
which is a grant from the Simons Foundation (651440, A.V.). AY was sponsored by the Army Research Office under award W911NF-21-2-0147 and by the Gordon and Betty Moore Foundation grant GBMF 12762. K.W. and T.T. acknowledge support from the JSPS KAKENHI (Grant Nos. 21H05233 and 23H02052), the CREST (JPMJCR24A5), JST and World Premier International Research Center Initiative (WPI), MEXT, Japan. Nanofabrication was performed at the Center for Nanoscale Systems at Harvard, supported in part by an NSF NNIN award ECS- 00335765.\\

\textbf{Author contributions.} C.E.H., J.R.E., T.W., and P.K. conceived the experiment. R.F. prepared the stack. K.W. and T.T. provided the hBN crystals. C.E.H. fabricated the device. J.R.E, T.W., M.E.W., and A.Y. built the measurement circuit. C.E.H. conducted the measurements with inputs from J.R.E.. C.E.H. and P.K. analyzed the data with inputs from J.R.E.. R.F., J.D., and B.I.H. performed the theoretical analysis. T.W., M.E.W., A.V., and A.Y. collaborated on discussions and analysis. C.E.H. and P.K. wrote the paper, with input from all authors.

\bibliography{main}

%apsrev4-2.bst 2019-01-14 (MD) hand-edited version of apsrev4-1.bst
%Control: key (0)
%Control: author (8) initials jnrlst
%Control: editor formatted (1) identically to author
%Control: production of article title (0) allowed
%Control: page (0) single
%Control: year (1) truncated
%Control: production of eprint (0) enabled
\begin{thebibliography}{64}%
\makeatletter
\providecommand \@ifxundefined [1]{%
 \@ifx{#1\undefined}
}%
\providecommand \@ifnum [1]{%
 \ifnum #1\expandafter \@firstoftwo
 \else \expandafter \@secondoftwo
 \fi
}%
\providecommand \@ifx [1]{%
 \ifx #1\expandafter \@firstoftwo
 \else \expandafter \@secondoftwo
 \fi
}%
\providecommand \natexlab [1]{#1}%
\providecommand \enquote  [1]{``#1''}%
\providecommand \bibnamefont  [1]{#1}%
\providecommand \bibfnamefont [1]{#1}%
\providecommand \citenamefont [1]{#1}%
\providecommand \href@noop [0]{\@secondoftwo}%
\providecommand \href [0]{\begingroup \@sanitize@url \@href}%
\providecommand \@href[1]{\@@startlink{#1}\@@href}%
\providecommand \@@href[1]{\endgroup#1\@@endlink}%
\providecommand \@sanitize@url [0]{\catcode `\\12\catcode `\$12\catcode `\&12\catcode `\#12\catcode `\^12\catcode `\_12\catcode `\%12\relax}%
\providecommand \@@startlink[1]{}%
\providecommand \@@endlink[0]{}%
\providecommand \url  [0]{\begingroup\@sanitize@url \@url }%
\providecommand \@url [1]{\endgroup\@href {#1}{\urlprefix }}%
\providecommand \urlprefix  [0]{URL }%
\providecommand \Eprint [0]{\href }%
\providecommand \doibase [0]{https://doi.org/}%
\providecommand \selectlanguage [0]{\@gobble}%
\providecommand \bibinfo  [0]{\@secondoftwo}%
\providecommand \bibfield  [0]{\@secondoftwo}%
\providecommand \translation [1]{[#1]}%
\providecommand \BibitemOpen [0]{}%
\providecommand \bibitemStop [0]{}%
\providecommand \bibitemNoStop [0]{.\EOS\space}%
\providecommand \EOS [0]{\spacefactor3000\relax}%
\providecommand \BibitemShut  [1]{\csname bibitem#1\endcsname}%
\let\auto@bib@innerbib\@empty
%</preamble>
\bibitem [{\citenamefont {Leinaas}\ and\ \citenamefont {Myrheim}(1977)}]{leinaas1977theory}%
  \BibitemOpen
  \bibfield  {author} {\bibinfo {author} {\bibfnamefont {J.~M.}\ \bibnamefont {Leinaas}}\ and\ \bibinfo {author} {\bibfnamefont {J.}~\bibnamefont {Myrheim}},\ }\bibfield  {title} {\bibinfo {title} {On the theory of identical particles},\ }\href@noop {} {\bibfield  {journal} {\bibinfo  {journal} {Il Nuovo Cimento B (1971-1996)}\ }\textbf {\bibinfo {volume} {37}},\ \bibinfo {pages} {1} (\bibinfo {year} {1977})}\BibitemShut {NoStop}%
\bibitem [{\citenamefont {Laughin}(1983)}]{laughlin1}%
  \BibitemOpen
  \bibfield  {author} {\bibinfo {author} {\bibfnamefont {R.~B.}\ \bibnamefont {Laughin}},\ }\bibfield  {title} {\bibinfo {title} {Anomalous quantum hall effect: An incompressible quantum fluid with fractionally charged excitations},\ }\href@noop {} {\bibfield  {journal} {\bibinfo  {journal} {Phys. Rev. Lett.}\ }\textbf {\bibinfo {volume} {50}},\ \bibinfo {pages} {1395} (\bibinfo {year} {1983})}\BibitemShut {NoStop}%
\bibitem [{\citenamefont {Halperin}(1984)}]{halperin1}%
  \BibitemOpen
  \bibfield  {author} {\bibinfo {author} {\bibfnamefont {B.~I.}\ \bibnamefont {Halperin}},\ }\bibfield  {title} {\bibinfo {title} {Statistics of quasiparticles and the hierarchy of fractional quantized hall states},\ }\href@noop {} {\bibfield  {journal} {\bibinfo  {journal} {Phys. Rev. Lett.}\ }\textbf {\bibinfo {volume} {52}},\ \bibinfo {pages} {1583} (\bibinfo {year} {1984})}\BibitemShut {NoStop}%
\bibitem [{\citenamefont {Arovas}\ \emph {et~al.}(1984)\citenamefont {Arovas}, \citenamefont {Schrieffer},\ and\ \citenamefont {Wilczek}}]{wilczek1}%
  \BibitemOpen
  \bibfield  {author} {\bibinfo {author} {\bibfnamefont {D.}~\bibnamefont {Arovas}}, \bibinfo {author} {\bibfnamefont {J.}~\bibnamefont {Schrieffer}},\ and\ \bibinfo {author} {\bibfnamefont {F.}~\bibnamefont {Wilczek}},\ }\bibfield  {title} {\bibinfo {title} {Fractional statisics and the quantum hall effect},\ }\href@noop {} {\bibfield  {journal} {\bibinfo  {journal} {Phys. Rev. Lett.}\ }\textbf {\bibinfo {volume} {53}},\ \bibinfo {pages} {722} (\bibinfo {year} {1984})}\BibitemShut {NoStop}%
\bibitem [{\citenamefont {Jeon}\ \emph {et~al.}(2003)\citenamefont {Jeon}, \citenamefont {Graham},\ and\ \citenamefont {Jain}}]{jain1}%
  \BibitemOpen
  \bibfield  {author} {\bibinfo {author} {\bibfnamefont {G.~S.}\ \bibnamefont {Jeon}}, \bibinfo {author} {\bibfnamefont {K.~L.}\ \bibnamefont {Graham}},\ and\ \bibinfo {author} {\bibfnamefont {J.~K.}\ \bibnamefont {Jain}},\ }\bibfield  {title} {\bibinfo {title} {Fractional statistics in the fractional quantum hall effect},\ }\href@noop {} {\bibfield  {journal} {\bibinfo  {journal} {Phys. Rev. Lett.}\ }\textbf {\bibinfo {volume} {91}},\ \bibinfo {pages} {036801} (\bibinfo {year} {2003})}\BibitemShut {NoStop}%
\bibitem [{\citenamefont {Stern}(2008)}]{stern1}%
  \BibitemOpen
  \bibfield  {author} {\bibinfo {author} {\bibfnamefont {A.}~\bibnamefont {Stern}},\ }\bibfield  {title} {\bibinfo {title} {Anyons and the quantum hall effect—a pedagogical review},\ }\href@noop {} {\bibfield  {journal} {\bibinfo  {journal} {Annals of Physics}\ }\textbf {\bibinfo {volume} {323}},\ \bibinfo {pages} {204} (\bibinfo {year} {2008})}\BibitemShut {NoStop}%
\bibitem [{\citenamefont {Feldman}\ and\ \citenamefont {Halperin}(2021{\natexlab{a}})}]{halperin2}%
  \BibitemOpen
  \bibfield  {author} {\bibinfo {author} {\bibfnamefont {D.~E.}\ \bibnamefont {Feldman}}\ and\ \bibinfo {author} {\bibfnamefont {B.~I.}\ \bibnamefont {Halperin}},\ }\bibfield  {title} {\bibinfo {title} {Fractional charge and fractional statistics in the quantum hall effects},\ }\href@noop {} {\bibfield  {journal} {\bibinfo  {journal} {Reports on Progress in Physics}\ }\textbf {\bibinfo {volume} {84}},\ \bibinfo {pages} {076501} (\bibinfo {year} {2021}{\natexlab{a}})}\BibitemShut {NoStop}%
\bibitem [{\citenamefont {Nakamura}\ \emph {et~al.}(2020)\citenamefont {Nakamura}, \citenamefont {Liang}, \citenamefont {Gardner},\ and\ \citenamefont {Manfra}}]{manfra1}%
  \BibitemOpen
  \bibfield  {author} {\bibinfo {author} {\bibfnamefont {J.}~\bibnamefont {Nakamura}}, \bibinfo {author} {\bibfnamefont {S.}~\bibnamefont {Liang}}, \bibinfo {author} {\bibfnamefont {G.~C.}\ \bibnamefont {Gardner}},\ and\ \bibinfo {author} {\bibfnamefont {M.~J.}\ \bibnamefont {Manfra}},\ }\bibfield  {title} {\bibinfo {title} {Direct observation of anyonic braiding statistics},\ }\href@noop {} {\bibfield  {journal} {\bibinfo  {journal} {Nature Physics}\ }\textbf {\bibinfo {volume} {16}},\ \bibinfo {pages} {931} (\bibinfo {year} {2020})}\BibitemShut {NoStop}%
\bibitem [{\citenamefont {Bartolomei}\ \emph {et~al.}(2020)\citenamefont {Bartolomei}, \citenamefont {Kumar}, \citenamefont {Bisognin}, \citenamefont {Marguerite}, \citenamefont {Berroir}, \citenamefont {Bocquillon}, \citenamefont {Placais}, \citenamefont {Cavanna}, \citenamefont {Dong}, \citenamefont {Gennser}, \citenamefont {Jin},\ and\ \citenamefont {Feve}}]{feve1}%
  \BibitemOpen
  \bibfield  {author} {\bibinfo {author} {\bibfnamefont {H.}~\bibnamefont {Bartolomei}}, \bibinfo {author} {\bibfnamefont {M.}~\bibnamefont {Kumar}}, \bibinfo {author} {\bibfnamefont {R.}~\bibnamefont {Bisognin}}, \bibinfo {author} {\bibfnamefont {A.}~\bibnamefont {Marguerite}}, \bibinfo {author} {\bibfnamefont {J.-M.}\ \bibnamefont {Berroir}}, \bibinfo {author} {\bibfnamefont {E.}~\bibnamefont {Bocquillon}}, \bibinfo {author} {\bibfnamefont {B.}~\bibnamefont {Placais}}, \bibinfo {author} {\bibfnamefont {A.}~\bibnamefont {Cavanna}}, \bibinfo {author} {\bibfnamefont {Q.}~\bibnamefont {Dong}}, \bibinfo {author} {\bibfnamefont {U.}~\bibnamefont {Gennser}}, \bibinfo {author} {\bibfnamefont {Y.}~\bibnamefont {Jin}},\ and\ \bibinfo {author} {\bibfnamefont {G.}~\bibnamefont {Feve}},\ }\bibfield  {title} {\bibinfo {title} {Fractional statistics in anyon collisions},\ }\href@noop {} {\bibfield  {journal} {\bibinfo  {journal} {Science}\ }\textbf {\bibinfo {volume} {368}},\ \bibinfo {pages} {173} (\bibinfo {year}
  {2020})}\BibitemShut {NoStop}%
\bibitem [{\citenamefont {Werkmeister}\ \emph {et~al.}(2025)\citenamefont {Werkmeister}, \citenamefont {Ehrets}, \citenamefont {Wesson}, \citenamefont {Najafabadi}, \citenamefont {Watanabe}, \citenamefont {Taniguchi}, \citenamefont {Halperin}, \citenamefont {Yacoby},\ and\ \citenamefont {Kim}}]{tom_james_braiding}%
  \BibitemOpen
  \bibfield  {author} {\bibinfo {author} {\bibfnamefont {T.}~\bibnamefont {Werkmeister}}, \bibinfo {author} {\bibfnamefont {J.~R.}\ \bibnamefont {Ehrets}}, \bibinfo {author} {\bibfnamefont {M.~E.}\ \bibnamefont {Wesson}}, \bibinfo {author} {\bibfnamefont {D.~H.}\ \bibnamefont {Najafabadi}}, \bibinfo {author} {\bibfnamefont {K.}~\bibnamefont {Watanabe}}, \bibinfo {author} {\bibfnamefont {T.}~\bibnamefont {Taniguchi}}, \bibinfo {author} {\bibfnamefont {B.~I.}\ \bibnamefont {Halperin}}, \bibinfo {author} {\bibfnamefont {A.}~\bibnamefont {Yacoby}},\ and\ \bibinfo {author} {\bibfnamefont {P.}~\bibnamefont {Kim}},\ }\bibfield  {title} {\bibinfo {title} {Anyon braiding and telegraph noise in a graphene interferometer},\ }\href@noop {} {\bibfield  {journal} {\bibinfo  {journal} {Science}\ }\textbf {\bibinfo {volume} {388}},\ \bibinfo {pages} {730} (\bibinfo {year} {2025})}\BibitemShut {NoStop}%
\bibitem [{\citenamefont {Samuelson}\ \emph {et~al.}(2025{\natexlab{a}})\citenamefont {Samuelson}, \citenamefont {Cohen}, \citenamefont {Wang}, \citenamefont {Blanch}, \citenamefont {Taniguchi}, \citenamefont {Watanabe}, \citenamefont {Zaletel},\ and\ \citenamefont {Young}}]{andrea_third}%
  \BibitemOpen
  \bibfield  {author} {\bibinfo {author} {\bibfnamefont {N.~L.}\ \bibnamefont {Samuelson}}, \bibinfo {author} {\bibfnamefont {L.~A.}\ \bibnamefont {Cohen}}, \bibinfo {author} {\bibfnamefont {W.}~\bibnamefont {Wang}}, \bibinfo {author} {\bibfnamefont {S.}~\bibnamefont {Blanch}}, \bibinfo {author} {\bibfnamefont {T.}~\bibnamefont {Taniguchi}}, \bibinfo {author} {\bibfnamefont {K.}~\bibnamefont {Watanabe}}, \bibinfo {author} {\bibfnamefont {M.~P.}\ \bibnamefont {Zaletel}},\ and\ \bibinfo {author} {\bibfnamefont {A.~F.}\ \bibnamefont {Young}},\ }\href {https://arxiv.org/abs/2403.19628} {\bibinfo {title} {Slow quasiparticle dynamics and anyonic statistics in a fractional quantum hall fabry-p\'erot interferometer}} (\bibinfo {year} {2025}{\natexlab{a}}),\ \Eprint {https://arxiv.org/abs/2403.19628} {arXiv:2403.19628 [cond-mat.mes-hall]} \BibitemShut {NoStop}%
\bibitem [{\citenamefont {Kim}\ \emph {et~al.}(2024)\citenamefont {Kim}, \citenamefont {Dev}, \citenamefont {Kumar}, \citenamefont {Ilin}, \citenamefont {Haug}, \citenamefont {Bhardwaj}, \citenamefont {Hong}, \citenamefont {Watanabe}, \citenamefont {Taniguchi}, \citenamefont {Stern},\ and\ \citenamefont {Ronen}}]{yuval_const_fill}%
  \BibitemOpen
  \bibfield  {author} {\bibinfo {author} {\bibfnamefont {J.}~\bibnamefont {Kim}}, \bibinfo {author} {\bibfnamefont {H.}~\bibnamefont {Dev}}, \bibinfo {author} {\bibfnamefont {R.}~\bibnamefont {Kumar}}, \bibinfo {author} {\bibfnamefont {A.}~\bibnamefont {Ilin}}, \bibinfo {author} {\bibfnamefont {A.}~\bibnamefont {Haug}}, \bibinfo {author} {\bibfnamefont {V.}~\bibnamefont {Bhardwaj}}, \bibinfo {author} {\bibfnamefont {C.}~\bibnamefont {Hong}}, \bibinfo {author} {\bibfnamefont {K.}~\bibnamefont {Watanabe}}, \bibinfo {author} {\bibfnamefont {T.}~\bibnamefont {Taniguchi}}, \bibinfo {author} {\bibfnamefont {A.}~\bibnamefont {Stern}},\ and\ \bibinfo {author} {\bibfnamefont {Y.}~\bibnamefont {Ronen}},\ }\bibfield  {title} {\bibinfo {title} {Aharonov–bohm interference and statistical phase-jump evolution in fractional quantum hall states in bilayer graphene},\ }\href@noop {} {\bibfield  {journal} {\bibinfo  {journal} {Nature Nanotechnology}\ }\textbf {\bibinfo {volume} {19}},\ \bibinfo {pages} {1619} (\bibinfo
  {year} {2024})}\BibitemShut {NoStop}%
\bibitem [{\citenamefont {Kim}\ \emph {et~al.}(2026)\citenamefont {Kim}, \citenamefont {Dev}, \citenamefont {Shaer}, \citenamefont {Kumar}, \citenamefont {Ilin}, \citenamefont {Haug}, \citenamefont {Iskoz}, \citenamefont {Watanabe}, \citenamefont {Taniguchi}, \citenamefont {Mross}, \citenamefont {Stern},\ and\ \citenamefont {Ronen}}]{yuval_even}%
  \BibitemOpen
  \bibfield  {author} {\bibinfo {author} {\bibfnamefont {J.}~\bibnamefont {Kim}}, \bibinfo {author} {\bibfnamefont {H.}~\bibnamefont {Dev}}, \bibinfo {author} {\bibfnamefont {A.}~\bibnamefont {Shaer}}, \bibinfo {author} {\bibfnamefont {R.}~\bibnamefont {Kumar}}, \bibinfo {author} {\bibfnamefont {A.}~\bibnamefont {Ilin}}, \bibinfo {author} {\bibfnamefont {A.}~\bibnamefont {Haug}}, \bibinfo {author} {\bibfnamefont {S.}~\bibnamefont {Iskoz}}, \bibinfo {author} {\bibfnamefont {K.}~\bibnamefont {Watanabe}}, \bibinfo {author} {\bibfnamefont {T.}~\bibnamefont {Taniguchi}}, \bibinfo {author} {\bibfnamefont {D.~F.}\ \bibnamefont {Mross}}, \bibinfo {author} {\bibfnamefont {A.}~\bibnamefont {Stern}},\ and\ \bibinfo {author} {\bibfnamefont {Y.}~\bibnamefont {Ronen}},\ }\bibfield  {title} {\bibinfo {title} {Aharonov–bohm interference in even-denominator fractional quantum hall states},\ }\href@noop {} {\bibfield  {journal} {\bibinfo  {journal} {Nature}\ }\textbf {\bibinfo {volume} {649}},\ \bibinfo {pages} {323}
  (\bibinfo {year} {2026})}\BibitemShut {NoStop}%
\bibitem [{\citenamefont {Willett}\ \emph {et~al.}(2023)\citenamefont {Willett}, \citenamefont {Shtengel}, \citenamefont {Nayak}, \citenamefont {Pfeiffer}, \citenamefont {Chung}, \citenamefont {Peabody}, \citenamefont {Baldwin},\ and\ \citenamefont {West}}]{willett}%
  \BibitemOpen
  \bibfield  {author} {\bibinfo {author} {\bibfnamefont {R.~L.}\ \bibnamefont {Willett}}, \bibinfo {author} {\bibfnamefont {K.}~\bibnamefont {Shtengel}}, \bibinfo {author} {\bibfnamefont {C.}~\bibnamefont {Nayak}}, \bibinfo {author} {\bibfnamefont {L.~N.}\ \bibnamefont {Pfeiffer}}, \bibinfo {author} {\bibfnamefont {Y.~J.}\ \bibnamefont {Chung}}, \bibinfo {author} {\bibfnamefont {M.~L.}\ \bibnamefont {Peabody}}, \bibinfo {author} {\bibfnamefont {K.~W.}\ \bibnamefont {Baldwin}},\ and\ \bibinfo {author} {\bibfnamefont {K.~W.}\ \bibnamefont {West}},\ }\bibfield  {title} {\bibinfo {title} {Interference measurements of non-abelian $e/4$ \& abelian $e/2$ quasiparticle braiding},\ }\href@noop {} {\bibfield  {journal} {\bibinfo  {journal} {Physical Review X}\ }\textbf {\bibinfo {volume} {13}},\ \bibinfo {pages} {011028} (\bibinfo {year} {2023})}\BibitemShut {NoStop}%
\bibitem [{\citenamefont {Kitaev}(2006)}]{kitaev1}%
  \BibitemOpen
  \bibfield  {author} {\bibinfo {author} {\bibfnamefont {A.}~\bibnamefont {Kitaev}},\ }\bibfield  {title} {\bibinfo {title} {Anyons in an exactly solved model and beyond},\ }\href@noop {} {\bibfield  {journal} {\bibinfo  {journal} {Annals of Physics}\ }\textbf {\bibinfo {volume} {321}},\ \bibinfo {pages} {2} (\bibinfo {year} {2006})}\BibitemShut {NoStop}%
\bibitem [{\citenamefont {Nayak}\ \emph {et~al.}(2008)\citenamefont {Nayak}, \citenamefont {Simon}, \citenamefont {Stern}, \citenamefont {Freedman},\ and\ \citenamefont {Sarma}}]{nayak1}%
  \BibitemOpen
  \bibfield  {author} {\bibinfo {author} {\bibfnamefont {C.}~\bibnamefont {Nayak}}, \bibinfo {author} {\bibfnamefont {S.~H.}\ \bibnamefont {Simon}}, \bibinfo {author} {\bibfnamefont {A.}~\bibnamefont {Stern}}, \bibinfo {author} {\bibfnamefont {M.}~\bibnamefont {Freedman}},\ and\ \bibinfo {author} {\bibfnamefont {S.~D.}\ \bibnamefont {Sarma}},\ }\bibfield  {title} {\bibinfo {title} {Non-abelian anyons and topological quantum computation},\ }\href@noop {} {\bibfield  {journal} {\bibinfo  {journal} {Rev. Mod. Phys.}\ }\textbf {\bibinfo {volume} {80}},\ \bibinfo {pages} {1083} (\bibinfo {year} {2008})}\BibitemShut {NoStop}%
\bibitem [{\citenamefont {Sarma}\ \emph {et~al.}(2015)\citenamefont {Sarma}, \citenamefont {Freedman},\ and\ \citenamefont {Nayak}}]{nayak2}%
  \BibitemOpen
  \bibfield  {author} {\bibinfo {author} {\bibfnamefont {S.}~\bibnamefont {Sarma}}, \bibinfo {author} {\bibfnamefont {M.}~\bibnamefont {Freedman}},\ and\ \bibinfo {author} {\bibfnamefont {C.}~\bibnamefont {Nayak}},\ }\bibfield  {title} {\bibinfo {title} {Majorana zero modes and topological quantum computation},\ }\href@noop {} {\bibfield  {journal} {\bibinfo  {journal} {npj Quantum Inf}\ }\textbf {\bibinfo {volume} {1}},\ \bibinfo {pages} {15001} (\bibinfo {year} {2015})}\BibitemShut {NoStop}%
\bibitem [{\citenamefont {Stern}\ and\ \citenamefont {Lindner}(2013)}]{stern2}%
  \BibitemOpen
  \bibfield  {author} {\bibinfo {author} {\bibfnamefont {A.}~\bibnamefont {Stern}}\ and\ \bibinfo {author} {\bibfnamefont {N.~H.}\ \bibnamefont {Lindner}},\ }\bibfield  {title} {\bibinfo {title} {Topological quantum computation—from basic concepts to first experiments},\ }\href@noop {} {\bibfield  {journal} {\bibinfo  {journal} {Science}\ }\textbf {\bibinfo {volume} {339}},\ \bibinfo {pages} {1179} (\bibinfo {year} {2013})}\BibitemShut {NoStop}%
\bibitem [{\citenamefont {Yazdani}\ \emph {et~al.}(2023)\citenamefont {Yazdani}, \citenamefont {Oppen}, \citenamefont {Halperin},\ and\ \citenamefont {Yacoby}}]{yazdani1}%
  \BibitemOpen
  \bibfield  {author} {\bibinfo {author} {\bibfnamefont {A.}~\bibnamefont {Yazdani}}, \bibinfo {author} {\bibfnamefont {F.~V.}\ \bibnamefont {Oppen}}, \bibinfo {author} {\bibfnamefont {B.~I.}\ \bibnamefont {Halperin}},\ and\ \bibinfo {author} {\bibfnamefont {A.}~\bibnamefont {Yacoby}},\ }\bibfield  {title} {\bibinfo {title} {Hunting for majoranas},\ }\href@noop {} {\bibfield  {journal} {\bibinfo  {journal} {Science}\ }\textbf {\bibinfo {volume} {380}},\ \bibinfo {pages} {eade0850} (\bibinfo {year} {2023})}\BibitemShut {NoStop}%
\bibitem [{\citenamefont {Jain}(1989)}]{jain_CF}%
  \BibitemOpen
  \bibfield  {author} {\bibinfo {author} {\bibfnamefont {J.~K.}\ \bibnamefont {Jain}},\ }\bibfield  {title} {\bibinfo {title} {Composite-fermion approach for the fractional quantum hall effect},\ }\href@noop {} {\bibfield  {journal} {\bibinfo  {journal} {Phys. Rev. Lett.}\ }\textbf {\bibinfo {volume} {63}},\ \bibinfo {pages} {199} (\bibinfo {year} {1989})}\BibitemShut {NoStop}%
\bibitem [{\citenamefont {Wen}\ and\ \citenamefont {Zee}(1992)}]{wen2}%
  \BibitemOpen
  \bibfield  {author} {\bibinfo {author} {\bibfnamefont {X.~G.}\ \bibnamefont {Wen}}\ and\ \bibinfo {author} {\bibfnamefont {A.}~\bibnamefont {Zee}},\ }\bibfield  {title} {\bibinfo {title} {Classification of abelian quantum hall states and matrix formulation of topological fluids},\ }\href@noop {} {\bibfield  {journal} {\bibinfo  {journal} {Phys. Rev. B}\ }\textbf {\bibinfo {volume} {46}},\ \bibinfo {pages} {2290} (\bibinfo {year} {1992})}\BibitemShut {NoStop}%
\bibitem [{\citenamefont {Zibrov}\ \emph {et~al.}(2017)\citenamefont {Zibrov}, \citenamefont {Kometter}, \citenamefont {Zhou}, \citenamefont {Spanton}, \citenamefont {Taniguchi}, \citenamefont {Watanabe}, \citenamefont {Zaletel},\ and\ \citenamefont {Young}}]{zibrov}%
  \BibitemOpen
  \bibfield  {author} {\bibinfo {author} {\bibfnamefont {A.}~\bibnamefont {Zibrov}}, \bibinfo {author} {\bibfnamefont {C.}~\bibnamefont {Kometter}}, \bibinfo {author} {\bibfnamefont {H.}~\bibnamefont {Zhou}}, \bibinfo {author} {\bibfnamefont {E.}~\bibnamefont {Spanton}}, \bibinfo {author} {\bibfnamefont {T.}~\bibnamefont {Taniguchi}}, \bibinfo {author} {\bibfnamefont {K.}~\bibnamefont {Watanabe}}, \bibinfo {author} {\bibfnamefont {M.}~\bibnamefont {Zaletel}},\ and\ \bibinfo {author} {\bibfnamefont {A.}~\bibnamefont {Young}},\ }\bibfield  {title} {\bibinfo {title} {Tunable interacting composite fermion phases in a half-filled bilayer-graphene landau level.},\ }\href@noop {} {\bibfield  {journal} {\bibinfo  {journal} {Nature}\ }\textbf {\bibinfo {volume} {549}},\ \bibinfo {pages} {360} (\bibinfo {year} {2017})}\BibitemShut {NoStop}%
\bibitem [{\citenamefont {Li}\ \emph {et~al.}(2017)\citenamefont {Li}, \citenamefont {Tan}, \citenamefont {Chen}, \citenamefont {Zeng}, \citenamefont {Taniguchi}, \citenamefont {Watanabe}, \citenamefont {Hone},\ and\ \citenamefont {Dean}}]{dean3}%
  \BibitemOpen
  \bibfield  {author} {\bibinfo {author} {\bibfnamefont {J.}~\bibnamefont {Li}}, \bibinfo {author} {\bibfnamefont {C.}~\bibnamefont {Tan}}, \bibinfo {author} {\bibfnamefont {S.}~\bibnamefont {Chen}}, \bibinfo {author} {\bibfnamefont {Y.}~\bibnamefont {Zeng}}, \bibinfo {author} {\bibfnamefont {T.}~\bibnamefont {Taniguchi}}, \bibinfo {author} {\bibfnamefont {K.}~\bibnamefont {Watanabe}}, \bibinfo {author} {\bibfnamefont {J.}~\bibnamefont {Hone}},\ and\ \bibinfo {author} {\bibfnamefont {C.}~\bibnamefont {Dean}},\ }\bibfield  {title} {\bibinfo {title} {Even denominator fractional quantum hall state in bilayer graphene},\ }\href@noop {} {\bibfield  {journal} {\bibinfo  {journal} {Science}\ }\textbf {\bibinfo {volume} {358}},\ \bibinfo {pages} {648} (\bibinfo {year} {2017})}\BibitemShut {NoStop}%
\bibitem [{\citenamefont {Huang}\ \emph {et~al.}(2022)\citenamefont {Huang}, \citenamefont {Fu}, \citenamefont {Hickey}, \citenamefont {Alem}, \citenamefont {Lin}, \citenamefont {Watanabe}, \citenamefont {Taniguchi},\ and\ \citenamefont {Zhu}}]{zhu2}%
  \BibitemOpen
  \bibfield  {author} {\bibinfo {author} {\bibfnamefont {K.}~\bibnamefont {Huang}}, \bibinfo {author} {\bibfnamefont {H.}~\bibnamefont {Fu}}, \bibinfo {author} {\bibfnamefont {D.~R.}\ \bibnamefont {Hickey}}, \bibinfo {author} {\bibfnamefont {N.}~\bibnamefont {Alem}}, \bibinfo {author} {\bibfnamefont {X.}~\bibnamefont {Lin}}, \bibinfo {author} {\bibfnamefont {K.}~\bibnamefont {Watanabe}}, \bibinfo {author} {\bibfnamefont {T.}~\bibnamefont {Taniguchi}},\ and\ \bibinfo {author} {\bibfnamefont {J.}~\bibnamefont {Zhu}},\ }\bibfield  {title} {\bibinfo {title} {Valley isospin controlled fractional quantum hall states in bilayer graphene},\ }\href@noop {} {\bibfield  {journal} {\bibinfo  {journal} {Physical Review X}\ }\textbf {\bibinfo {volume} {12}},\ \bibinfo {pages} {031019} (\bibinfo {year} {2022})}\BibitemShut {NoStop}%
\bibitem [{\citenamefont {Assouline}\ \emph {et~al.}(2024)\citenamefont {Assouline}, \citenamefont {Wang}, \citenamefont {Zhou}, \citenamefont {Cohen}, \citenamefont {Yang}, \citenamefont {Zhang}, \citenamefont {Taniguchi}, \citenamefont {Watanabe}, \citenamefont {Mong}, \citenamefont {Zaletel},\ and\ \citenamefont {Young}}]{assouline}%
  \BibitemOpen
  \bibfield  {author} {\bibinfo {author} {\bibfnamefont {A.}~\bibnamefont {Assouline}}, \bibinfo {author} {\bibfnamefont {T.}~\bibnamefont {Wang}}, \bibinfo {author} {\bibfnamefont {H.}~\bibnamefont {Zhou}}, \bibinfo {author} {\bibfnamefont {L.~A.}\ \bibnamefont {Cohen}}, \bibinfo {author} {\bibfnamefont {F.}~\bibnamefont {Yang}}, \bibinfo {author} {\bibfnamefont {R.}~\bibnamefont {Zhang}}, \bibinfo {author} {\bibfnamefont {T.}~\bibnamefont {Taniguchi}}, \bibinfo {author} {\bibfnamefont {K.}~\bibnamefont {Watanabe}}, \bibinfo {author} {\bibfnamefont {R.~S.}\ \bibnamefont {Mong}}, \bibinfo {author} {\bibfnamefont {M.~P.}\ \bibnamefont {Zaletel}},\ and\ \bibinfo {author} {\bibfnamefont {A.~F.}\ \bibnamefont {Young}},\ }\bibfield  {title} {\bibinfo {title} {Energy gap of the even-denominator fractional quantum hall state in bilayer graphene},\ }\href@noop {} {\bibfield  {journal} {\bibinfo  {journal} {Physical Review Letters}\ }\textbf {\bibinfo {volume} {132}},\ \bibinfo {pages} {046603} (\bibinfo {year}
  {2024})}\BibitemShut {NoStop}%
\bibitem [{\citenamefont {Hu}\ \emph {et~al.}(2025)\citenamefont {Hu}, \citenamefont {Tsui}, \citenamefont {He}, \citenamefont {Kamber}, \citenamefont {Wang}, \citenamefont {Mohammadi}, \citenamefont {Watanabe}, \citenamefont {Taniguchi}, \citenamefont {Papić}, \citenamefont {Zaletel},\ and\ \citenamefont {Yazdani}}]{yazdani2}%
  \BibitemOpen
  \bibfield  {author} {\bibinfo {author} {\bibfnamefont {Y.}~\bibnamefont {Hu}}, \bibinfo {author} {\bibfnamefont {Y.-C.}\ \bibnamefont {Tsui}}, \bibinfo {author} {\bibfnamefont {M.}~\bibnamefont {He}}, \bibinfo {author} {\bibfnamefont {U.}~\bibnamefont {Kamber}}, \bibinfo {author} {\bibfnamefont {T.}~\bibnamefont {Wang}}, \bibinfo {author} {\bibfnamefont {A.~S.}\ \bibnamefont {Mohammadi}}, \bibinfo {author} {\bibfnamefont {K.}~\bibnamefont {Watanabe}}, \bibinfo {author} {\bibfnamefont {T.}~\bibnamefont {Taniguchi}}, \bibinfo {author} {\bibfnamefont {Z.}~\bibnamefont {Papić}}, \bibinfo {author} {\bibfnamefont {M.~P.}\ \bibnamefont {Zaletel}},\ and\ \bibinfo {author} {\bibfnamefont {A.}~\bibnamefont {Yazdani}},\ }\bibfield  {title} {\bibinfo {title} {High-resolution tunnelling spectroscopy of fractional quantum hall states},\ }\href@noop {} {\bibfield  {journal} {\bibinfo  {journal} {Nature Physics}\ }\textbf {\bibinfo {volume} {21}},\ \bibinfo {pages} {716} (\bibinfo {year} {2025})}\BibitemShut {NoStop}%
\bibitem [{\citenamefont {Kumar}\ \emph {et~al.}(2025)\citenamefont {Kumar}, \citenamefont {Haug}, \citenamefont {Kim}, \citenamefont {Yutushui}, \citenamefont {Khudiakov}, \citenamefont {Bhardwaj}, \citenamefont {Ilin}, \citenamefont {Watanabe}, \citenamefont {Taniguchi}, \citenamefont {Mross},\ and\ \citenamefont {Ronen}}]{ronen_quarter}%
  \BibitemOpen
  \bibfield  {author} {\bibinfo {author} {\bibfnamefont {R.}~\bibnamefont {Kumar}}, \bibinfo {author} {\bibfnamefont {A.}~\bibnamefont {Haug}}, \bibinfo {author} {\bibfnamefont {J.}~\bibnamefont {Kim}}, \bibinfo {author} {\bibfnamefont {M.}~\bibnamefont {Yutushui}}, \bibinfo {author} {\bibfnamefont {K.}~\bibnamefont {Khudiakov}}, \bibinfo {author} {\bibfnamefont {V.}~\bibnamefont {Bhardwaj}}, \bibinfo {author} {\bibfnamefont {A.}~\bibnamefont {Ilin}}, \bibinfo {author} {\bibfnamefont {K.}~\bibnamefont {Watanabe}}, \bibinfo {author} {\bibfnamefont {T.}~\bibnamefont {Taniguchi}}, \bibinfo {author} {\bibfnamefont {D.~F.}\ \bibnamefont {Mross}},\ and\ \bibinfo {author} {\bibfnamefont {Y.}~\bibnamefont {Ronen}},\ }\bibfield  {title} {\bibinfo {title} {Quarter-and half-filled quantum hall states and their topological orders revealed by daughter states in bilayer graphene},\ }\href@noop {} {\bibfield  {journal} {\bibinfo  {journal} {Nature Communications}\ }\textbf {\bibinfo {volume} {16}},\ \bibinfo {pages} {7255}
  (\bibinfo {year} {2025})}\BibitemShut {NoStop}%
\bibitem [{\citenamefont {Déprez}\ \emph {et~al.}(2021)\citenamefont {Déprez}, \citenamefont {Veyrat}, \citenamefont {Vignaud}, \citenamefont {Nayak}, \citenamefont {Watanabe}, \citenamefont {Taniguchi}, \citenamefont {Gay}, \citenamefont {Sellier},\ and\ \citenamefont {Sacépé}}]{sacepe1}%
  \BibitemOpen
  \bibfield  {author} {\bibinfo {author} {\bibfnamefont {C.}~\bibnamefont {Déprez}}, \bibinfo {author} {\bibfnamefont {L.}~\bibnamefont {Veyrat}}, \bibinfo {author} {\bibfnamefont {H.}~\bibnamefont {Vignaud}}, \bibinfo {author} {\bibfnamefont {G.}~\bibnamefont {Nayak}}, \bibinfo {author} {\bibfnamefont {K.}~\bibnamefont {Watanabe}}, \bibinfo {author} {\bibfnamefont {T.}~\bibnamefont {Taniguchi}}, \bibinfo {author} {\bibfnamefont {F.}~\bibnamefont {Gay}}, \bibinfo {author} {\bibfnamefont {H.}~\bibnamefont {Sellier}},\ and\ \bibinfo {author} {\bibfnamefont {B.}~\bibnamefont {Sacépé}},\ }\bibfield  {title} {\bibinfo {title} {A tunable fabry–pérot quantum hall interferometer in graphene},\ }\href@noop {} {\bibfield  {journal} {\bibinfo  {journal} {Nature nanotechnology}\ }\textbf {\bibinfo {volume} {16}},\ \bibinfo {pages} {555} (\bibinfo {year} {2021})}\BibitemShut {NoStop}%
\bibitem [{\citenamefont {Ronen}\ \emph {et~al.}(2021)\citenamefont {Ronen}, \citenamefont {Werkmeister}, \citenamefont {Najafabadi}, \citenamefont {Pierce}, \citenamefont {Anderson}, \citenamefont {Shin}, \citenamefont {Lee}, \citenamefont {Lee}, \citenamefont {Johnson}, \citenamefont {Watanabe}, \citenamefont {Taniguchi}, \citenamefont {Yacoby},\ and\ \citenamefont {Kim}}]{tom_yuval}%
  \BibitemOpen
  \bibfield  {author} {\bibinfo {author} {\bibfnamefont {Y.}~\bibnamefont {Ronen}}, \bibinfo {author} {\bibfnamefont {T.}~\bibnamefont {Werkmeister}}, \bibinfo {author} {\bibfnamefont {D.~H.}\ \bibnamefont {Najafabadi}}, \bibinfo {author} {\bibfnamefont {A.~T.}\ \bibnamefont {Pierce}}, \bibinfo {author} {\bibfnamefont {L.~E.}\ \bibnamefont {Anderson}}, \bibinfo {author} {\bibfnamefont {Y.~J.}\ \bibnamefont {Shin}}, \bibinfo {author} {\bibfnamefont {S.~Y.}\ \bibnamefont {Lee}}, \bibinfo {author} {\bibfnamefont {Y.~H.}\ \bibnamefont {Lee}}, \bibinfo {author} {\bibfnamefont {B.}~\bibnamefont {Johnson}}, \bibinfo {author} {\bibfnamefont {K.}~\bibnamefont {Watanabe}}, \bibinfo {author} {\bibfnamefont {T.}~\bibnamefont {Taniguchi}}, \bibinfo {author} {\bibfnamefont {A.}~\bibnamefont {Yacoby}},\ and\ \bibinfo {author} {\bibfnamefont {P.}~\bibnamefont {Kim}},\ }\bibfield  {title} {\bibinfo {title} {Aharonov–bohm effect in graphene-based fabry–pérot quantum hall interferometers},\ }\href@noop {} {\bibfield
  {journal} {\bibinfo  {journal} {Nature Nanotechnology}\ }\textbf {\bibinfo {volume} {16}},\ \bibinfo {pages} {563} (\bibinfo {year} {2021})}\BibitemShut {NoStop}%
\bibitem [{\citenamefont {Zhao}\ \emph {et~al.}(2022)\citenamefont {Zhao}, \citenamefont {Arnault}, \citenamefont {Larson}, \citenamefont {andAndrew Seredinski}, \citenamefont {Fleming}, \citenamefont {Watanabe}, \citenamefont {Taniguchi}, \citenamefont {Amet},\ and\ \citenamefont {Finkelstein}}]{finkelstein}%
  \BibitemOpen
  \bibfield  {author} {\bibinfo {author} {\bibfnamefont {L.}~\bibnamefont {Zhao}}, \bibinfo {author} {\bibfnamefont {E.~G.}\ \bibnamefont {Arnault}}, \bibinfo {author} {\bibfnamefont {T.~F.~Q.}\ \bibnamefont {Larson}}, \bibinfo {author} {\bibfnamefont {Z.~I.}\ \bibnamefont {andAndrew Seredinski}}, \bibinfo {author} {\bibfnamefont {T.}~\bibnamefont {Fleming}}, \bibinfo {author} {\bibfnamefont {K.}~\bibnamefont {Watanabe}}, \bibinfo {author} {\bibfnamefont {T.}~\bibnamefont {Taniguchi}}, \bibinfo {author} {\bibfnamefont {F.}~\bibnamefont {Amet}},\ and\ \bibinfo {author} {\bibfnamefont {G.}~\bibnamefont {Finkelstein}},\ }\bibfield  {title} {\bibinfo {title} {Graphene-based quantum hall interferometer with self-aligned side gates},\ }\href@noop {} {\bibfield  {journal} {\bibinfo  {journal} {Nano Letters}\ }\textbf {\bibinfo {volume} {22}},\ \bibinfo {pages} {9645} (\bibinfo {year} {2022})}\BibitemShut {NoStop}%
\bibitem [{\citenamefont {Fu}\ \emph {et~al.}(2023)\citenamefont {Fu}, \citenamefont {Huang}, \citenamefont {Watanabe}, \citenamefont {Taniguchi}, \citenamefont {Kayyalha},\ and\ \citenamefont {Zhu}}]{zhu}%
  \BibitemOpen
  \bibfield  {author} {\bibinfo {author} {\bibfnamefont {H.}~\bibnamefont {Fu}}, \bibinfo {author} {\bibfnamefont {K.}~\bibnamefont {Huang}}, \bibinfo {author} {\bibfnamefont {K.}~\bibnamefont {Watanabe}}, \bibinfo {author} {\bibfnamefont {T.}~\bibnamefont {Taniguchi}}, \bibinfo {author} {\bibfnamefont {M.}~\bibnamefont {Kayyalha}},\ and\ \bibinfo {author} {\bibfnamefont {J.}~\bibnamefont {Zhu}},\ }\bibfield  {title} {\bibinfo {title} {Aharonov–bohm oscillations in bilayer graphene quantum hall edge state fabry–pérot interferometers},\ }\href@noop {} {\bibfield  {journal} {\bibinfo  {journal} {Nano Letters}\ }\textbf {\bibinfo {volume} {23}},\ \bibinfo {pages} {718} (\bibinfo {year} {2023})}\BibitemShut {NoStop}%
\bibitem [{\citenamefont {Werkmeister}\ \emph {et~al.}(2024)\citenamefont {Werkmeister}, \citenamefont {Ehrets}, \citenamefont {Ronen}, \citenamefont {Wesson}, \citenamefont {Najafabadi}, \citenamefont {Wei}, \citenamefont {Watanabe}, \citenamefont {Taniguchi}, \citenamefont {Feldman}, \citenamefont {Halperin}, \citenamefont {Yacoby},\ and\ \citenamefont {Kim}}]{tom_james_coupling}%
  \BibitemOpen
  \bibfield  {author} {\bibinfo {author} {\bibfnamefont {T.}~\bibnamefont {Werkmeister}}, \bibinfo {author} {\bibfnamefont {J.~R.}\ \bibnamefont {Ehrets}}, \bibinfo {author} {\bibfnamefont {Y.}~\bibnamefont {Ronen}}, \bibinfo {author} {\bibfnamefont {M.~E.}\ \bibnamefont {Wesson}}, \bibinfo {author} {\bibfnamefont {D.}~\bibnamefont {Najafabadi}}, \bibinfo {author} {\bibfnamefont {Z.}~\bibnamefont {Wei}}, \bibinfo {author} {\bibfnamefont {K.}~\bibnamefont {Watanabe}}, \bibinfo {author} {\bibfnamefont {T.}~\bibnamefont {Taniguchi}}, \bibinfo {author} {\bibfnamefont {D.}~\bibnamefont {Feldman}}, \bibinfo {author} {\bibfnamefont {B.~I.}\ \bibnamefont {Halperin}}, \bibinfo {author} {\bibfnamefont {A.}~\bibnamefont {Yacoby}},\ and\ \bibinfo {author} {\bibfnamefont {P.}~\bibnamefont {Kim}},\ }\bibfield  {title} {\bibinfo {title} {Strongly coupled edge states in a graphene quantum hall interferometer},\ }\href@noop {} {\bibfield  {journal} {\bibinfo  {journal} {Nature Communications}\ }\textbf {\bibinfo {volume}
  {15}},\ \bibinfo {pages} {6533} (\bibinfo {year} {2024})}\BibitemShut {NoStop}%
\bibitem [{\citenamefont {Yang}\ \emph {et~al.}(2024)\citenamefont {Yang}, \citenamefont {Perconte}, \citenamefont {Déprez}, \citenamefont {Watanabe}, \citenamefont {Taniguchi}, \citenamefont {Dumont}, \citenamefont {Wagner}, \citenamefont {Gay}, \citenamefont {Safi}, \citenamefont {Sellier},\ and\ \citenamefont {Sacépé}}]{sacepe2}%
  \BibitemOpen
  \bibfield  {author} {\bibinfo {author} {\bibfnamefont {W.}~\bibnamefont {Yang}}, \bibinfo {author} {\bibfnamefont {D.}~\bibnamefont {Perconte}}, \bibinfo {author} {\bibfnamefont {C.}~\bibnamefont {Déprez}}, \bibinfo {author} {\bibfnamefont {K.}~\bibnamefont {Watanabe}}, \bibinfo {author} {\bibfnamefont {T.}~\bibnamefont {Taniguchi}}, \bibinfo {author} {\bibfnamefont {S.}~\bibnamefont {Dumont}}, \bibinfo {author} {\bibfnamefont {E.}~\bibnamefont {Wagner}}, \bibinfo {author} {\bibfnamefont {F.}~\bibnamefont {Gay}}, \bibinfo {author} {\bibfnamefont {I.}~\bibnamefont {Safi}}, \bibinfo {author} {\bibfnamefont {H.}~\bibnamefont {Sellier}},\ and\ \bibinfo {author} {\bibfnamefont {B.}~\bibnamefont {Sacépé}},\ }\bibfield  {title} {\bibinfo {title} {Evidence for correlated electron pairs and triplets in quantum hall interferometers},\ }\href@noop {} {\bibfield  {journal} {\bibinfo  {journal} {Nature Communications}\ }\textbf {\bibinfo {volume} {15}},\ \bibinfo {pages} {10064} (\bibinfo {year} {2024})}\BibitemShut
  {NoStop}%
\bibitem [{\citenamefont {Samuelson}\ \emph {et~al.}(2025{\natexlab{b}})\citenamefont {Samuelson}, \citenamefont {Cohen}, \citenamefont {Wang}, \citenamefont {Blanch}, \citenamefont {Taniguchi}, \citenamefont {Watanabe}, \citenamefont {Zaletel},\ and\ \citenamefont {Young}}]{andrea_hard_soft}%
  \BibitemOpen
  \bibfield  {author} {\bibinfo {author} {\bibfnamefont {N.~L.}\ \bibnamefont {Samuelson}}, \bibinfo {author} {\bibfnamefont {L.~A.}\ \bibnamefont {Cohen}}, \bibinfo {author} {\bibfnamefont {W.}~\bibnamefont {Wang}}, \bibinfo {author} {\bibfnamefont {S.}~\bibnamefont {Blanch}}, \bibinfo {author} {\bibfnamefont {T.}~\bibnamefont {Taniguchi}}, \bibinfo {author} {\bibfnamefont {K.}~\bibnamefont {Watanabe}}, \bibinfo {author} {\bibfnamefont {M.~P.}\ \bibnamefont {Zaletel}},\ and\ \bibinfo {author} {\bibfnamefont {A.~F.}\ \bibnamefont {Young}},\ }\bibfield  {title} {\bibinfo {title} {Hard and soft phase slips in a fabry-pérot quantum hall interferometer},\ }\href@noop {} {\bibfield  {journal} {\bibinfo  {journal} {Phys. Rev. Lett.}\ }\textbf {\bibinfo {volume} {134}},\ \bibinfo {pages} {256301} (\bibinfo {year} {2025}{\natexlab{b}})}\BibitemShut {NoStop}%
\bibitem [{\citenamefont {Zeng}\ \emph {et~al.}(2019)\citenamefont {Zeng}, \citenamefont {Li}, \citenamefont {Dietrich}, \citenamefont {Ghosh}, \citenamefont {Watanabe}, \citenamefont {Taniguchi}, \citenamefont {Hone},\ and\ \citenamefont {Dean}}]{dean1}%
  \BibitemOpen
  \bibfield  {author} {\bibinfo {author} {\bibfnamefont {Y.}~\bibnamefont {Zeng}}, \bibinfo {author} {\bibfnamefont {J.}~\bibnamefont {Li}}, \bibinfo {author} {\bibfnamefont {S.}~\bibnamefont {Dietrich}}, \bibinfo {author} {\bibfnamefont {O.}~\bibnamefont {Ghosh}}, \bibinfo {author} {\bibfnamefont {K.}~\bibnamefont {Watanabe}}, \bibinfo {author} {\bibfnamefont {T.}~\bibnamefont {Taniguchi}}, \bibinfo {author} {\bibfnamefont {J.}~\bibnamefont {Hone}},\ and\ \bibinfo {author} {\bibfnamefont {C.}~\bibnamefont {Dean}},\ }\bibfield  {title} {\bibinfo {title} {High-quality magnetotransport in graphene using the edge-free corbino geometry},\ }\href@noop {} {\bibfield  {journal} {\bibinfo  {journal} {Physical Review Letters}\ }\textbf {\bibinfo {volume} {122}},\ \bibinfo {pages} {137701} (\bibinfo {year} {2019})}\BibitemShut {NoStop}%
\bibitem [{\citenamefont {Ribeiro-Palau}\ \emph {et~al.}(2019)\citenamefont {Ribeiro-Palau}, \citenamefont {Chen}, \citenamefont {Zeng}, \citenamefont {Watanabe}, \citenamefont {Taniguchi}, \citenamefont {Hone},\ and\ \citenamefont {Dean}}]{dean2}%
  \BibitemOpen
  \bibfield  {author} {\bibinfo {author} {\bibfnamefont {R.}~\bibnamefont {Ribeiro-Palau}}, \bibinfo {author} {\bibfnamefont {S.}~\bibnamefont {Chen}}, \bibinfo {author} {\bibfnamefont {Y.}~\bibnamefont {Zeng}}, \bibinfo {author} {\bibfnamefont {K.}~\bibnamefont {Watanabe}}, \bibinfo {author} {\bibfnamefont {T.}~\bibnamefont {Taniguchi}}, \bibinfo {author} {\bibfnamefont {J.}~\bibnamefont {Hone}},\ and\ \bibinfo {author} {\bibfnamefont {C.~R.}\ \bibnamefont {Dean}},\ }\bibfield  {title} {\bibinfo {title} {High-quality electrostatically defined hall bars in monolayer graphene},\ }\href@noop {} {\bibfield  {journal} {\bibinfo  {journal} {Nano letters}\ }\textbf {\bibinfo {volume} {19}},\ \bibinfo {pages} {2583} (\bibinfo {year} {2019})}\BibitemShut {NoStop}%
\bibitem [{\citenamefont {de~C~Chamon}\ \emph {et~al.}(1997)\citenamefont {de~C~Chamon}, \citenamefont {Freed}, \citenamefont {Kivelson}, \citenamefont {Sondhi},\ and\ \citenamefont {Wen}}]{wen1}%
  \BibitemOpen
  \bibfield  {author} {\bibinfo {author} {\bibfnamefont {C.}~\bibnamefont {de~C~Chamon}}, \bibinfo {author} {\bibfnamefont {D.}~\bibnamefont {Freed}}, \bibinfo {author} {\bibfnamefont {S.}~\bibnamefont {Kivelson}}, \bibinfo {author} {\bibfnamefont {S.}~\bibnamefont {Sondhi}},\ and\ \bibinfo {author} {\bibfnamefont {X.}~\bibnamefont {Wen}},\ }\bibfield  {title} {\bibinfo {title} {Two point-contact interferometer for quantum hall systems},\ }\href@noop {} {\bibfield  {journal} {\bibinfo  {journal} {Physical Review B}\ }\textbf {\bibinfo {volume} {55}},\ \bibinfo {pages} {2331} (\bibinfo {year} {1997})}\BibitemShut {NoStop}%
\bibitem [{\citenamefont {Halperin}\ \emph {et~al.}(2011)\citenamefont {Halperin}, \citenamefont {Stern}, \citenamefont {Neder},\ and\ \citenamefont {Rosenow}}]{halperin_theory_of_fp}%
  \BibitemOpen
  \bibfield  {author} {\bibinfo {author} {\bibfnamefont {B.~I.}\ \bibnamefont {Halperin}}, \bibinfo {author} {\bibfnamefont {A.}~\bibnamefont {Stern}}, \bibinfo {author} {\bibfnamefont {I.}~\bibnamefont {Neder}},\ and\ \bibinfo {author} {\bibfnamefont {B.}~\bibnamefont {Rosenow}},\ }\bibfield  {title} {\bibinfo {title} {Theory of the fabry-pérot quantum hall interferometer},\ }\href@noop {} {\bibfield  {journal} {\bibinfo  {journal} {Physical Review B—Condensed Matter and Materials Physics}\ }\textbf {\bibinfo {volume} {83}},\ \bibinfo {pages} {155440} (\bibinfo {year} {2011})}\BibitemShut {NoStop}%
\bibitem [{\citenamefont {Feldman}\ and\ \citenamefont {Halperin}(2022)}]{halperin_robustness}%
  \BibitemOpen
  \bibfield  {author} {\bibinfo {author} {\bibfnamefont {D.}~\bibnamefont {Feldman}}\ and\ \bibinfo {author} {\bibfnamefont {B.~I.}\ \bibnamefont {Halperin}},\ }\bibfield  {title} {\bibinfo {title} {Robustness of quantum hall interferometry},\ }\href@noop {} {\bibfield  {journal} {\bibinfo  {journal} {Physical Review B}\ }\textbf {\bibinfo {volume} {105}},\ \bibinfo {pages} {165310} (\bibinfo {year} {2022})}\BibitemShut {NoStop}%
\bibitem [{\citenamefont {Carrega}\ \emph {et~al.}(2021)\citenamefont {Carrega}, \citenamefont {Chirolli}, \citenamefont {Heun},\ and\ \citenamefont {Sorba}}]{sorba1}%
  \BibitemOpen
  \bibfield  {author} {\bibinfo {author} {\bibfnamefont {M.}~\bibnamefont {Carrega}}, \bibinfo {author} {\bibfnamefont {L.}~\bibnamefont {Chirolli}}, \bibinfo {author} {\bibfnamefont {S.}~\bibnamefont {Heun}},\ and\ \bibinfo {author} {\bibfnamefont {L.}~\bibnamefont {Sorba}},\ }\bibfield  {title} {\bibinfo {title} {Anyons in quantum hall interferometry},\ }\href@noop {} {\bibfield  {journal} {\bibinfo  {journal} {Nature Reviews Physics}\ }\textbf {\bibinfo {volume} {3}},\ \bibinfo {pages} {698} (\bibinfo {year} {2021})}\BibitemShut {NoStop}%
\bibitem [{\citenamefont {Kim}(2006)}]{eunah_kim}%
  \BibitemOpen
  \bibfield  {author} {\bibinfo {author} {\bibfnamefont {E.-A.}\ \bibnamefont {Kim}},\ }\bibfield  {title} {\bibinfo {title} {Aharanov-bohm interference and fractional statistics in a quantum hall interferometer},\ }\href@noop {} {\bibfield  {journal} {\bibinfo  {journal} {Physical Review Letters}\ }\textbf {\bibinfo {volume} {97}},\ \bibinfo {pages} {216404} (\bibinfo {year} {2006})}\BibitemShut {NoStop}%
\bibitem [{\citenamefont {Read}\ and\ \citenamefont {Sarma}(2024)}]{dassarma1}%
  \BibitemOpen
  \bibfield  {author} {\bibinfo {author} {\bibfnamefont {N.}~\bibnamefont {Read}}\ and\ \bibinfo {author} {\bibfnamefont {S.~D.}\ \bibnamefont {Sarma}},\ }\bibfield  {title} {\bibinfo {title} {Clarification of braiding statistics in fabry–perot interferometry},\ }\href@noop {} {\bibfield  {journal} {\bibinfo  {journal} {Nature Physics}\ }\textbf {\bibinfo {volume} {20}},\ \bibinfo {pages} {381} (\bibinfo {year} {2024})}\BibitemShut {NoStop}%
\bibitem [{\citenamefont {Kivelson}\ and\ \citenamefont {Murthy}(2025)}]{kivelson1}%
  \BibitemOpen
  \bibfield  {author} {\bibinfo {author} {\bibfnamefont {S.~A.}\ \bibnamefont {Kivelson}}\ and\ \bibinfo {author} {\bibfnamefont {C.}~\bibnamefont {Murthy}},\ }\bibfield  {title} {\bibinfo {title} {Modified interferometer to measure anyonic braiding statistics},\ }\href@noop {} {\bibfield  {journal} {\bibinfo  {journal} {Physical Review Letters}\ }\textbf {\bibinfo {volume} {135}},\ \bibinfo {pages} {126605} (\bibinfo {year} {2025})}\BibitemShut {NoStop}%
\bibitem [{\citenamefont {Stern}\ and\ \citenamefont {Halperin}(2006)}]{halperin_evenodd}%
  \BibitemOpen
  \bibfield  {author} {\bibinfo {author} {\bibfnamefont {A.}~\bibnamefont {Stern}}\ and\ \bibinfo {author} {\bibfnamefont {B.~I.}\ \bibnamefont {Halperin}},\ }\bibfield  {title} {\bibinfo {title} {Proposed experiments to probe the non-abelian $\nu=5/2$ quantum hall state},\ }\href@noop {} {\bibfield  {journal} {\bibinfo  {journal} {Physical Review Letters}\ }\textbf {\bibinfo {volume} {96}},\ \bibinfo {pages} {016802} (\bibinfo {year} {2006})}\BibitemShut {NoStop}%
\bibitem [{\citenamefont {Parsa~Bonderson}(2006)}]{shtengel}%
  \BibitemOpen
  \bibfield  {author} {\bibinfo {author} {\bibfnamefont {K.~S.}\ \bibnamefont {Parsa~Bonderson}, \bibfnamefont {Alexei~Kitaev}},\ }\bibfield  {title} {\bibinfo {title} {Detecting non-abelian statistics in the $\nu=5/2$ fractional quantum hall state},\ }\href@noop {} {\bibfield  {journal} {\bibinfo  {journal} {Physical review letters}\ }\textbf {\bibinfo {volume} {96}},\ \bibinfo {pages} {016803} (\bibinfo {year} {2006})}\BibitemShut {NoStop}%
\bibitem [{\citenamefont {Bishara}\ \emph {et~al.}(2009)\citenamefont {Bishara}, \citenamefont {Bonderson}, \citenamefont {Nayak}, \citenamefont {Shtengel},\ and\ \citenamefont {Slingerland}}]{slingerland}%
  \BibitemOpen
  \bibfield  {author} {\bibinfo {author} {\bibfnamefont {W.}~\bibnamefont {Bishara}}, \bibinfo {author} {\bibfnamefont {P.}~\bibnamefont {Bonderson}}, \bibinfo {author} {\bibfnamefont {C.}~\bibnamefont {Nayak}}, \bibinfo {author} {\bibfnamefont {K.}~\bibnamefont {Shtengel}},\ and\ \bibinfo {author} {\bibfnamefont {J.}~\bibnamefont {Slingerland}},\ }\bibfield  {title} {\bibinfo {title} {Interferometric signature of non-abelian anyons},\ }\href@noop {} {\bibfield  {journal} {\bibinfo  {journal} {Physical Review B—Condensed Matter and Materials Physics}\ }\textbf {\bibinfo {volume} {80}},\ \bibinfo {pages} {155303} (\bibinfo {year} {2009})}\BibitemShut {NoStop}%
\bibitem [{\citenamefont {Das~Sarma}\ \emph {et~al.}(2005)\citenamefont {Das~Sarma}, \citenamefont {Freedman},\ and\ \citenamefont {Nayak}}]{nayak3}%
  \BibitemOpen
  \bibfield  {author} {\bibinfo {author} {\bibfnamefont {S.}~\bibnamefont {Das~Sarma}}, \bibinfo {author} {\bibfnamefont {M.}~\bibnamefont {Freedman}},\ and\ \bibinfo {author} {\bibfnamefont {C.}~\bibnamefont {Nayak}},\ }\bibfield  {title} {\bibinfo {title} {Topologically protected qubits from a possible non-abelian fractional quantum hall state},\ }\href@noop {} {\bibfield  {journal} {\bibinfo  {journal} {Phys. Rev. Lett.}\ }\textbf {\bibinfo {volume} {94}},\ \bibinfo {pages} {166802} (\bibinfo {year} {2005})}\BibitemShut {NoStop}%
\bibitem [{\citenamefont {Nakamura}\ \emph {et~al.}(2023)\citenamefont {Nakamura}, \citenamefont {Liang}, \citenamefont {Gardner},\ and\ \citenamefont {Manfra}}]{manfra2}%
  \BibitemOpen
  \bibfield  {author} {\bibinfo {author} {\bibfnamefont {J.}~\bibnamefont {Nakamura}}, \bibinfo {author} {\bibfnamefont {S.}~\bibnamefont {Liang}}, \bibinfo {author} {\bibfnamefont {G.~C.}\ \bibnamefont {Gardner}},\ and\ \bibinfo {author} {\bibfnamefont {M.~J.}\ \bibnamefont {Manfra}},\ }\bibfield  {title} {\bibinfo {title} {Fabry-p\'erot interferometry at the $\nu=2/5$ fractional quantum hall state},\ }\href@noop {} {\bibfield  {journal} {\bibinfo  {journal} {Physical Review X}\ }\textbf {\bibinfo {volume} {13}},\ \bibinfo {pages} {041012} (\bibinfo {year} {2023})}\BibitemShut {NoStop}%
\bibitem [{\citenamefont {Goldman}\ and\ \citenamefont {Su}(1995)}]{goldman1}%
  \BibitemOpen
  \bibfield  {author} {\bibinfo {author} {\bibfnamefont {V.~J.}\ \bibnamefont {Goldman}}\ and\ \bibinfo {author} {\bibfnamefont {B.}~\bibnamefont {Su}},\ }\bibfield  {title} {\bibinfo {title} {Resonant tunneling in the quantum hall regime: Measurement of fractional charge},\ }\href@noop {} {\bibfield  {journal} {\bibinfo  {journal} {Science}\ }\textbf {\bibinfo {volume} {267}},\ \bibinfo {pages} {1010} (\bibinfo {year} {1995})}\BibitemShut {NoStop}%
\bibitem [{\citenamefont {Goldman}\ \emph {et~al.}(2005)\citenamefont {Goldman}, \citenamefont {Liu},\ and\ \citenamefont {Zaslavsky}}]{goldman2}%
  \BibitemOpen
  \bibfield  {author} {\bibinfo {author} {\bibfnamefont {V.~J.}\ \bibnamefont {Goldman}}, \bibinfo {author} {\bibfnamefont {J.}~\bibnamefont {Liu}},\ and\ \bibinfo {author} {\bibfnamefont {A.}~\bibnamefont {Zaslavsky}},\ }\bibfield  {title} {\bibinfo {title} {Fractional statistics of laughlin quasiparticles in quantum antidots},\ }\href@noop {} {\bibfield  {journal} {\bibinfo  {journal} {Phys. Rev. B}\ }\textbf {\bibinfo {volume} {71}},\ \bibinfo {pages} {153303} (\bibinfo {year} {2005})}\BibitemShut {NoStop}%
\bibitem [{\citenamefont {Kou}\ \emph {et~al.}(2012)\citenamefont {Kou}, \citenamefont {Marcus}, \citenamefont {Pfeiffer},\ and\ \citenamefont {West}}]{west1}%
  \BibitemOpen
  \bibfield  {author} {\bibinfo {author} {\bibfnamefont {A.}~\bibnamefont {Kou}}, \bibinfo {author} {\bibfnamefont {C.~M.}\ \bibnamefont {Marcus}}, \bibinfo {author} {\bibfnamefont {L.~N.}\ \bibnamefont {Pfeiffer}},\ and\ \bibinfo {author} {\bibfnamefont {K.~W.}\ \bibnamefont {West}},\ }\bibfield  {title} {\bibinfo {title} {Coulomb oscillations in antidots in the integer and fractional quantum hall regimes},\ }\href@noop {} {\bibfield  {journal} {\bibinfo  {journal} {Phys. Rev. Lett.}\ }\textbf {\bibinfo {volume} {108}},\ \bibinfo {pages} {256803} (\bibinfo {year} {2012})}\BibitemShut {NoStop}%
\bibitem [{\citenamefont {Mills}\ \emph {et~al.}(2020)\citenamefont {Mills}, \citenamefont {Averin},\ and\ \citenamefont {Du}}]{du1}%
  \BibitemOpen
  \bibfield  {author} {\bibinfo {author} {\bibfnamefont {S.~M.}\ \bibnamefont {Mills}}, \bibinfo {author} {\bibfnamefont {S.~V.}\ \bibnamefont {Averin}},\ and\ \bibinfo {author} {\bibfnamefont {X.}~\bibnamefont {Du}},\ }\bibfield  {title} {\bibinfo {title} {Localizing farctional quasiparticles on graphene quantum hall antidots},\ }\href@noop {} {\bibfield  {journal} {\bibinfo  {journal} {Phys. Rev. Lett.}\ }\textbf {\bibinfo {volume} {125}},\ \bibinfo {pages} {2227701} (\bibinfo {year} {2020})}\BibitemShut {NoStop}%
\bibitem [{\citenamefont {Luca}\ \emph {et~al.}(2025)\citenamefont {Luca}, \citenamefont {Hajigeorgiou}, \citenamefont {Zhou}, \citenamefont {Lotrič}, \citenamefont {Feng}, \citenamefont {Watanabe}, \citenamefont {Taniguchi}, \citenamefont {Simon},\ and\ \citenamefont {Banerjee}}]{diluca}%
  \BibitemOpen
  \bibfield  {author} {\bibinfo {author} {\bibfnamefont {M.~D.}\ \bibnamefont {Luca}}, \bibinfo {author} {\bibfnamefont {E.}~\bibnamefont {Hajigeorgiou}}, \bibinfo {author} {\bibfnamefont {Z.}~\bibnamefont {Zhou}}, \bibinfo {author} {\bibfnamefont {T.}~\bibnamefont {Lotrič}}, \bibinfo {author} {\bibfnamefont {T.}~\bibnamefont {Feng}}, \bibinfo {author} {\bibfnamefont {K.}~\bibnamefont {Watanabe}}, \bibinfo {author} {\bibfnamefont {T.}~\bibnamefont {Taniguchi}}, \bibinfo {author} {\bibfnamefont {S.~H.}\ \bibnamefont {Simon}},\ and\ \bibinfo {author} {\bibfnamefont {M.}~\bibnamefont {Banerjee}},\ }\href {https://arxiv.org/abs/2509.04209} {\bibinfo {title} {Quantum hall antidot as a fractional coulombmeter}} (\bibinfo {year} {2025}),\ \Eprint {https://arxiv.org/abs/2509.04209} {arXiv:2509.04209 [cond-mat.mes-hall]} \BibitemShut {NoStop}%
\bibitem [{\citenamefont {Ghosh}\ \emph {et~al.}(2025{\natexlab{a}})\citenamefont {Ghosh}, \citenamefont {Labendik}, \citenamefont {Musina}, \citenamefont {Umansky}, \citenamefont {Heiblum},\ and\ \citenamefont {Mross}}]{heiblum1}%
  \BibitemOpen
  \bibfield  {author} {\bibinfo {author} {\bibfnamefont {B.}~\bibnamefont {Ghosh}}, \bibinfo {author} {\bibfnamefont {M.}~\bibnamefont {Labendik}}, \bibinfo {author} {\bibfnamefont {L.}~\bibnamefont {Musina}}, \bibinfo {author} {\bibfnamefont {V.}~\bibnamefont {Umansky}}, \bibinfo {author} {\bibfnamefont {M.}~\bibnamefont {Heiblum}},\ and\ \bibinfo {author} {\bibfnamefont {D.~F.}\ \bibnamefont {Mross}},\ }\bibfield  {title} {\bibinfo {title} {Anyonic braiding in a chiral mach–zehnder interferometer},\ }\href@noop {} {\bibfield  {journal} {\bibinfo  {journal} {Nature Physics}\ }\textbf {\bibinfo {volume} {21}},\ \bibinfo {pages} {1392} (\bibinfo {year} {2025}{\natexlab{a}})}\BibitemShut {NoStop}%
\bibitem [{\citenamefont {Ghosh}\ \emph {et~al.}(2025{\natexlab{b}})\citenamefont {Ghosh}, \citenamefont {Labendik}, \citenamefont {Umansky}, \citenamefont {Heiblum},\ and\ \citenamefont {Mross}}]{heiblum2}%
  \BibitemOpen
  \bibfield  {author} {\bibinfo {author} {\bibfnamefont {B.}~\bibnamefont {Ghosh}}, \bibinfo {author} {\bibfnamefont {M.}~\bibnamefont {Labendik}}, \bibinfo {author} {\bibfnamefont {V.}~\bibnamefont {Umansky}}, \bibinfo {author} {\bibfnamefont {M.}~\bibnamefont {Heiblum}},\ and\ \bibinfo {author} {\bibfnamefont {D.~F.}\ \bibnamefont {Mross}},\ }\bibfield  {title} {\bibinfo {title} {Coherent bunching of anyons and dissociation in an interference experiment},\ }\href@noop {} {\bibfield  {journal} {\bibinfo  {journal} {Nature}\ }\textbf {\bibinfo {volume} {642}},\ \bibinfo {pages} {922} (\bibinfo {year} {2025}{\natexlab{b}})}\BibitemShut {NoStop}%
\bibitem [{\citenamefont {Gattu}\ and\ \citenamefont {Jain}(2025)}]{jain_molec}%
  \BibitemOpen
  \bibfield  {author} {\bibinfo {author} {\bibfnamefont {M.}~\bibnamefont {Gattu}}\ and\ \bibinfo {author} {\bibfnamefont {J.}~\bibnamefont {Jain}},\ }\bibfield  {title} {\bibinfo {title} {Molecular anyons in the fractional quantum hall effect},\ }\href@noop {} {\bibfield  {journal} {\bibinfo  {journal} {Physical Review Letters}\ }\textbf {\bibinfo {volume} {135}},\ \bibinfo {pages} {236601} (\bibinfo {year} {2025})}\BibitemShut {NoStop}%
\bibitem [{\citenamefont {Jeon}\ \emph {et~al.}(2004)\citenamefont {Jeon}, \citenamefont {Graham},\ and\ \citenamefont {Jain}}]{jain_AB}%
  \BibitemOpen
  \bibfield  {author} {\bibinfo {author} {\bibfnamefont {G.~S.}\ \bibnamefont {Jeon}}, \bibinfo {author} {\bibfnamefont {K.~L.}\ \bibnamefont {Graham}},\ and\ \bibinfo {author} {\bibfnamefont {J.~K.}\ \bibnamefont {Jain}},\ }\bibfield  {title} {\bibinfo {title} {Berry phases for composite fermions: Effective magnetic field and fractional statistics},\ }\href@noop {} {\bibfield  {journal} {\bibinfo  {journal} {Physical Review B}\ }\textbf {\bibinfo {volume} {70}},\ \bibinfo {pages} {125316} (\bibinfo {year} {2004})}\BibitemShut {NoStop}%
\bibitem [{\citenamefont {Jain}(1990)}]{jain_fqh}%
  \BibitemOpen
  \bibfield  {author} {\bibinfo {author} {\bibfnamefont {J.~K.}\ \bibnamefont {Jain}},\ }\bibfield  {title} {\bibinfo {title} {Theory of the fractional quantum hall effect},\ }\href@noop {} {\bibfield  {journal} {\bibinfo  {journal} {Phys. Rev. B}\ }\textbf {\bibinfo {volume} {41}},\ \bibinfo {pages} {7653} (\bibinfo {year} {1990})}\BibitemShut {NoStop}%
\bibitem [{\citenamefont {Venkatachalam}\ \emph {et~al.}(2011)\citenamefont {Venkatachalam}, \citenamefont {Yacoby}, \citenamefont {Pfeiffer},\ and\ \citenamefont {West}}]{yacoby_e4}%
  \BibitemOpen
  \bibfield  {author} {\bibinfo {author} {\bibfnamefont {V.}~\bibnamefont {Venkatachalam}}, \bibinfo {author} {\bibfnamefont {A.}~\bibnamefont {Yacoby}}, \bibinfo {author} {\bibfnamefont {L.}~\bibnamefont {Pfeiffer}},\ and\ \bibinfo {author} {\bibfnamefont {K.}~\bibnamefont {West}},\ }\bibfield  {title} {\bibinfo {title} {Local charge of the $\nu=5/2$ fractional quantum hall state},\ }\href@noop {} {\bibfield  {journal} {\bibinfo  {journal} {Nature}\ }\textbf {\bibinfo {volume} {469}},\ \bibinfo {pages} {185} (\bibinfo {year} {2011})}\BibitemShut {NoStop}%
\bibitem [{\citenamefont {Alkalai}\ \emph {et~al.}(2026)\citenamefont {Alkalai}, \citenamefont {Hajigeorgiou}, \citenamefont {Gupta}, \citenamefont {Senapati}, \citenamefont {Tiwari}, \citenamefont {Tai}, \citenamefont {Singh}, \citenamefont {Baldwin}, \citenamefont {Pfeiffer}, \citenamefont {Shayegan}, \citenamefont {Banerjee},\ and\ \citenamefont {Heiblum}}]{heiblum3}%
  \BibitemOpen
  \bibfield  {author} {\bibinfo {author} {\bibfnamefont {T.}~\bibnamefont {Alkalai}}, \bibinfo {author} {\bibfnamefont {E.}~\bibnamefont {Hajigeorgiou}}, \bibinfo {author} {\bibfnamefont {A.}~\bibnamefont {Gupta}}, \bibinfo {author} {\bibfnamefont {T.}~\bibnamefont {Senapati}}, \bibinfo {author} {\bibfnamefont {P.}~\bibnamefont {Tiwari}}, \bibinfo {author} {\bibfnamefont {C.-T.}\ \bibnamefont {Tai}}, \bibinfo {author} {\bibfnamefont {S.~K.}\ \bibnamefont {Singh}}, \bibinfo {author} {\bibfnamefont {K.~W.}\ \bibnamefont {Baldwin}}, \bibinfo {author} {\bibfnamefont {L.~N.}\ \bibnamefont {Pfeiffer}}, \bibinfo {author} {\bibfnamefont {M.}~\bibnamefont {Shayegan}}, \bibinfo {author} {\bibfnamefont {M.}~\bibnamefont {Banerjee}},\ and\ \bibinfo {author} {\bibfnamefont {M.}~\bibnamefont {Heiblum}},\ }\href {https://arxiv.org/abs/2602.08468} {\bibinfo {title} {Observation of e/4 charge at $\nu=1/2$ in gaas}} (\bibinfo {year} {2026}),\ \Eprint {https://arxiv.org/abs/2602.08468} {arXiv:2602.08468 [cond-mat.mes-hall]}
  \BibitemShut {NoStop}%
\bibitem [{\citenamefont {Nakamura}\ \emph {et~al.}(2019)\citenamefont {Nakamura}, \citenamefont {Fallahi}, \citenamefont {Sahasrabudhe}, \citenamefont {Rahman}, \citenamefont {Liang}, \citenamefont {Gardner},\ and\ \citenamefont {Manfra}}]{manfra3}%
  \BibitemOpen
  \bibfield  {author} {\bibinfo {author} {\bibfnamefont {J.}~\bibnamefont {Nakamura}}, \bibinfo {author} {\bibfnamefont {S.}~\bibnamefont {Fallahi}}, \bibinfo {author} {\bibfnamefont {H.}~\bibnamefont {Sahasrabudhe}}, \bibinfo {author} {\bibfnamefont {R.}~\bibnamefont {Rahman}}, \bibinfo {author} {\bibfnamefont {S.}~\bibnamefont {Liang}}, \bibinfo {author} {\bibfnamefont {G.~C.}\ \bibnamefont {Gardner}},\ and\ \bibinfo {author} {\bibfnamefont {M.~J.}\ \bibnamefont {Manfra}},\ }\bibfield  {title} {\bibinfo {title} {Aharonov–bohm interference of fractional quantum hall edge modes},\ }\href@noop {} {\bibfield  {journal} {\bibinfo  {journal} {Nature Physics}\ }\textbf {\bibinfo {volume} {15}},\ \bibinfo {pages} {563} (\bibinfo {year} {2019})}\BibitemShut {NoStop}%
\bibitem [{\citenamefont {Rosenow}\ and\ \citenamefont {Stern}(2020)}]{rosenow1}%
  \BibitemOpen
  \bibfield  {author} {\bibinfo {author} {\bibfnamefont {B.}~\bibnamefont {Rosenow}}\ and\ \bibinfo {author} {\bibfnamefont {A.}~\bibnamefont {Stern}},\ }\bibfield  {title} {\bibinfo {title} {Flux superperiods and periodicity transitions in quantum hall interferometers},\ }\href@noop {} {\bibfield  {journal} {\bibinfo  {journal} {Physical Review Letters}\ }\textbf {\bibinfo {volume} {124}},\ \bibinfo {pages} {106805} (\bibinfo {year} {2020})}\BibitemShut {NoStop}%
\bibitem [{\citenamefont {Nakamura}\ \emph {et~al.}(2022)\citenamefont {Nakamura}, \citenamefont {Liang}, \citenamefont {Gardner},\ and\ \citenamefont {Manfra}}]{manfra4}%
  \BibitemOpen
  \bibfield  {author} {\bibinfo {author} {\bibfnamefont {J.}~\bibnamefont {Nakamura}}, \bibinfo {author} {\bibfnamefont {S.}~\bibnamefont {Liang}}, \bibinfo {author} {\bibfnamefont {G.~C.}\ \bibnamefont {Gardner}},\ and\ \bibinfo {author} {\bibfnamefont {M.~J.}\ \bibnamefont {Manfra}},\ }\bibfield  {title} {\bibinfo {title} {Impact of bulk-edge coupling on observation of anyonic braiding statistics in quantum hall interferometers},\ }\href@noop {} {\bibfield  {journal} {\bibinfo  {journal} {Nature Communications}\ }\textbf {\bibinfo {volume} {13}},\ \bibinfo {pages} {344} (\bibinfo {year} {2022})}\BibitemShut {NoStop}%
\bibitem [{\citenamefont {Feldman}\ and\ \citenamefont {Halperin}(2021{\natexlab{b}})}]{halperin3}%
  \BibitemOpen
  \bibfield  {author} {\bibinfo {author} {\bibfnamefont {D.~E.}\ \bibnamefont {Feldman}}\ and\ \bibinfo {author} {\bibfnamefont {B.~I.}\ \bibnamefont {Halperin}},\ }\bibfield  {title} {\bibinfo {title} {Fractional charge and fractional statistics in the quantum hall effects},\ }\href@noop {} {\bibfield  {journal} {\bibinfo  {journal} {Rep. Prog. Phys.}\ }\textbf {\bibinfo {volume} {84}},\ \bibinfo {pages} {076501} (\bibinfo {year} {2021}{\natexlab{b}})}\BibitemShut {NoStop}%
\end{thebibliography}%

\onecolumngrid

\newpage

\section*{Extended Data Figures}
\setcounter{figure}{0}
\renewcommand{\figurename}{Extended Data Fig.}
\renewcommand{\thefigure}{\arabic{figure}}

\begin{figure}[H]
\centering
\makebox[\textwidth][c]{\includegraphics[width=1\textwidth]{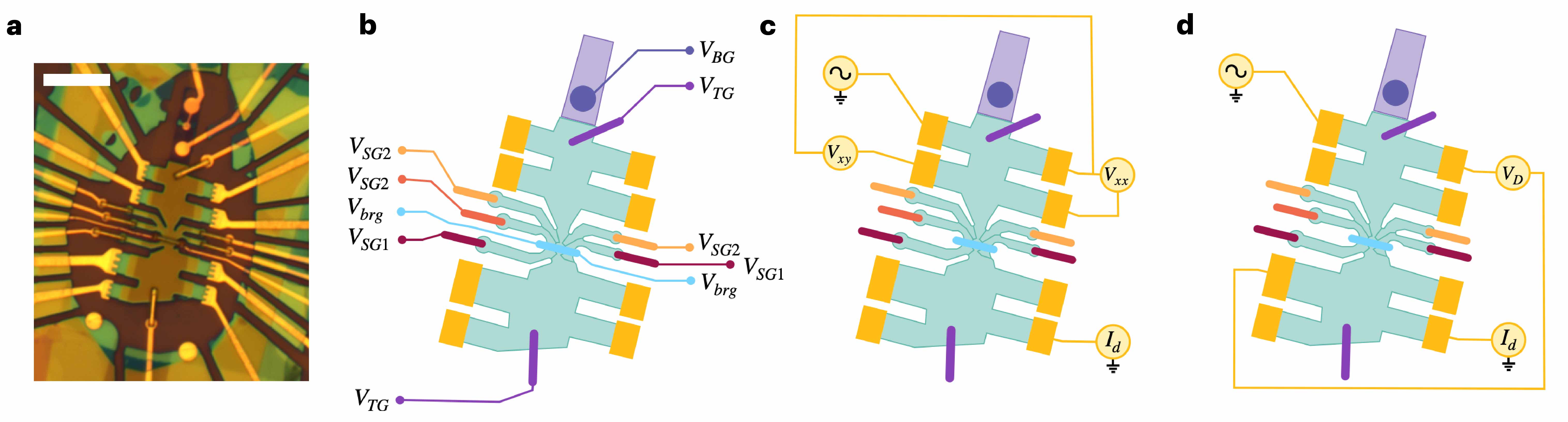}}
\caption{\textbf{Device contacts, gates, and measurement setup.} \textbf{a,} Optical image of measured device, with a scale bar of 10~$\mu\text{m}$. \textbf{b,} Schematic of the gate setup shown in Fig.~\ref{FIG:1}. Each gate is contacted via a metallic bridge. The top gate and bottom gate set the density in the graphene, while the split gates and plunger gate define the interferometry cavity. The bridge gate controls the AD in the center of the cavity. \textbf{c,} Schematic of the measurement of quantum Hall transport reported in Fig.~\ref{FIG:1}c and Extended Fig.~\ref{fig:BG_states}. The metallic contacts connect to the BLG in regions without top and bottom gates, with the silicon back gate doping this region. The voltage probes are labeled $V_{xx}$ and $V_{xy}$ for the longitudinal and Hall measurements, respectively. The source contact is indicated by the AC bias and the drain current is labeled $I_d$. \textbf{d,} Schematic of the  measurement for diagonal conductance $G_D = I_d/V_D$ reported throughout the paper. The diagonal voltage drop across the interferometry cavity measures the conductance through the QPCs.}
\label{fig:measurement_setup}
\end{figure}

\begin{figure}[H]
\centering
\makebox[\textwidth][c]{\includegraphics[width=1\textwidth]{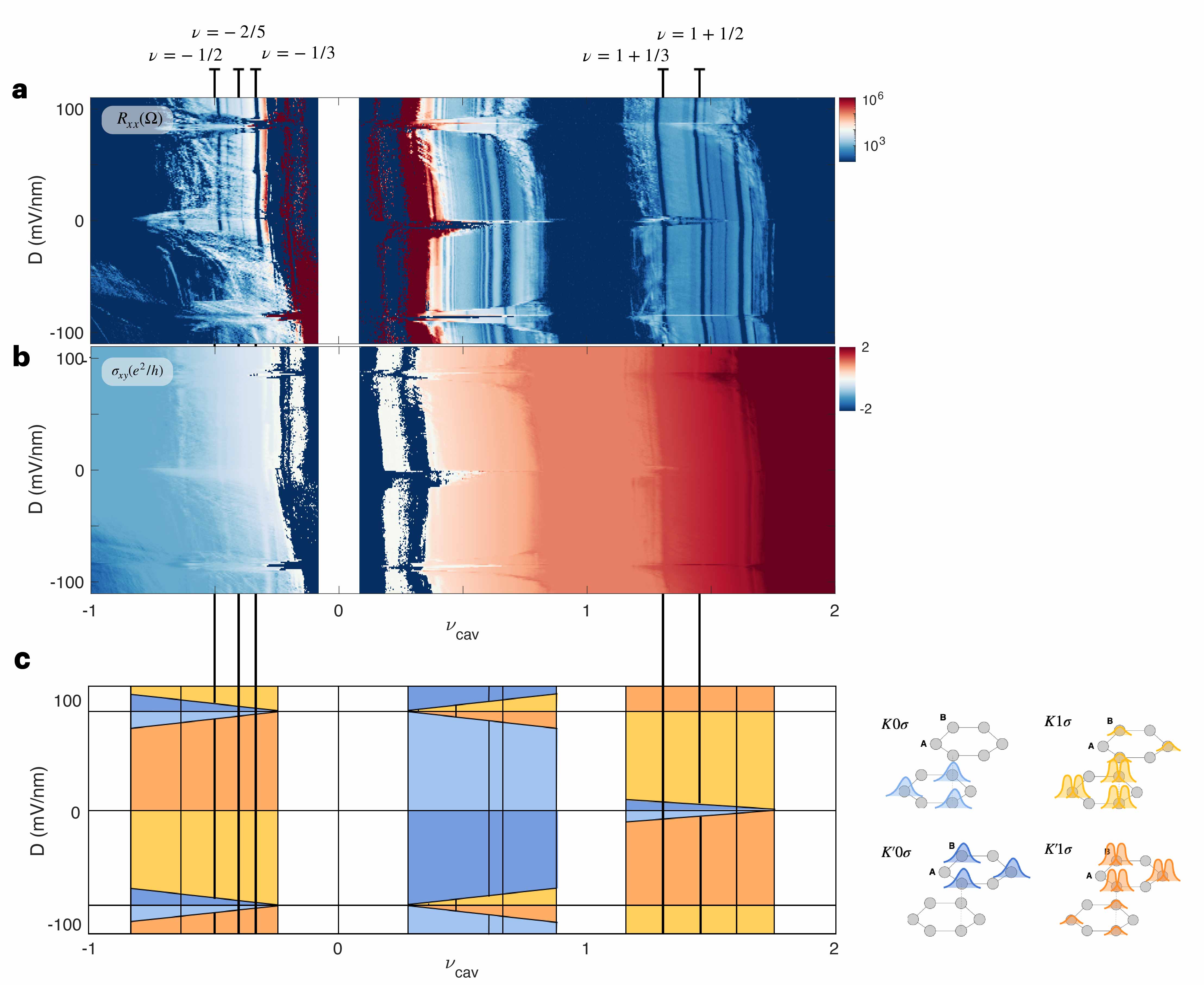}}
\caption{\textbf{BLG FQH states.} \textbf{a,} Longitudinal resistance $R_{xx}$ of the filling $\nu$ with respect to the displacement field $D$ for $-1<\nu<2$. The five states measured in this experiment $\nu = -1/2, -2/5, -2/3, 1+1/3, \text{ and } 1+1/2$ are persistent throughout large ranges of $D$. The obscuring of the hole-like states ($\nu<0$) for $D<0$ is attributed to issues with our metallic contacts to the BLG layer. \textbf{b,} Quantum Hall plateaus in $\sigma_{xy}$ in the same range. \textbf{c,} Schematic of states present in the measurement, with colors indicating the eight possible ground states of $K0\sigma, K'0\sigma, K1\sigma, K'1\sigma$ for the spin index $\sigma=\pm$, the valley index $K,K'$ and the orbital index $N=0,1$. The five states discussed in this work are indicated by thick black lines. Note that our experiments were carried out only in the cavity states with orbital index $N=1$ (yellow and orange), as these states assume a more complex wavefunction necessary for even-denominator states to develop. All measurements were performed at $T=20\, \si{mK}$. and $B=9.95\, \si{T}$.}
\label{fig:BG_states}
\end{figure}

\begin{figure}[H]
\centering
\makebox[\textwidth][c]{\includegraphics[width=1\textwidth]{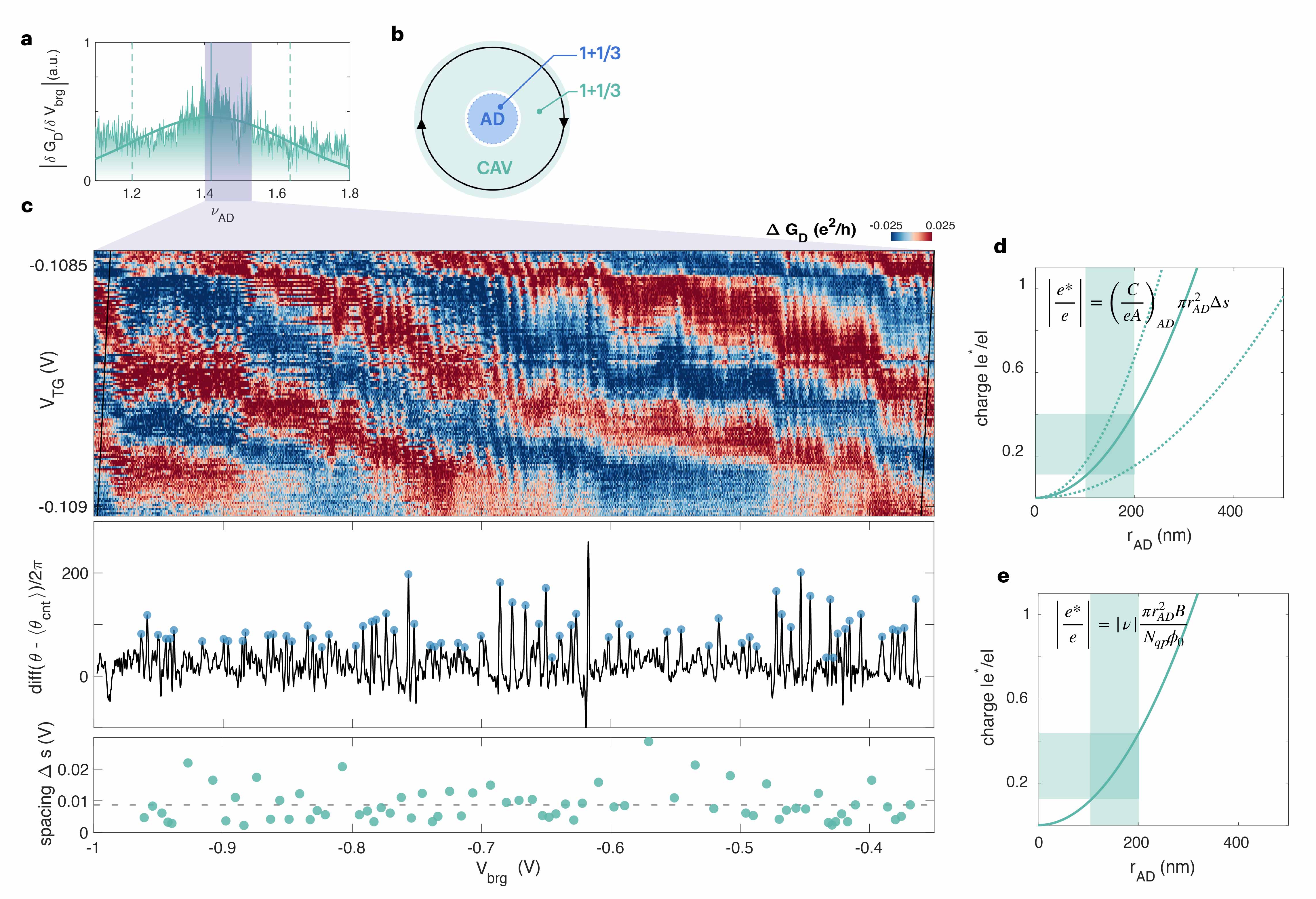}}
\caption{\textbf{Wide range of phase slips in $\nu_{cav}\approx\nu_{AD}=1+1/3$.} \textbf{a,} Zoom-in of Fig.\ref{FIG:4} of $|\delta G_D/\delta V_{brg}|$ as a function of $\nu_{AD}$ to show the width of the phase slip regime for $\nu_{cav}\approx\nu_{AD}=1+1/3$. \textbf{b,} Schematic of the interferometry cavity and AD to clarify fillings. \textbf{c,} Derivative of 1D FFT parallel to the phase slip slopes is plotted in the second panel after subtracting the continuous phase $\theta_{cnt}$ to show each phase slip as a peak in the signal. The spacing $\Delta s$ between each jump is extracted and plotted in the bottom panel. We count $N \geq 70$ consecutive jumps (extrapolated to $N_{ps} \geq 300$ over the entire jump range) and find an average spacing of $\Delta s = 0.0087 \pm 0.0055$. Note that for the less prominent phase slips, it is difficult to extract a phase slip size (the sizes of a phase slip $\Delta \theta_{ps}$ are the area under the peaks in the middle panel). For many less prominent jumps than those studied in the main text, $\Delta \theta_{ps}$ is often \textit{less} than the theoretically expected value. This hints towards bulk-edge coupling strength $\zeta = K_{IL}/K_{I}$ decreasing $\Delta \theta_{ps}$ by allowing the area of the interferometer to change slightly each time a new quasiparticle jumps into the cavity. For fractional fillings $\tilde{\nu}$, bulk-edge coupling decreases $\Delta \theta$ according to \cite{manfra4, halperin_theory_of_fp} $\Delta\theta/2\pi = \theta_a/2\pi - \zeta (e^*_{loc}/e)/\tilde{\nu}$. A similar decrease in $\Delta \theta_{ps}$ was attributed to bulk-edge coupling in previous interferometer experiments \cite{manfra1, manfra2, andrea_third}. \textbf{d--e,} Two relations between AD radius and charge jumping into the AD discussed in Methods based on \textbf{(d)} the gate voltage spacing between the phase slips and \textbf{(e)} the number of jumps. For reasonable AD radii values of $r_{AD} = 100-200\, \si{nm}$, we extract fractional charge consistent with $|e^*/e|= 1/3$ and inconsistent with $|e^*/e|= 1$. All measurements were performed at $T=20\, \si{mK}$. and $B=9.95\, \si{T}$.}
\label{fig:many_jumps}
\end{figure}

\begin{figure}[H]
\centering
\makebox[\textwidth][c]{\includegraphics[width=1\textwidth]{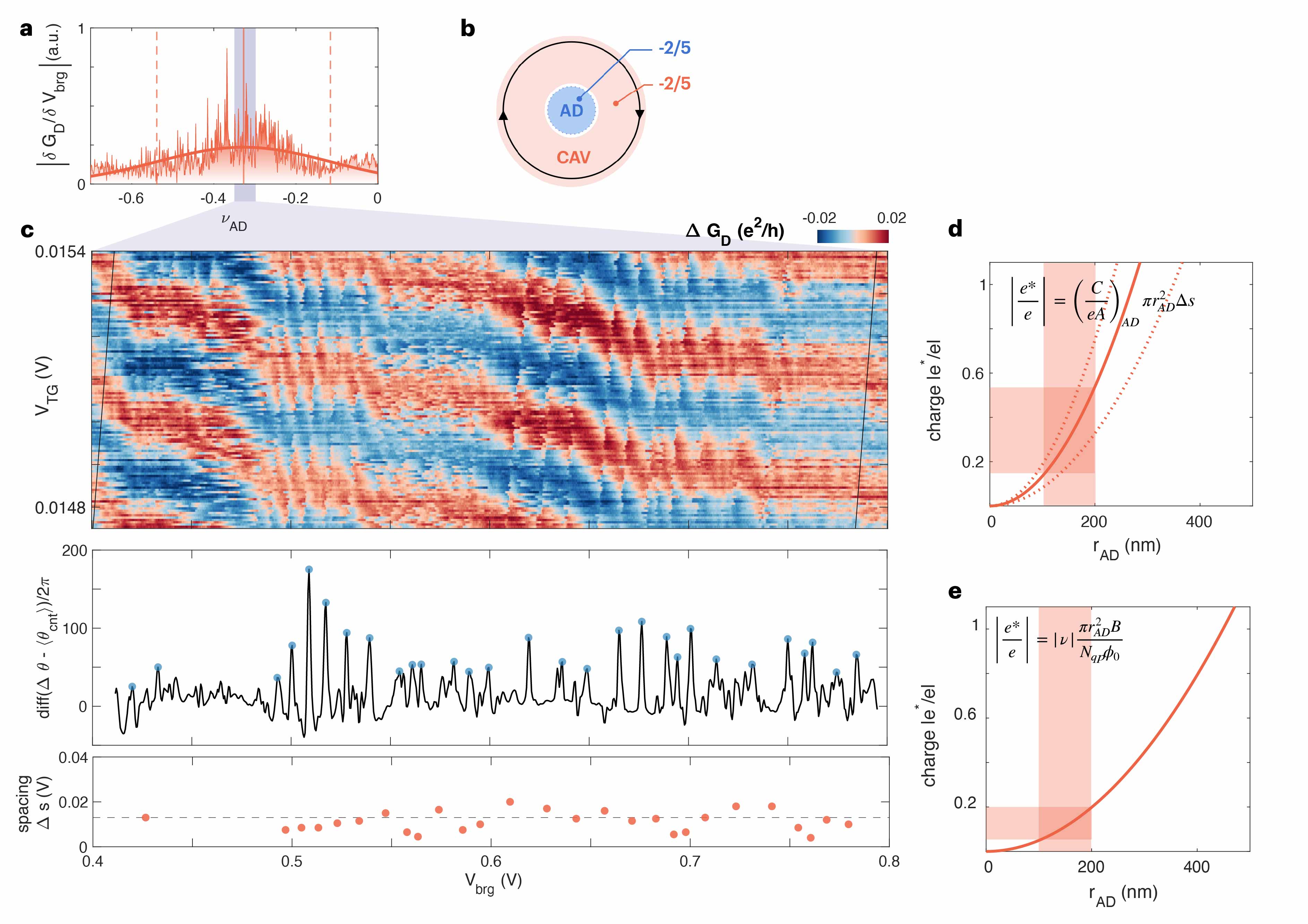}}
\caption{\textbf{Wide range of phase slips in $\nu_{cav}\approx\nu_{AD}=-2/5$.} \textbf{a,} Zoom-in of Fig.\ref{FIG:4} of $|\delta G_D/\delta V_{brg}|$ as a function of $\nu_{AD}$ to show the width of the phase slip regime for $\nu_{cav}\approx\nu_{AD}=-2/5$. \textbf{b,} Schematic of the interferometry cavity and AD to clarify fillings. \textbf{c,} 1D FFT parallel to the phase slip slopes is plotted in the second panel after subtracting the continuous phase $\theta_{cnt}$ to show each phase slip as a peak in the signal (see Method). The spacing $\Delta s$ between each phase slip is extracted and plotted in the bottom panel. We count $N \geq 29$ consecutive phase slips (extrapolated to $N_{ps} \geq 200$ over the entire jump range) and find an average spacing of $\Delta s = 0.0113 \pm 0.0157\, \si{V}$. See Extended Fig.~\ref{fig:many_jumps} for further discussion on less prominent phase slips. \textbf{d--e,} Two relations between AD radius and charge jumping into the AD discussed in Methods based on \textbf{(d)} the spacing of the phase slips and \textbf{(e)} the number of phase slips. For reasonable AD radii values of $r_{AD} = 100-200\, \si{nm}$, we extract fractional charge consistent with $|e^*/e|= 1/5$ and inconsistent with $|e^*/e|= 1$. All measurements were performed at $T=20\, \si{mK}$. and $B=9.95\, \si{T}$.}
\label{fig:many_jumps_25}
\end{figure}

\begin{figure}[H]
\centering
\makebox[\textwidth][c]{\includegraphics[width=1\textwidth]{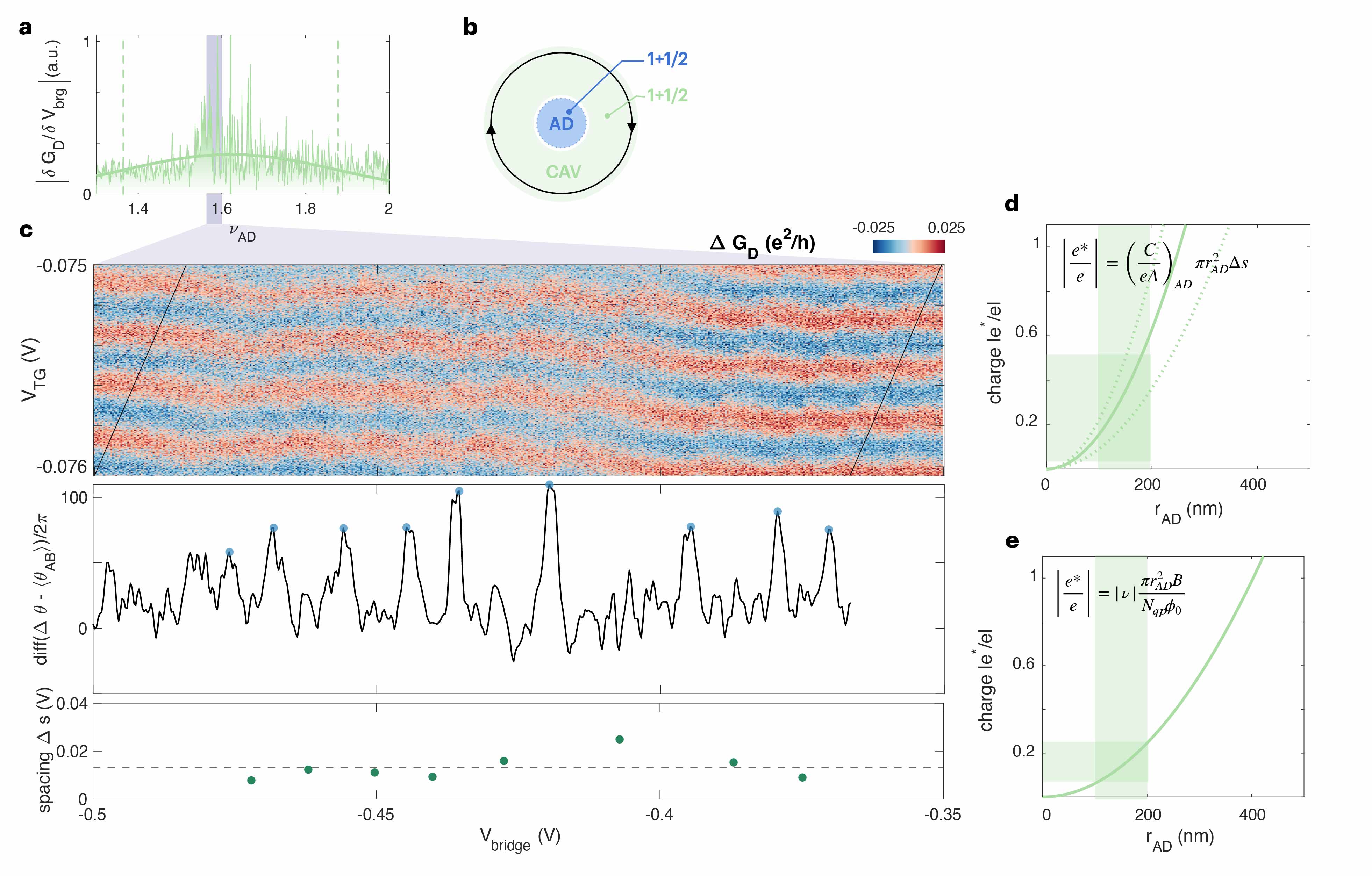}}
\caption{\textbf{Wide range of phase slips in $\nu_{cav}\approx\nu_{AD}=1+1/2$.} \textbf{a,} Zoom-in of Fig.\ref{FIG:4} of $|\delta G_D/\delta V_{brg}|$ as a function of $\nu_{AD}$ to show the width of the jump regime for $\nu_{cav}\approx\nu_{AD}=1+1/2$. \textbf{b,} Schematic of the interferometry cavity and AD to clarify fillings. \textbf{c,} 1D FFT parallel to the phase slip slopes is plotted in the second panel after subtracting the continuous phase $\theta_{cnt}$ to show each phase slip as a peak in the signal. The spacing $\Delta s$ between each phase slip is extracted and plotted in the bottom panel. We count $N \geq 9$ consecutive slips (extrapolated to $N_{ps} \geq 200$ over the entire jump range) and find an average spacing of $\Delta s = 0.0132 \pm 0.0056\, \si{V}$. See Extended Fig.~\ref{fig:many_jumps} for further discussion on less prominent phase slips. \textbf{d--e,} Two relations between AD radius and charge jumping into the AD discussed in Methods based on \textbf{(d)} the spacing of the phase slips and \textbf{(e)} the number of phase slips. For reasonable AD radii values of $r_{AD} = 100-200\, \si{nm}$, we extract fractional charge consistent with $|e^*/e|= 1/4$ and inconsistent with $|e^*/e|= 1$. All measurements were performed at $T=20\, \si{mK}$. and $B=9.95\, \si{T}$.}
\label{fig:many_jumps_32}
\end{figure}

\begin{figure}[H]
\centering
\makebox[\textwidth][c]{\includegraphics[width=1\textwidth]{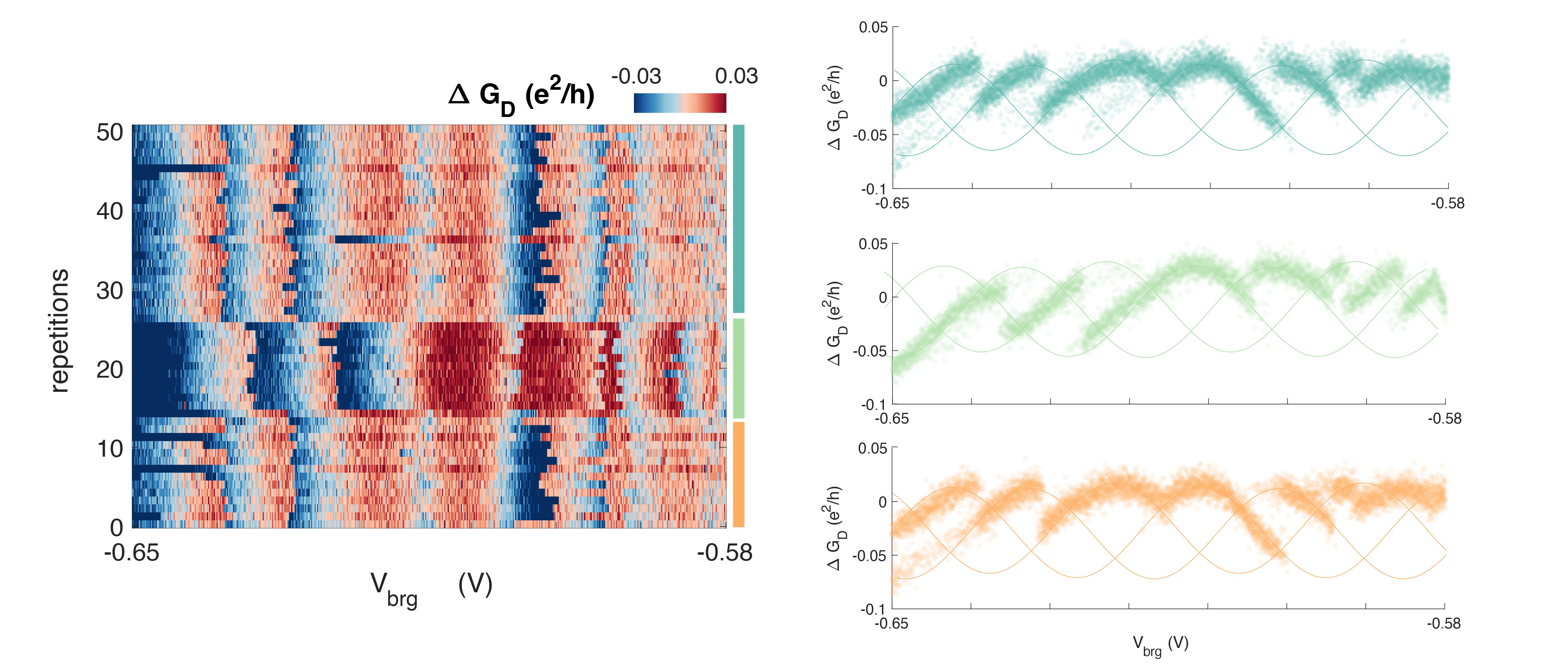}}
\caption{\textbf{Stability of the phase slips.} We sweep the bridge gate voltage repeatedly to show the stochastic nature of phase slip over the gate range corresponding to $\nu_{cav} = 1+1/3$ and $0.5571<\nu_{AD}<0.5682$. The diagonal conductance jumps between three sinusoids, as seen on the right panels. The full sinusoids are guides to the eye. Each line scan takes $t_{line} \sim 3.5\, \si{min}$. We observe that the jumps occur always around the same $V_{brg}$, indicating that they are indeed induced by a change in the potential landscape of the AD by the bridge gate. We note a change in the number of quasiparticles in the AD at around 15 repetitions, which may be due to random telegraph noise \cite{tom_james_braiding}. These switches are rare, appearing only on the order of hours or days, and were only observed in $\nu_{cav} = 1+1/3$. The system re-stabilizes itself to its original configuration within minutes, as seen by the second switch around 25 repetitions. All measurements were performed at $T=20\, \si{mK}$. and $B=9.95\, \si{T}$.}
\label{fig:repeated_sweep}
\end{figure}

\begin{figure}[H]
\centering
\makebox[\textwidth][c]{\includegraphics[width=1\textwidth]{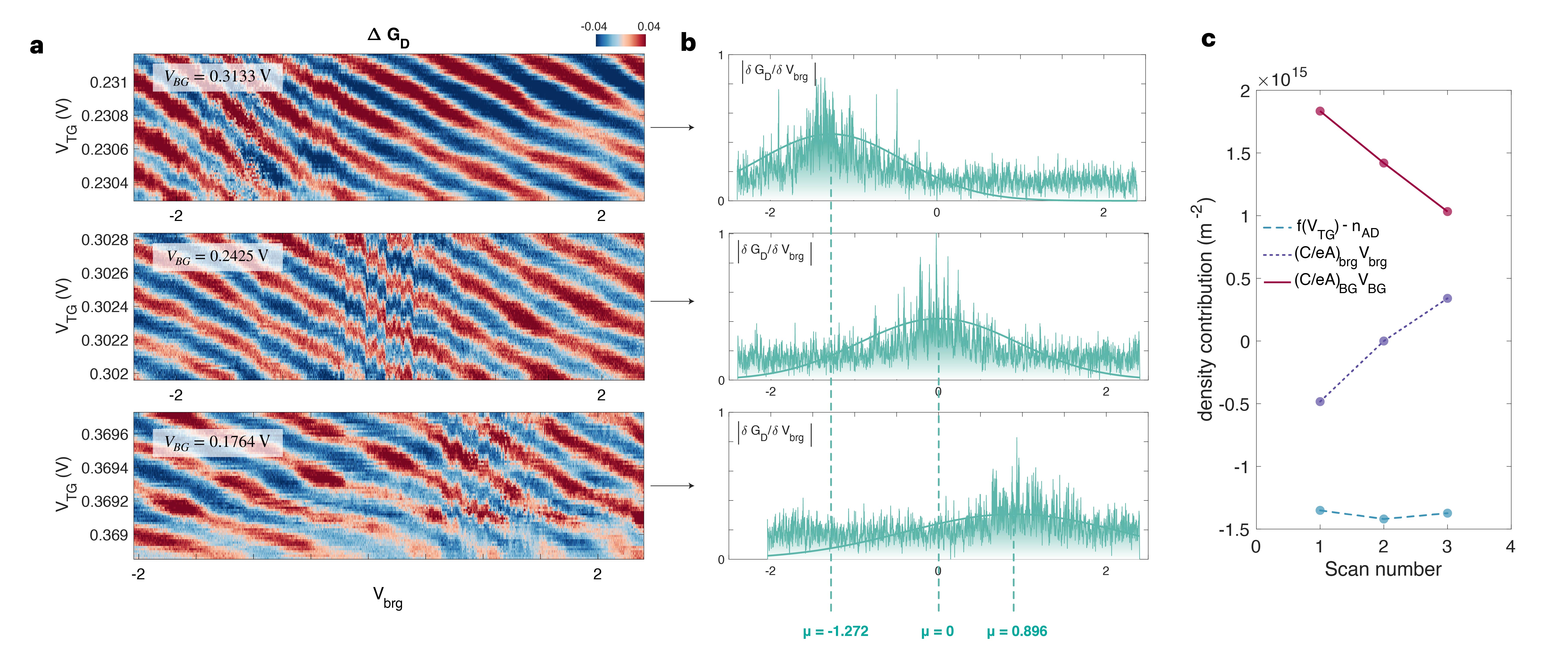}}
\caption{\textbf{Determination of effect of $f(V_{TG})$ on the density of the AD.} \textbf{a,} 2D scans of $V_{brg}$ vs. $V_{TG}$ at three different BG values $V_{BG} = 0.3133, 0.2425, 0.1764\, \si{V}$ showing the location of the jumps moving to different $V_{brg}$ values. \textbf{b,} Absolute value of the derivative per row of each scan in \textbf{(a)}. We fit a Gaussian distribution to each peak to extract the center of the jumps, which gives us $V_{brg}$ to keep $n_{AD}$ fixed. \textbf{c,} Plot of the contribution of the change bridge gate (purple) and the BG (pink) on the density, as well as the contribution of their sum (blue). We observe that $f( V_{TG}) - n_{AD} = -(C/eA)_{brg}V_{brg} - (C/eA)_{BG}V_{BG}$ remains flat across the scans, as $n_{AD}$ is set to remain constant. This means that $f(V_{TG})$ does not play a significant role on the density of the central part of the AD when compared to the BG and bridge. All measurements were performed at $T=20\, \si{mK}$. and $B=9.95\, \si{T}$.}
\label{fig:AD_filling_det}
\end{figure}

\begin{figure}[H]
\centering
\makebox[\textwidth][c]{\includegraphics[width=1\textwidth]{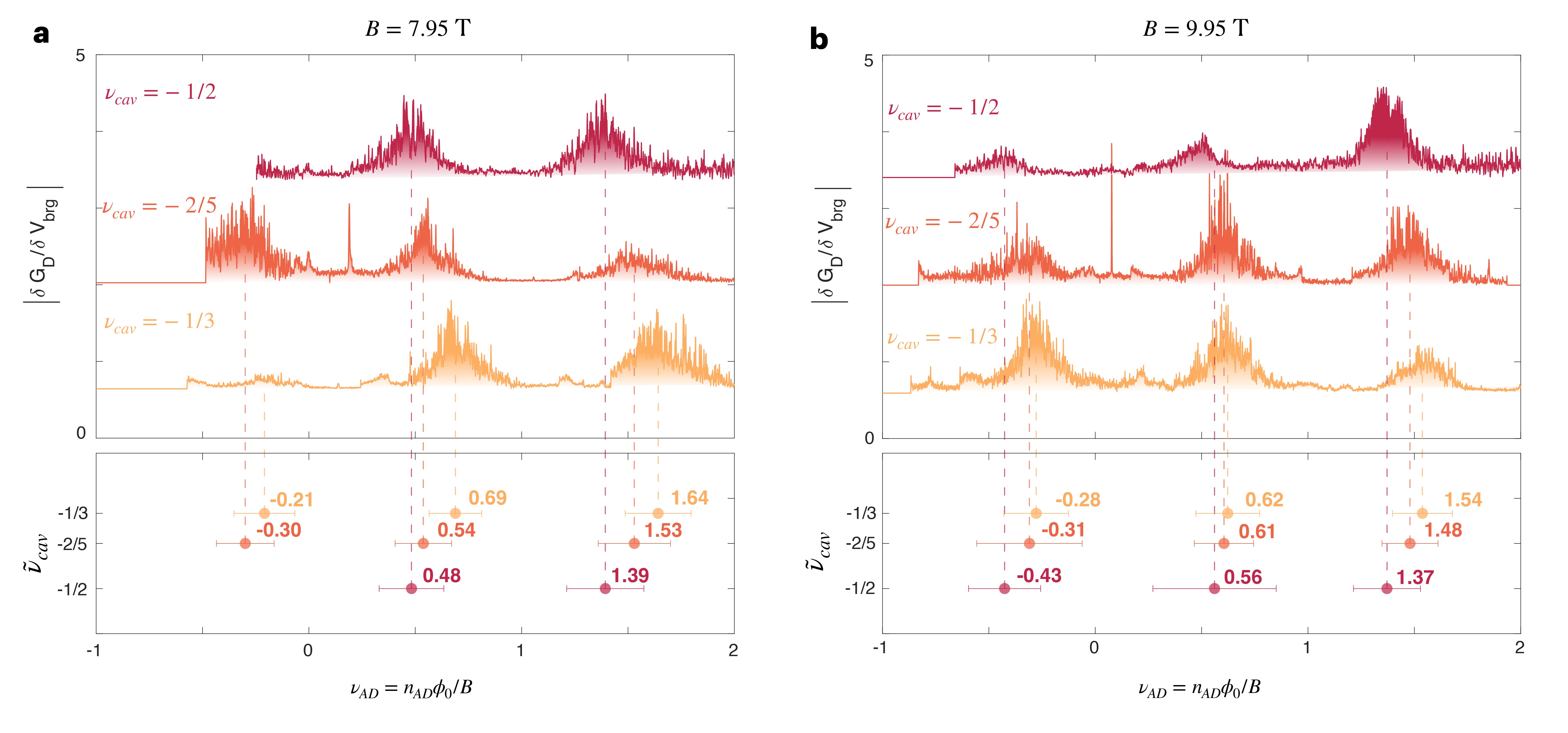}}
\caption{\textbf{Phase slip locations at $B=7.95\, \si{T}$ and $B=9.95\, \si{T}$.} \textbf{a,} Plot of the absolute value of the derivative if the conductance with respect to $V_{brg}$ for $B=7.95\, \si{T}$, similar to the method used in FIG.~\ref{FIG:4}. Lower panel shows mean and standard deviation of each peak found in upper panel. Phase slip peaks are found at $\nu_{AD} = -0.21, 0.69, 1.64$ for $\nu_{cav} = -1/2$,  $\nu_{AD} = -0.30, 0.54, 1.53$ for $\nu_{cav} = -2/5$, and $\nu_{AD} =0.48, 1.39$ for $\nu_{cav} = -1/3$. \textbf{b,} Same as \textbf{(a)} at $B=9.95\, \si{T}$. Phase slips are found at $\nu_{AD} = -0.28, 0.62, 1.54$ for $\nu_{cav} = -1/2$,  $\nu_{AD} = -0.31, 0.61, 1.48$ for $\nu_{cav} = -2/5$, and $\nu_{AD} =-0.43, 0.56, 1.37$ for $\nu_{cav} = -1/3$. We therefore observe the phase slips at the same fillings $\nu_{AD}$ at both fields, confirming our assignment of $\nu_{AD}$ using the electrostatic relation expressed in Eq.~(\ref{eq:nuAD}) ignoring $f(V_{TG})$.}
\label{fig:mult_field}
\end{figure}

\begin{figure}[H]
\centering
\makebox[\textwidth][c]{\includegraphics[width=1\textwidth]{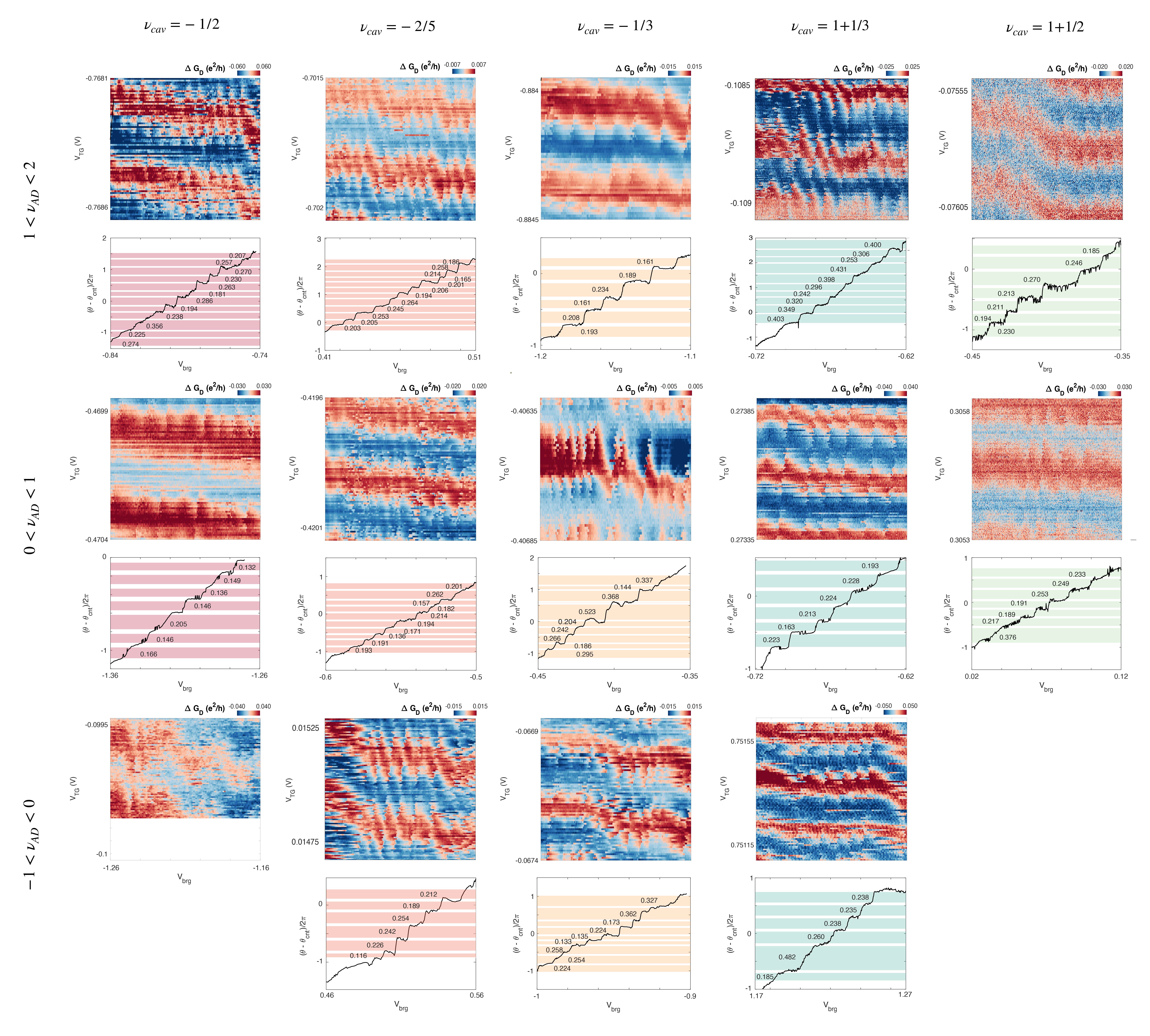}}
\caption{\textbf{Phase slips in all $\nu_{cav}, \nu_{AD}$ combinations.} 2D phase maps of $V_{brg}$ as a function of $V_{TG}$ and their phase slip magnitude FFT analysis for all AD and cavity fillings discussed in the main text. The FFT analysis follows the same procedure outlined in Methods and demonstrated in Fig.~\ref{FIG:3}. All measurements were performed at $T=20\, \si{mK}$. and $B=9.95\, \si{T}$.}
\label{fig:all_jumps}
\end{figure}

\begin{figure}[H]
\centering
\makebox[\textwidth][c]{\includegraphics[width=1\textwidth]{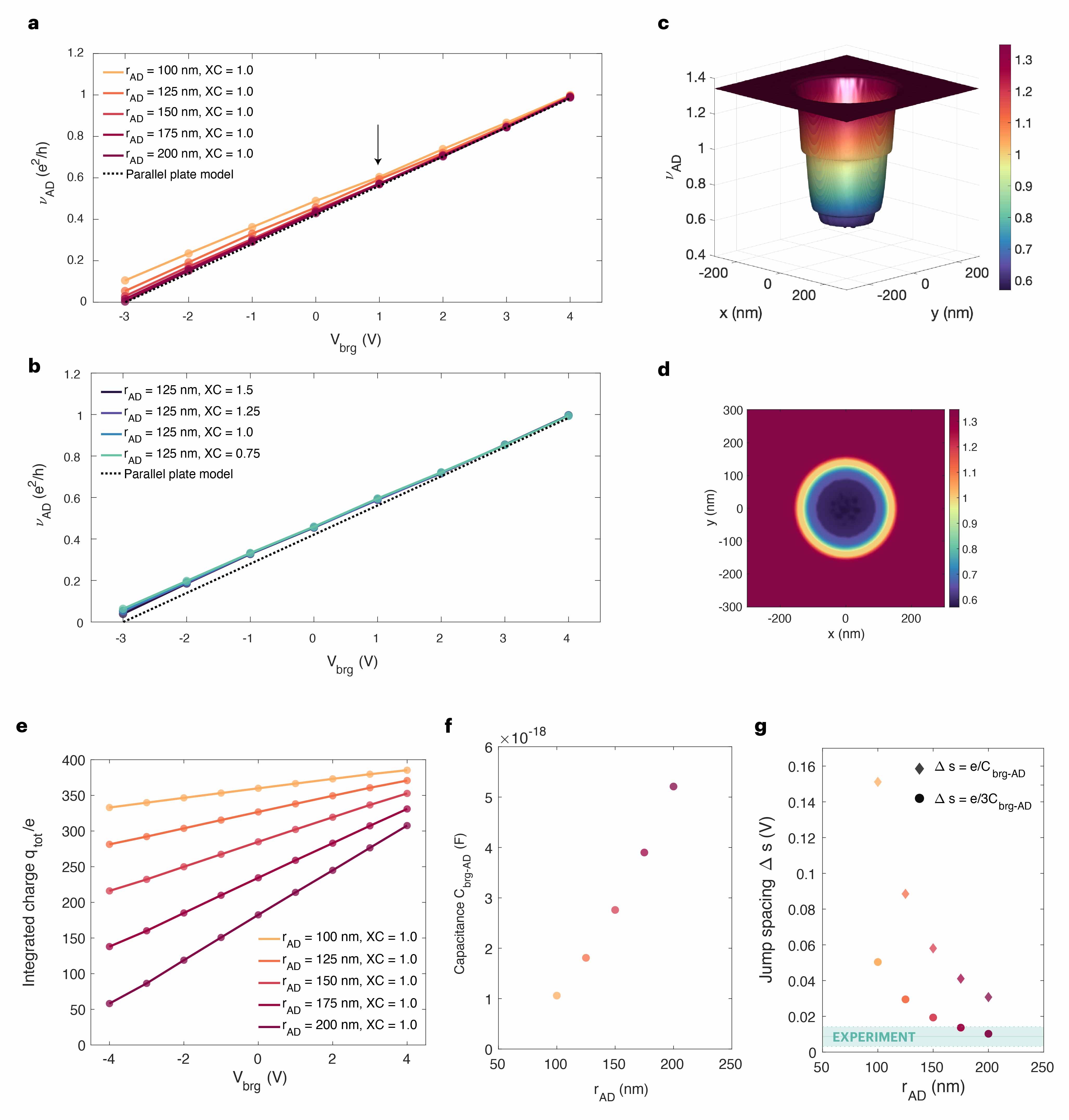}}
\caption{\textbf{Density profile simulations of the AD in the cavity in $\nu_{cav} = 1+1/3$ as a function of the bridge gate voltage.} \textbf{a,} Calculated filling at the center of the AD $\nu_{AD}$ as the bridge gate voltage $V_{brg}$ is swept. Simulations are performed at various reasonable AD sizes $r_{AD} = 100-200\, \si{nm}$ with exchange correlation factor $XC = 1$ (See SI more details). We observe that the calculated $\nu_{AD}$ agrees well with those obtained from the parallel plate capacitor model used in the main text across all parameters. \textbf{b,} Calculated $\nu_{AD}$ as $V_{brg}$ is swept at various $XC = 0.75-1.5$ with $r_{AD} = 125\, \si{nm}$. We observe that the calculated $\nu_{AD}$ agrees well with those obtained from the parallel plate capacitor model in the main text across all parameters. \textbf{c--d,} Example density profile for $V_{brg}=1\, \si{V}$, $r_{AD} = 150\, \si{nm}$, $XC=1.0$,  as indicated by the black arrow in \textbf{(a)}. We observe two compressible rings
for  $\nu_{AD}$ between $1$ and $4/3$ and for  $\nu_{AD}$ between $2/3$ and $1$.
\textbf{e,} Total charge integrated over the AD area in the Thomas-Fermi simulation as a function of $V_{brg}$ at various $r_{AD} = 100-200\, \si{nm}$ and $XC=1$. \textbf{f,} Capacitance extracted from the slopes in \textbf{(a)} for each AD radius. \textbf{g,} Phase slips spacing $\Delta s$ calculated from the capacitance at each AD radius for both charges $\Delta Q=e$ and $\Delta Q=e/3$. The jump spacing values agree with the experimental data for $\Delta Q=e/3$ and $r_{AD} \approx 150-200\, \si{nm}$, further confirming that fractional charge is entering the AD.
}
\label{fig:rik1}
\end{figure}

\newpage
\section{Supplementary Information}

\setcounter{figure}{0}
\renewcommand{\figurename}{SI Fig.}
\renewcommand{\thefigure}{\arabic{figure}}

\subsection{Sign Conventions}

Throughout the paper our conventions are the following:
\begin{itemize}
    \item $e=1.6\times 10^{-19}\, \si{C} > 0$ is the {\it opposite} charge of the electron.
    \item The filling $\nu>0$ when electrons are doped, and $\nu<0$ when holes are doped.
    \item $B>0$ refers to a magnetic field that points in the $-\hat{z}$ direction. Below we assume $B>0$.
    \item The cyclotron orbits for the electrons are clockwise.
    \item  Charges on the edge move counterclockwise for positive integer $\nu_{cav}$ when interfaced with electron vacuum or with a region of lower $\nu$.  Charges on the edge move clockwise for a cavity with negative integer $\nu_{cav}$ interfaced with the vacuum or a region of positive $\nu$.
    \item The flux quantum is defined as $\phi_0=h/e$. Therefore, the Aharonov-Bohm phase when a particle with charge $q$ moves around an area $A$ with magnetic field $B_z=-B$ in the counterclockwise direction will be \begin{equation}
        \theta_{AB}=\frac{qAB_z}{\hbar} = -2\pi\frac{q}{e} \frac{AB}{\phi_0}.
    \end{equation}
    \item $\theta_a(q,q')$ denotes the mutual braiding phase an anyon of charge $q$ encircles another with charge $q'$ in the counterclockwise direction.
    For quasiholes carrying charge $q=e/3$ in a $\nu=1/3$ Laughlin state, $\theta_a(q,q)=2\pi/3$.
    \item The number of quasiparticles $N_{qp}$ with charge $q'$ in the cavity is given by $Q_{qp}/q'$, where $Q_{qp}$ is the total charge associated with quasiparticles.
    \item Thus, when a charge $q$ particle encircles an area $A$ pierced by a magnetic field $B$ with a total of $N_{qp}$ quasiparticles of charge $q'$ in the counterclockwise direction, it obtains the phase
    \begin{equation}
        \theta' = -2\pi\frac{q}{e} \frac{AB}{\phi_0} + N_{qp}\theta_a(q,q').
    \end{equation}
    \item Experimentally, we can only measure $\cos(\theta')$, which is an even function of $\theta'$. We thus choose to study $\theta_{tot}=-\theta'$ and identify $q=e^{*}_{ie}$, $q'=e^{*}_{loc}$, which reduces to Eq.~(\ref{eq:1}). An equivalent physical interpretation is that the interference patterns are caused by quasiparticles and quasiholes on the edge, which contribute opposite phases. We will choose the species of quasiparticle that satisfies the relation $e^{*}_{ie}=\tilde{\nu}_{cav} e$.
\end{itemize}

\subsection{Density profiles for AD fillings in $\nu_{cav}=1+1/3$}
\begin{figure}[H]
\centering
\makebox[\textwidth][c]{\includegraphics[width=1\textwidth]{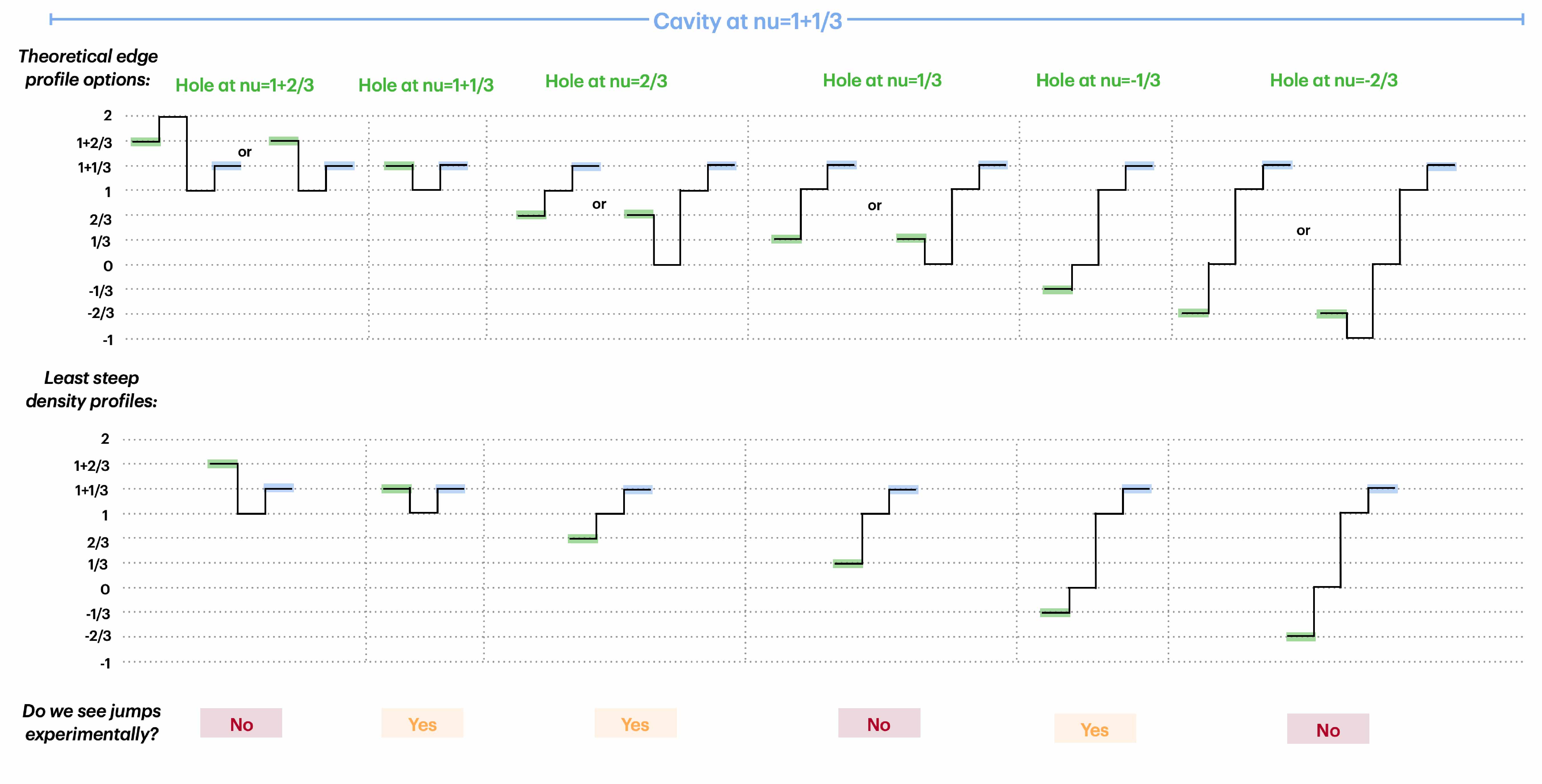}}
\caption{\textbf{Density profile sketches of one side of the AD for $\nu_{cav}=1+1/3$ and $-1<\nu_{AD}<2$.} In the upper panel, we sketch the various possible density profiles on one side of the AD for the experimental range of $\nu_{cav}=1+1/3$ (blue end) and $-1<\nu_{AD}<2$ (green end). In the middle panel, we then pick the density profiles with the least total variation from the upper panel for each $\nu_{AD}$, as strong variations in charge density are penalized electrostatically. In the lower panels, we then note that we observe jumps experimentally only in $\nu_{AD} \approx -1/3, 2/3, 1+1/3$, which corresponds exclusively to the fillings where there is a 1/3 edge both inside the AD and around it, in the cavity.}
\label{fig:density_sketches}
\end{figure}

We sketch a possible explanation for the observation of jumps in $\nu_{cav}=1+1/3$ and $\nu_{AD}\approx-1/3, 2/3, 1+1/3$ shown in Fig. 4 in the main text below it. This explanation relies on the \textit{difference} in density between the AD and the cavity. In SI Fig.~\ref{fig:density_sketches}, we sketch possible AD filling profiles for various combinations of $\nu_{AD}$ with $\nu_{cav}=4/3$.  In cases where several alternative profiles appear, we select the ones with the least steep profile. We notice that the fillings where jumps are visible correspond to the least steep density profiles that have a 1/3 edge both in the AD and in the cavity. If this simple picture is correct, this would imply that phase slips only occur when there are two well-formed 1/3 edges both inside the AD and outside the AD (in the surrounding cavity), carrying charges quantized in $e/3$. However, they do not occur when the edge inside the AD reconstructs to a 2/3 edge, carrying charge quantized to $2e/3$. The microscopic details of the edge state reconstruction at the interface between two adjacent fractional states are not yet well understood, for instance between 1/3 and 2/3 edges.

\subsection{Numerical simulations of the density profile of the AD}
\label{sec:simulations}

We consider an effective classical problem of determining the electron density profile in the integer and fractional quantum Hall regimes by combining a three-dimensional finite-element simulation using the numerical software package provided by COMSOL Multiphysics with additional consideration of a two-dimensional Thomas-Fermi calculation, extending the work presented in Ref.~\cite{assouline}. The former is used to compute a realistic electrostatic external potential, which is difficult to obtain analytically, by modeling the actual device used in the experiment, while the latter is used to obtain the density profile on the BLG surface through the effective classical model described below. To address the subtleties of the overall self-consistency problem, we iterate between the COMSOL and Thomas-Fermi simulations so that the converged density distribution simultaneously satisfies the convergence conditions of the Poisson equation in COMSOL and the Thomas-Fermi minimization.

\subsubsection{COMSOL simulation}
To obtain a realistic external electric potential, we used the commercial finite-element simulation tool COMSOL Multiphysics with the electrostatics package. We replicate the gate structure of the measured interferometer. The modeled hBN has a perpendicular dielectric constant of 3.0, and a parallel one of 6.6. The top and bottom BNs are $25, 30\, \si{nm}$, respectively, close to the measured values of the BNs used in the device via atomic force microscope.
The bridge gate hanging over the hole in the top graphite gate is modeled as $100\, \si{nm}$-thick, $300\, \si{nm}$-wide gold slab that has a fixed potential on its surface, separated by $150\, \si{nm}$ from the top graphite gate. The top and bottom graphite gates are modeled with a finite thickness of $5\, \si{nm}$, and the top graphite has a hole of radius $r_{AD, sim}$, which we set close to the lithographically defined size.

For graphene, we use the surface charge density module, imposing the boundary condition of
\begin{equation}
    \hat{n}\cdot(\mathbf{D}_{{\rm above}}(\mathbf{r})-\mathbf{D}_{{\rm below}}(\mathbf{r}))=(-e)n(\mathbf{r})
\end{equation}
where $\hat{n}$ is the unit vector perpendicular to the graphene, $\mathbf{D}_{{\rm above}}$ and $\mathbf{D}_{{\rm below}}$ are the electric displacement fields above and below the graphene surface, and $n(\mathbf{r})$ is the fixed carrier density of graphene, which is updated iteratively in a self-consistent manner. We also impose the periodic boundary condition on the boundary planes to avoid unrealistic divergence. We then independently set the voltages of the bridge gate, the top graphite gate, and the bottom graphite gate, run the electrostatic simulation, and extract the electric potential on the graphene surface. This potential is used as the external potential input for the Thomas-Fermi simulation.

\subsubsection{Thomas-Fermi simulation}
We minimize the following energy functional as a function of the local density in two dimensions to obtain an electron density profile in the integer and fractional quantum Hall regimes:

\begin{equation} \mathcal{E}[n(\mathbf{r})]=\mathcal{E}_{{\rm H}}+\mathcal{E}_{{\rm ext}}+\mathcal{E}_{{\rm XC}}+\mathcal{E}_{{\rm smooth}}
\end{equation}

The first term is the Hartree energy, the second term is the external potential energy, the third term is the exchange-correlation energy, and the fourth term is the effective smoothing energy. For the Hartree term, we use the analytically derived dual gate screened Coulomb interaction,
\begin{equation}
    \mathcal{E}_{{\rm H}}[n(\mathbf{r})] = \frac{e^2}{2}\int d\mathbf{r}_1 d\mathbf{r}_2 \ n(\mathbf{r}_1)V_{{\rm H}}(\mathbf{r}_1,\mathbf{r}_2)n(\mathbf{r}_2),
\end{equation}
 where the Fourier transformed Coulomb kernel is
\begin{equation}
    V_{{\rm H}}(\mathbf{q})=\frac{e^2}{4\pi\epsilon_0\epsilon_{hBN}}\frac{ 4\pi\sinh{(\beta d_t|\mathbf{q}})|\sinh{(\beta d_b|\mathbf{q}|)}}{\sinh{(\beta(d_t+d_b)|\mathbf{q}|)}|\mathbf{q}|},
\label{eq:gate_screened}
\end{equation}
with $\epsilon_{hBN}$ as the the geometric mean of the perpendicular relative permittivity $\epsilon_{\perp}$ and the parallel relative permittivity $\epsilon_{\parallel}$ of hBN, which we take to be 3.0 and 6.6, respectively; $\beta = \sqrt{\epsilon_{\parallel}/\epsilon_{\perp}}$; $d_t$ and $d_b$ are the thicknesses of the top and bottom hBN layers respectively. This gate screened Coulomb kernel is derived analytically from the setting where two perfectly metallic plates are placed above and below the sample at distances $d_t$ and $d_b$, so there should be discrepancies caused by the existence of the hole in the top graphite gate, although this is gradually taken into account by the iteration procedure described below.

For the second term, the external one-body electric potential term is given by
\begin{equation}
    \mathcal{E}_{{\rm ext}}[n(\mathbf{r})] = e\int  d\mathbf{r} \ \Phi(\mathbf{r})n(\mathbf{r}).
\end{equation}
This potential should contain two different elements: first, this potential should express the external electric potential from the gate voltages, and second, it must contain the negative of the screening effect to compensate for the over-accounted screening effect in Eq.~(\ref{eq:gate_screened}). Therefore, since this term should not contain the Hartree potential in the form of Eq.~(\ref{eq:gate_screened}), as the iterations progress, it should be subtracted as
\begin{equation}
\Phi(\mathbf{r}) = \Phi_{{\rm COMSOL}}(\mathbf{r}) - \frac{1}{e}\int d\mathbf{r}' \ V_{\rm H}(\mathbf{r}, \mathbf{r}')n_{{\rm old}}(\mathbf{r}')    
\end{equation}
where $\Phi_{{\rm COMSOL}}(\mathbf{r})$ is the external potential computed by COMSOL in the presence of the fixed charge density $n_{{\rm old}}(\mathbf{r})$ on the graphene surface.

The third term $\mathcal{E}_{{\rm XC}}[n(\mathbf{r})]$ is taken from the experimental thermodynamic measurements, using data kindly provided by the authors of Ref.~\cite{assouline}. 
(See Fig. 1 of that reference.)
This data contains the exchange-correlation energy and single-particle contributions, but do not include the geometric capacitance contributions. Provided that the density varies slowly on the scale of the magnetic length $l_B$, we may use the local density approximation (LDA)
\begin{equation}
    \mathcal{E}_{{\rm XC}}[n(\mathbf{r})]=XC\int d\mathbf{r}\ E_{{\rm XC}}(n(\mathbf{r})),
\end{equation}
where the exchange-correlation functional $E_{\rm{XC}}$ is defined by the experimentally obtained chemical potential $\mu(n)$ as
\begin{equation}
    E_{\rm{XC}}[n(\mathbf{r})]=\int_0^{n(\mathbf{r})}dn' \ \mu(n').
\end{equation}
and $XC$ is a scaling factor close to $1$. 

The fourth term, $\mathcal{E}_{smooth}$, is the effective energy that penalizes rapidly varying density distributions and suppresses the high-frequency contribution
\begin{equation}
    \mathcal{E}_{smooth}[n(\mathbf{r})] = \frac{\kappa}{2}\int d\mathbf{r} \ (\nabla \nu(\mathbf{r}))^2.
\end{equation}
This term reflects the fact that the density cannot vary rapidly within a scale smaller than $l_B$. $\kappa$ is empirically taken to be $1\, \si{meV}$, which is not too large to erase the fractional density plateau, but large enough to suppress unrealistic density fluctuations. A more detailed determination of this value is left for future work.\\
\\
Based on the equations above, we discretize the system on a square grid whose total area is identical to that used in COMSOL, with the same periodic boundary conditions, and choose the lattice constant to be much smaller than $l_B$. In evaluating $\mathcal{E}_{\rm smooth}$, the finite difference method is used to calculate the density gradients. In the following simulations, we used the $600\, \si{nm}\times600\, \si{nm}$ plane, with a $256 \times 256$ grid, so that the lattice constant is small enough compared to the magnetic length at $B=9.95\,\si{T}$. We employ a basin-hopping global optimizer with local L-BFGS-B minimization to vary $n(\mathbf{r})$ within this minimization and find the lowest energy configuration.

\subsubsection{Iterative procedure}
To obtain the entirely self-consistent result, we iteratively solved the three-dimensional classical electrostatics problem and the two-dimensional Thomas-Fermi problem using the COMSOL module ${\rm Livelink^{TM}} $ for MATLAB. All computations in this work were performed on the FASRC Cannon cluster supported by the FAS Division of Science Research Computing Group at Harvard University. The overall procedure is summarized in the Table~\ref{tab:table1}. 
In the $k$-th iteration, COMSOL solves the Poisson equation with the input of the surface charge density on the graphene from the $(k-1)$th Thomas Fermi calculation. We extract the electric potential at the graphene surface which contains the contribution from the surface charge itself, so before updating the external potential, we subtract the Hartree contribution as described above 
\begin{equation}
    \Phi_{{\rm H},k}(\mathbf{r}) = \Phi_{{\rm COMSOL}}(\mathbf{r})-\frac{1}{e}\int d \mathbf{r'}\ V_{\rm H}(\mathbf{r}, \mathbf{r}')n_{{\rm out},k-1}(\mathbf{r}').
\end{equation}
After this subtraction, we update the output external electric potential via a linear mixing procedure with damping factor $\alpha=0.1$ in this simulation. Then we compute the surface charge density through the Thomas Fermi calculation with the input from the previous COMSOL simulation, and return the output charge density through the same mixing procedure.

\begin{table}[h]
    \centering
    \caption{Iteration Procedure}
    \label{tab:table1}
    \begin{tabular}{|c|c|c|c|} \hline \hline
    \multicolumn{2}{|c|}{COMSOL} & \multicolumn{2}{|c|}{Thomas-Fermi}  \\ \hline
    \multicolumn{2}{|c|}{$k=1$}   &
    \multicolumn{2}{|c|}{$k=1$}\\ \hline
    Input & $n_0(\mathbf{r})=0$ & Input & $\Phi_{{\rm out},1}(\mathbf{r})$ \\
    Compute & $\Phi_{{\rm COMSOL},1}(\mathbf{r})$ &Compute & $n_{{\rm TF},1}(\mathbf{r})$ \\ 
    Update & $\Phi_{{\rm out},1}(\mathbf{r})= \Phi_{{\rm COMSOL},1}(\mathbf{r})$ & Update & $n_{{\rm out},1}(\mathbf{r})=n_{{\rm TF},1}(\mathbf{r})$ \\ \hline
    \multicolumn{2}{|c|}{$k>1$}   &
    \multicolumn{2}{|c|}{$k>1$}\\ \hline
    Input & $n_{{\rm out},k-1}(\mathbf{r})$ & Input & $\Phi_{{\rm out},k}(\mathbf{r})$ \\
    Compute & $\Phi_{{\rm COMSOL},k}(\mathbf{r})$ & Compute & $n_{{\rm TF},k}(\mathbf{r})$ \\ 
    Subtract & $\Phi_{{\rm sub},k}(\mathbf{r})=\Phi_{{\rm COMSOL},k}(\mathbf{r})-\Phi_{{\rm H},k}(\mathbf{r})$ &  \diagbox[]{}{} & \diagbox[]{}{} \\
    Update & 
    \begin{tabular}{c}
         $\Phi_{{\rm out},k}(\mathbf{r})=$  \\
         $(1-\alpha)\Phi_{{\rm out},k-1}+\alpha\Phi_{{\rm sub},k}(\mathbf{r})$
    \end{tabular} 
    & Update & 
    \begin{tabular}{c}
    $n_{{\rm out},k}(\mathbf{r})=$ \\
    $(1-\alpha)n_{{\rm out},k-1}(\mathbf{r})+\alpha n_{{\rm TF},k}(\mathbf{r})$
    \end{tabular}
    \\ \hline
    
    \end{tabular}  
\end{table}

\subsubsection{Density profile results}
We simulate the density profiles with voltage configurations used in the experiment. The goal is to calculate the filling in the center if the AD, and to see if the parallel-plate capacitor approximation used in the main text is valid. We set the cavity to $\nu_{cav}=4/3$ using the voltages in the experiment $V_{{TG}}=0.3630\, \si{V}$, $V_{{BG}}=0.1730\, \si{V}$. We then sweep the bridge gate from $-3 \, \si{V}< V_{brg} < 4 \, \si{V}$ and observe the change in the simulated filling factor of the AD $\nu_{AD}$ in the center of the density profile. In our simulations, we have two unknowns: (1) the exact radius of the AD, as the radius should be larger on the graphene plane than the lithographically defined hole in the graphite gate ($r_{litho} = 100\, \si{nm}$) due to the electric field lines penetrating through $25\, \si{nm}$ hBN, and since this radius can change slightly in different electrostatic configurations in each state, and (2) the scaling factor of the exchange correlation term. To verify the robustness of our results, we vary each of these parameters independently. We plot the filling factor obtained at the center of the AD with respect to the bridge gate voltage at various AD sizes from $r_{AD} = 100-200\, \si{nm}$ (Extended Fig. \ref{fig:rik1}a) as well as at various exchange correlation scaling factors $XC=0.75-1.5$ (Extended Fig. \ref{fig:rik1}b).\\
\\
We calculate the filling obtained through the parallel plate capacitor model following:
\begin{align*}
    \nu_{AD} = \frac{\left(\frac{C}{eA}\right)_{brg} V_{brg} + \left(\frac{C}{eA}\right)_{BG} V_{BG}}{B/\phi_0} \quad \text{ where } \quad \left(\frac{C}{eA}\right)_{brg} = \frac{\epsilon_0}{e(d_{TBN}\epsilon_0/\epsilon_{BN}+d_{vac})}
\end{align*}
where $\epsilon_0, \epsilon_{BN} = 3\epsilon_0$ are the vacuum and BN permittivities and $d_{TBN} = 25 \text{ nm}, d_{vac} = 150 $ nm are the thickness of the top BN and of the vacuum between the top BN and the bridge. We plot the filling obtained through the parallel plate capacitor model with the black dashed line (Extended Fig. \ref{fig:rik1}a, b) and observe that across all the swept parameter ranges, the simulated filling $\nu_{AD}$ agrees well with a simple geometrical capacitor model presented in the method section. Therefore, the assignment of the filling in the main text is supported through our Thomas-Fermi calculations.\\
\\
An example density profile is plotted for the settings $V_{brg}=1\, \si{V}$, $B=10\, \si{T}$, which correspond approximately to the experimental settings used for $\nu_{AD} = 2/3$. We observe two compressible rings: one between $\nu_{AD}\sim4/3$ and $\nu_{AD}\sim 1$, and one between $\nu=1$ and $\nu=2/3$ (Extended Fig. \ref{fig:rik1}c,d).

\subsubsection{Bridge/graphene capacitance calculation}
To confirm our experimental findings of the graphene/bridge capacitance, we extract it with our simulations. The total charge is integrated over the AD to extract the number of electrons populating the AD as we sweep the bridge gate voltage. Extended Fig.~\ref{fig:rik1}e shows the relationship between the bridge voltage and the number of electrons contained in the AD. We extract the capacitance from the slope, according to $C_{brg-AD} = \Delta q_{tot}/\Delta V_{brg}$ (Extended Fig.~\ref{fig:rik1}f). This procedure is repeated over various possible AD radii $r_{AD} = 100-200\, \si{nm}$. To compare the simulation with our experimental data, we calculate the change in voltage required to add either an electron ($\Delta Q=e$) or a fractional quasiparticle ($\Delta Q=e/3$) with the simulated capacitance according to $\Delta s = \Delta Q/C_{brg-AD}$. We plot the result and observe that the simulated jump spacings $\Delta s$ are in good agreement with our experimental results for $\Delta Q=e/3$ and $r_{AD} \sim 150-200\, \si{nm}$. This radius range is also the range that had the highest agreement with the parallel-plate capacitor model above (Extended Fig.~\ref{fig:rik1}a, ~\ref{fig:rik1}b). We therefore conclude that the simulations agree with fractional charge $\Delta Q=e/3$ jumping into the AD.

\end{document}